\documentclass[aps,prx,twocolumn,superscriptaddress]{revtex4-2}
\usepackage{graphicx}
\usepackage{color}
\usepackage{amsfonts}
\usepackage[normalem]{ulem}
\usepackage{easyReview}
\usepackage{siunitx}
\usepackage{amsmath}

\definecolor{brown}{RGB}{200,100,0}

\def\WS2{WS$_2$}
\def\MoS2{MoS$_2$}{

\begin{document}

\title{Ultrafast electronic line width broadening in the {C} $1s$ core level of graphene}

\author{ Davide Curcio}
\affiliation{Department of Physics and Astronomy, Interdisciplinary Nanoscience Center, Aarhus University, Ny Munkegade 120, 8000 Aarhus C, Denmark}

\author{Sahar Pakdel}
\affiliation{Department of Physics and Astronomy, Interdisciplinary Nanoscience Center, Aarhus University, Ny Munkegade 120, 8000 Aarhus C, Denmark}

\author{Klara Volckaert}
\affiliation{Department of Physics and Astronomy, Interdisciplinary Nanoscience Center, Aarhus University, Ny Munkegade 120, 8000 Aarhus C, Denmark}

\author{Jill A. Miwa}
\affiliation{Department of Physics and Astronomy, Interdisciplinary Nanoscience Center, Aarhus University, Ny Munkegade 120, 8000 Aarhus C, Denmark}

\author{S\o ren Ulstrup}
\affiliation{Department of Physics and Astronomy, Interdisciplinary Nanoscience Center, Aarhus University, Ny Munkegade 120, 8000 Aarhus C, Denmark}

\author{Nicola Lanat\`a}
\affiliation{Department of Physics and Astronomy, Aarhus University, 8000, Aarhus C, Denmark}
\affiliation{Nordita, KTH Royal Institute of Technology and Stockholm University, Roslagstullsbacken 23, 10691 Stockholm, Sweden}

\author{Marco Bianchi}
\affiliation{Department of Physics and Astronomy, Interdisciplinary Nanoscience Center, Aarhus University, Ny Munkegade 120, 8000 Aarhus C, Denmark}

\author{Dmytro Kutnyakhov}
\affiliation{Deutsches Elektronen-Synchrotron DESY, Notkestr. 85, 22607 Hamburg, Germany}

\author{Federico Pressacco}
\affiliation{Deutsches Elektronen-Synchrotron DESY, Notkestr. 85, 22607 Hamburg, Germany}

\author{G\"unter Brenner}
\affiliation{Deutsches Elektronen-Synchrotron DESY, Notkestr. 85, 22607 Hamburg, Germany}

\author{Siarhei Dziarzhytski}
\affiliation{Deutsches Elektronen-Synchrotron DESY, Notkestr. 85, 22607 Hamburg, Germany}

\author{Harald Redlin}
\affiliation{Deutsches Elektronen-Synchrotron DESY, Notkestr. 85, 22607 Hamburg, Germany}

\author{Steinn Agustsson}
\affiliation{Johannes Gutenberg-Universit\"at, Institut f\"ur Physik, 55099, Mainz, Germany}

\author{Katerina Medjanik}
\affiliation{Johannes Gutenberg-Universit\"at, Institut f\"ur Physik, 55099, Mainz, Germany}

\author{Dmitry Vasilyev}
\affiliation{Johannes Gutenberg-Universit\"at, Institut f\"ur Physik, 55099, Mainz, Germany}

\author{Hans-Joachim Elmers}
\affiliation{Johannes Gutenberg-Universit\"at, Institut f\"ur Physik, 55099, Mainz, Germany}

\author{Gerd Sch\"onhense}
\affiliation{Johannes Gutenberg-Universit\"at, Institut f\"ur Physik, 55099, Mainz, Germany}

\author{Christian Tusche}
\affiliation{Peter Gr\"unberg Institut PGI, Forschungszentrum J\"ulich and JARA- Fundamentals of Future Information Technologies, 52425 J\"ulich, Germany}

\author{Ying-Jiun Chen}
\affiliation{Peter Gr\"unberg Institut (PGI-6), Forschungszentrum J\"ulich, 52425
J\"ulich, Germany}
\affiliation{Fakult\"at f\"ur Physik, Universit\"at Duisburg-Essen, 47057 Duisburg,
Germany}

\author{Florian Speck}
\affiliation{Institut f\"ur Physik, Technische Universit\"at Chemnitz, Reichenhainer Str. 70, Chemnitz 09126, Germany}

\author{Thomas Seyller}
\affiliation{Institut f\"ur Physik, Technische Universit\"at Chemnitz, Reichenhainer Str. 70, Chemnitz 09126, Germany}

\author{Kevin B\"uhlmann}
\affiliation{Department of Physics, Laboratory for Solid State Physics, ETH Z\"urich, Otto-Stern-Weg 1, 8093 Z\"urich, Switzerland}

\author{Rafael Gort}
\affiliation{Department of Physics, Laboratory for Solid State Physics, ETH Z\"urich, Otto-Stern-Weg 1, 8093 Z\"urich, Switzerland}

\author{Florian Diekmann}
\affiliation{Institute of Experimental and Applied Physics, Kiel University, 24098 Kiel, Germany}

\author{Kai Rossnagel}
\affiliation{Institute of Experimental and Applied Physics, Kiel University, 24098 Kiel, Germany}
\affiliation{Ruprecht Haensel Laboratory, Deutsches Elektronen-Synchrotron DESY, 22607 Hamburg, Germany}

\author{Yves Acremann}
\affiliation{Department of Physics, Laboratory for Solid State Physics, ETH Z\"urich, Otto-Stern-Weg 1, 8093 Z\"urich, Switzerland}

\author{Jure Demsar}
\affiliation{Johannes Gutenberg-Universit\"at, Institut f\"ur Physik, 55099, Mainz, Germany}

\author{Wilfried Wurth}
\affiliation{Deutsches Elektronen-Synchrotron DESY, Notkestr. 85, 22607 Hamburg, Germany}
\affiliation{Center for Free-Electron Laser Science CFEL, Hamburg University, Luruper Chausee 149, 22761 Hamburg, Germany}
\email{Deceased.}

\author{Daniel Lizzit}
\affiliation{Elettra-Sincrotrone Trieste S.C.p.A., Strada Statale 14 - km 163.5 in AREA Science Park, 34149 Trieste, Italy}

\author{Luca Bignardi}
\affiliation{Elettra-Sincrotrone Trieste S.C.p.A., Strada Statale 14 - km 163.5 in AREA Science Park, 34149 Trieste, Italy}
\affiliation{Current affiliation: Department of Physics, University of Trieste, Via Valerio 2, 34127 Trieste, Italy}

\author{Paolo Lacovig}
\affiliation{Elettra-Sincrotrone Trieste S.C.p.A., Strada Statale 14 - km 163.5 in AREA Science Park, 34149 Trieste, Italy}

\author{Silvano Lizzit}
\affiliation{Elettra-Sincrotrone Trieste S.C.p.A., Strada Statale 14 - km 163.5 in AREA Science Park, 34149 Trieste, Italy}

\author{Charlotte E. Sanders}
\affiliation{Central Laser Facility, STFC Rutherford Appleton Laboratory, Harwell OX11 0QX, United Kingdom}

\author{ Philip Hofmann}
\affiliation{Department of Physics and Astronomy, Interdisciplinary Nanoscience Center, Aarhus University, Ny Munkegade 120, 8000 Aarhus C, Denmark}
\email{philip@phys.au.dk}

\begin{abstract}
    Core level binding energies and absorption edges are at the heart of many experimental techniques concerned with element-specific structure, electronic structure, chemical reactivity, elementary excitations and magnetism.
    X-ray photoemission spectroscopy (XPS) in particular, can provide information about the electronic and vibrational many-body interactions in a solid as these are reflected in the detailed energy distribution of the photoelectrons.
    Ultrafast pump-probe techniques add a new dimension to such studies, introducing the ability to probe a  transient state of the many-body system.
    Here we use a free electron laser to investigate the effect of a transiently excited hot electron gas on the core level spectrum of graphene, showing that it leads to a substantial broadening of the C~$1s$ line shape.
    Confirming a decade-old prediction, the broadening is found to be caused by an exchange of energy and momentum between the photoemitted core electron and the hot electron system, rather than by vibrational excitations.
    This interpretation is supported by a quantitative line shape analysis that accounts for the presence of the excited electrons.
    Fitting the spectra to this model directly yields the electronic temperature of the system, in good agreement with electronic temperature values obtained from valence band data.
    Furthermore, making use of time- and momentum-resolved C~$1s$ spectra, we illustrate how the momentum change of the outgoing core electrons leads to a small but detectable change in the time-resolved photoelectron diffraction pattern and to a nearly complete elimination of the core level binding energy variation associated with the presence of a narrow $\sigma$-band in the C~$1s$ state.
    The results demonstrate that the XPS line shape can be used as an element-specific and local probe of the excited electron system and that X-ray photoelectron diffraction investigations remain feasible at very high electronic temperatures.
\end{abstract}
\maketitle

\begin{figure*}[t!]
    \begin{center}
        \includegraphics[width=\textwidth]{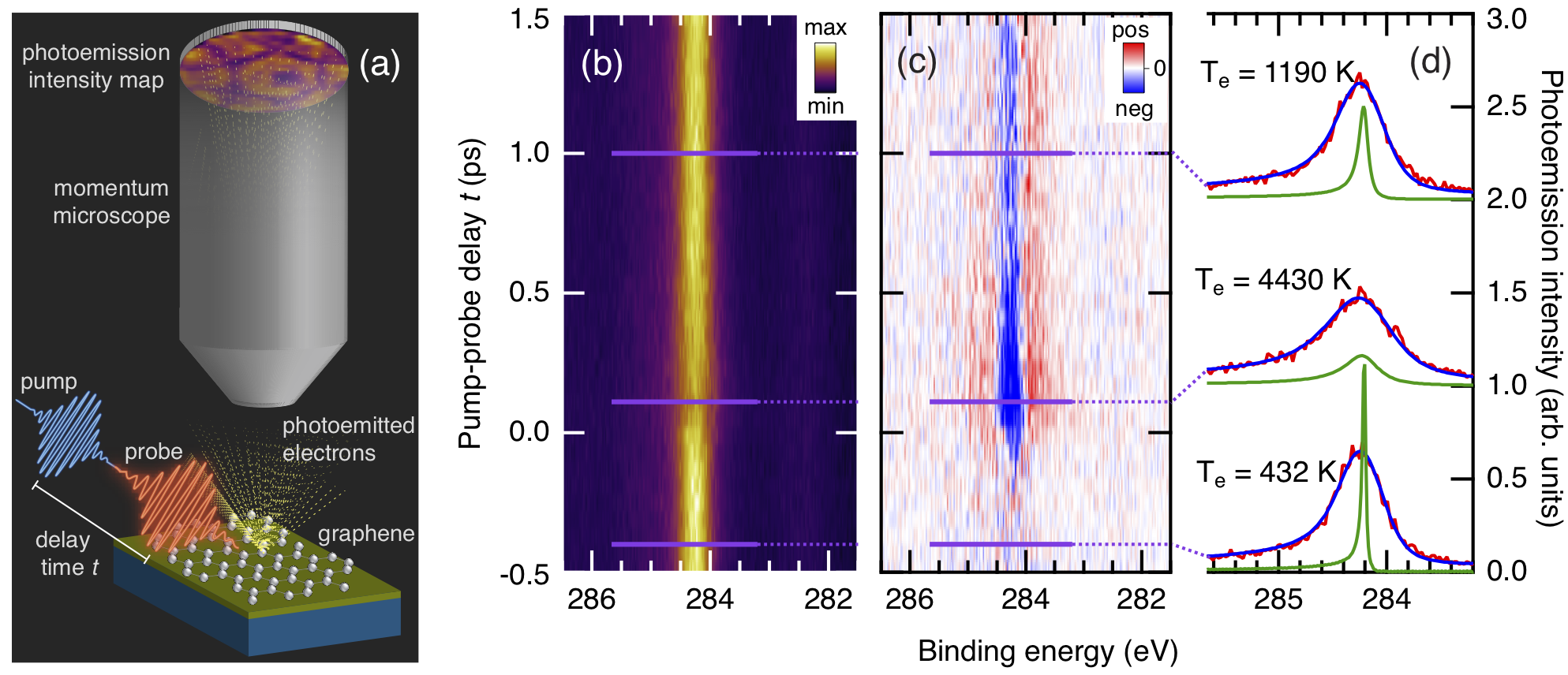}
        \caption{(a) Sketch of the setup used in the experiment.
            The time-resolved photoemission intensity is measured from the graphene C~$1s$ core level and valence band by combining an infrared pump pulse with a soft X-ray probe pulse from FLASH.
            The electrons are detected over a wide angular range using a momentum microscope.
            (b) C~$1s$ photoemission intensity as a function of pump-probe delay and binding energy.
            (c) Same as in (b) but with the average spectrum before excitation subtracted.
            Red and blue indicate an increase and decrease in electron counts, respectively.
            (d) Selected spectra (cuts at specific time delays) along with best fit to the model described in the text (blue line).
            The temperature-dependent asymmetry kernel for each fit is shown in green.
            The electronic temperatures resulting from the fit are noted close to the spectra.}
        \label{fig:1}
    \end{center}
\end{figure*}

\section{Introduction}
XPS is an experimental tool capable of accurately determining the core level binding energies of atoms~\cite{Nordling:1957aa}.
This energy is not only element-specific, but also provides detailed information on the atoms' oxidation state and their chemical environment, making XPS an essential tool for studying a wide range of phenomena, from band bending in semiconductors to chemical analysis and heterogeneous catalysis.

Due to the relative simplicity of the core level photoemission process, XPS line shapes have also attracted considerable theoretical interest.
In the 1970s, XPS was used as a testing ground for new many-body theories, with the objective of understanding the detailed XPS line shape in terms of a solid's electronic and vibrational many-body response to the removal of the photoelectron (see, \textit{e.g.}, Ref.~\cite{Citrin:1977aa} and references therein).
In the resulting simplified picture, the XPS line shape has a Lorentzian contribution stemming from the finite core hole lifetime and a Gaussian contribution due to vibrational excitations in the photoemission process.
The Gaussian width depends on the initial state phonon population and increases with temperature.
Electronic excitations can result in satellites, such as replicas of the main XPS peak displaced by the plasmon energy, and in asymmetric line shapes due to the excitation of electron-hole pairs in metals~\cite{Doniach:1970aa}.
A temperature-dependence of the latter process is usually neglected because $k_B T$ is typically much smaller than the XPS line width.
However, it has been predicted almost 40 years ago that such electron-hole pair excitations could affect the XPS line width and shape for high (electronic) temperatures~\cite{Satpathy:1982aa}.
This has, to the best of our knowledge, never been confirmed, presumably because such electronic effects are masked by vibrational broadening at high sample temperatures.

The advent of ultrafast pump-probe XPS at free electron laser (FEL) sources now offers the opportunity to separate electronic and vibrational contributions to the XPS line shape since either the electronic or specific vibrational degrees of freedom can be addressed by choosing an appropriate pump energy \cite{Torre:2021wg,Zong:2021ui}.
On the other hand, high-resolution line shape studies are highly challenging at FELs due to the low repetition rate and the resulting need to balance an acceptable count rate against space charge effects~\cite{Passlack:2006aa,Pietzsch:2008aa,Hellmann:2009aa,Hellmann:2012ab,Schonhense:2021aa}.
Recently, a detailed XPS line shape investigation has been presented for pumped WSe$_2$, tracking an excitonic Mott transition in the XPS line shape~\cite{Dendzik:2020aa}.

Here we study the ultrafast evolution of the C~$1s$ line shape, intensity, and binding energy in graphene upon pumping the electronic system to high temperatures.
We benefit from the fact that this transient electronic state is well understood from ultrafast valence band spectroscopy~\cite{Gierz:2013aa,Johannsen:2013ab,Johannsen:2015aa,Aeschlimann:2017aa,Rohde:2018ab} and theory~\cite{Caruso:2020aa}.
In short, after excitation with an infrared pump pulse, the electrons thermalize within a few tens of~fs.
The hot electron gas in the Dirac cone of graphene initially loses energy via strongly coupled optical phonons but this process quickly ceases to be efficient due to the Dirac point representing a bottleneck for the hot electron decay.
This opens the possibility of studying the effect of a hot electron gas on the core level spectrum while the lattice remains essentially at equilibrium, a regime that cannot be reached by static heating~\cite{Pozzo:2011aa}.

This paper is structured as follows: After a brief description of the experimental details, we show how the ultrafast excitation of the electrons leads to a broadening of the C~$1s$ core level line.
We extend a theoretical description of the core level line shape to allow for the exchange of energy and momentum between the excited core electron and a hot electron gas.
This leads to an excellent fit of the data, confirming a decade-old prediction of such a mechanism~\cite{Satpathy:1982aa}.
The model even permits a determination of the electronic temperature from the core level spectra and we demonstrate that the results agree with electronic temperature values obtained from the broadening of the Fermi-Dirac distribution in the valence band.
We then show that the novel electronic broadening mechanism has a surprisingly small effect on the momentum-resolved C~$1s$ intensity distribution, the so called X-ray photoelectron diffraction (XPD) pattern.
Finally, we investigate the time-dependent changes of the momentum-resolved C~$1s$ binding energy.
For localized core states, the binding energy is usually assumed to be independent of the crystal momentum but for graphene this is not the case due to a narrow $\sigma$-band formed by the C~$1s$ states~\cite{Lizzit:2010aa}.
We argue that the electronic final state scattering can be expected to remove the subtle binding energy variations associated with this band dispersion and we show that this is indeed the case.

\section{Experiment}

Time- and angle-resolved XPS and valence band photoemission experiments were performed on the PG2 beamline of the FLASH FEL~\cite{Kutnyakhov:2020aa} using hydrogen-intercalated quasi free-standing monolayer graphene on SiC~\cite{Riedl:2009aa,Speck:2011aa} with a hole doping density of \SI{5e12}{cm^{-2}}.
The sample was annealed in ultra high vacuum (base pressure better than \SI{5e-10}{mbar}) at \SI{470}{K} for \SI{10}{min}.
The sample was held at room temperature during all measurements.
Photoelectrons were detected using a momentum microscope covering a momentum range of \SI{6.2}{\angstrom^{-1}} with a resolution of \SI{0.06}{\angstrom^{-1}}~\cite{Kutnyakhov:2020aa} (see sketch in Fig.~\ref{fig:1}(a)).
The analysis of the raw data followed the procedures outlined in Ref.~\cite{Xian:2020aa} and in Appendix~\ref{appendix:datacube}.
The pump and probe photon energies were \SI{1.55}{eV} and \SI{337.5}{eV} (the third harmonic of the FEL), respectively, for the C~$1s$ data.
The valence band was probed using the first harmonic of the FEL at \SI{112.5}{eV}.
The pump fluence was $\approx$~\SI{0.5}{mJcm^{-2}}, and the polarization for both the pump and the probe was linear p-type, with an incidence angle of \SI{68}{\degree} off-normal, resulting in negligible laser-assisted photoemission replicas with relative intensities of \SI{1.3}{\percent} and \SI{2.8}{\percent}~\cite{Keunecke:2020aa} for the C~$1s$ and valence band spectra, respectively.
The overall time and energy resolution were \SI{210}{fs} and \SI{190}{meV}.
Static XPS and XPD data were collected as references at the SuperESCA beamline of Elettra,~\cite{Abrami:1995aa} using the same sample and same excitation photon energy, and with an overall energy and angular resolution of better than 50 meV and $\approx$ \SI{3}{\degree}, respectively.
The momentum microscope data is known to suffer from space-charge induced energy shifts, especially close to the normal emission axis:
These can be corrected based on such reference data~\cite{Schonhense:2018aa}.

\section{Results and Discussion}
Figure~\ref{fig:1}(b)-(d) show the time-dependent C~$1s$ core level intensity as a function of binding energy and pump-probe time delay $t$, the difference between the time-dependent photoemission intensity and the average spectrum before the excitation, and spectra at selected time delays.
The data shown in the figure was collected from a $k$-range of \SI{0.35}{\angstrom^{-1}} around the $M$ point of graphene at ${\bf k}=(-2.50 $~\AA$^{-1},-0.05$~\AA$^{-1})$ (location marked in Figure~\ref{fig:S5} in Appendix~\ref{appendix:Mpnts}), where space-charge effects are minimal due to the distance from normal emission~\cite{Schonhense:2018aa}.
The C~$1s$ spectrum from quasi free-standing monolayer graphene on SiC shows two components, separated by \SI{1.79}{eV} in binding energy: They are found at \SI{282.59(1)}{eV} and at \SI{284.38(1)}{eV} (see Appendix~\ref{appendix:spectraC1s}).
The low binding energy component can be assigned to the carbon atoms in the SiC substrate and the high binding energy component to the graphene layer~\cite{Coletti:2011aa}.
The chosen X-ray photon energy renders the experiment very surface-sensitive, suppressing the SiC C~$1s$ peak.
In the following, we will only discuss the graphene C~$1s$ peak.
The pump-induced excitation of graphene around $t=0$ is clearly reflected in a broadening of this peak.
This is most evident in the difference plot and the individual spectra; see Figure~\ref{fig:1}(c) and (d).
The pump-induced broadening is a large effect: it increases the C~$1s$ line width by almost \SI{280}{meV}, from \SI{560}{meV} to \SI{840}{meV} at peak excitation, recovering to \SI{580}{meV} in our measured pump-probe delay window.

In order to quantify the pump-induced line shape changes, we fit a model for the core level photoemission intensity to the data.
For insulators at equilibrium, an appropriate choice of line profile is a Voigt function in which the Lorentzian line width reflects the core hole lifetime while the phonon contribution gives rise to a Gaussian broadening.
The experimental energy resolution is also assumed to lead to a Gaussian broadening and can thus be treated along with the phonon contribution.
In metals, the photoemission intensity shows a tail to lower kinetic energies (high binding energies) due to the possibility of exciting electron-hole pairs upon the sudden removal of the core electron.
For simple metals with a roughly constant density of states near $E_F$, this is captured in the frequently used Doniach-\v Sunji\' c line shape~\cite{Doniach:1970aa}.
The C~$1s$ line shape of graphene also shows such an asymmetry.

An appropriate model for the data in Fig.~\ref{fig:1}, and in general for out-of-equilibrium metals, should take into account that high electronic temperatures are reached in the pump process, leading to a smearing of the Fermi-Dirac distribution over an energy range of more than one electron volt~\cite{Johannsen:2013ab}.
This has two important consequences: (1) Given the substantial population of states above $E_F$, the possible excitation of electrons in the photoemission process is no longer restricted to electrons below $E_F$, and (2) the outgoing electron can not only lose energy by exciting an electron-hole pair, but also \textit{gain} energy upon electron-hole recombination~\cite{Satpathy:1982aa}.
Thus, the asymmetric line-shape in equilibrium  is expected to broaden and become more symmetric in the highly excited state.

We include these effects into the model for the core level photoemission intensity proposed by Hughes and Scarfe~\cite{Hughes:1996ab}, of which the Doniach-\v Sunji\' c line shape is a special case.
Their expression for the energy-dependent photoemission intensity $I(E)$ of a core level at energy $E_0$ is
\begin{align}
    I(E) \propto \int_{-\infty}^{+\infty} e^{-iEt}e^{-iE_0t}e^{-\lambda |t|}e^{-\frac{\sigma^2t^2}{2}} \label{equ:1} \\
    \exp \left( \int_{0}^{+\infty} J(E') \frac{e^{iE't}-1}{E'^2} dE' \right)dt,\nonumber
\end{align}
where $\lambda$ and $\sigma$ are the parameters controlling the Lorentzian and Gaussian width.
The last integral in the exponential describes the possible energy losses, resulting from the density of possible excitations $J(E')$ given by
\begin{equation}
    J(E')=a^2\int_{-\infty}^{+\infty}{D_{filled}(\epsilon)D_{empty}(\epsilon+E')} d\epsilon,
\end{equation}
with $D_{filled/empty}(\epsilon)$ being the electronic density of filled/empty states, and $a$ the asymmetry parameter.
We now include finite temperature effects by setting the lower integration limit in the last integral of equation (\ref{equ:1}) to $-\infty$ in order to also allow for energy gains, and modifying the density of excitations to account for the temperature dependence of transitions between the filled initial and the empty final states:
\begin{multline}
    J(E')=\\
    a^2\int_{-\infty}^{+\infty}{D(\epsilon)f(\epsilon,T_e)D(\epsilon+E')(1 - f(\epsilon + E',T_e))d\epsilon},
    \label{equ:3}
\end{multline}
where $T_e$ is the electronic temperature and $f(\epsilon,T_e)$ the Fermi-Dirac distribution.

The fit to the data in Figure~\ref{fig:1} is performed by fixing $\lambda$ and $a$ obtained from the un-pumped data in order to minimize space charge effects, $\sigma$ and $E_0$ from data at negative time delays, and using the electronic temperature $T_e$ as the only line shape fit parameter for the entire series of pumped data.
The density of states $D(\epsilon)$ is taken to be constant.
The simple model leads to an excellent fit to the data, as shown for the example spectra in Fig.~\ref{fig:1}(d).
We not only show the actual fit but also the temperature-dependent asymmetry kernel (shifted by $E_0$, green curves in Figure~\ref{fig:1}(d)), which is the Fourier transform of the last term in equation~(\ref{equ:1}) and the only part of the expression that depends on $T_e$.
For the un-pumped system, the asymmetry kernel is strongly peaked and its intensity is largely found at energies higher than the peak binding energy, giving rise to the characteristic asymmetric line shape.
This is expected because the inelastic processes are completely dominated by energy losses.
At peak excitation, on the other hand, the electronic temperature reaches \SI{4200}{K} and the asymmetry kernel is much broader.
Also, its median moves to lower binding energy, emphasizing the contribution of processes involving an energy gain.

Note that an asymmetric XPS line shape for weakly doped graphene is commonly observed~\cite{Riedl:2009aa,Lizzit:2010aa} but not naively expected.
After all, the density of states at the Fermi level is approaching zero, leaving only a very small phase space for the inelastic excitations needed to generate a low energy tail in the spectrum.
The observed asymmetry has been explained by the fact that in graphene not only electron-hole pair excitations but also plasmon generation can contribute to the asymmetric tail of the C~$1s$ line~\cite{Sernelius:2015ab}.
Our assumed constant $D(\epsilon)$ and the resulting $J(E')$ should therefore not be taken as the true density of electronic states in graphene but rather as a phenomenological way of establishing the most simple model for a joint density of possible excitations able to fit the entire data set.
Indeed, as we show in Appendix ~\ref{appendix:dos}, it is not possible to obtain a reasonable fit the the C~$1s$ spectrum when using a linear $D(\epsilon)$, not even for the high-resolution equilibrium spectrum.

\begin{figure}[htb!]
    \begin{center}
        \includegraphics[width=0.45\textwidth]{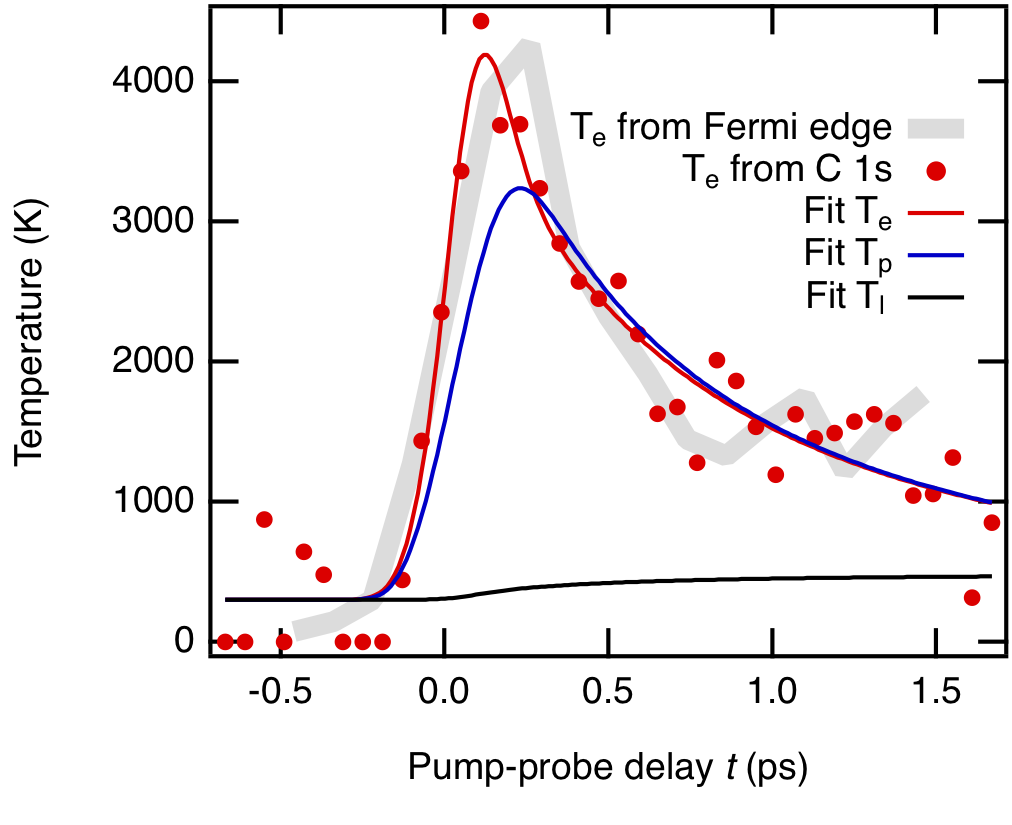}
        \caption{
            Time-dependent electronic temperature extracted from the fit of the C~$1s$ core level data in Figure~\ref{fig:1}(b) (red markers).
            The red solid line is a fit to the electronic temperature $T_e$ using a three-temperature model, giving the temperatures of the strongly coupled phonons $T_p$ (blue line) and the remaining lattice $T_l$ (black line).
            The gray, thick line, corresponds to the $T_e$ obtained from an analysis of the Fermi-Dirac distribution in the valence band.
        }
        \label{fig:2}
    \end{center}
\end{figure}

Figure~\ref{fig:2} shows the time-dependent electronic temperature resulting from the fit.
It is qualitatively similar to previous experimental~\cite{Gierz:2013aa,Johannsen:2013ab} and theoretical~\cite{Caruso:2020aa} results from the same system.
This can be further confirmed by the electronic temperature directly obtained from the Fermi-Dirac distribution in the valence band, using the approach outlined in Refs.~\cite{Johannsen:2013ab,Ulstrup:2014aa} (gray line in Fig.~\ref{fig:2}).
Note that due to the high photon energy used here, the valence band data quality is somewhat inferior to the results in Refs.~\cite{Gierz:2013aa,Johannsen:2013ab}.
The electronic temperature after excitation obtained from the core level data is well-described by a double exponential decay with time constants of $\tau_1 =$\SI{170(50)}{fs} and $\tau_2 =$\SI{1.3(2)}{ps}, also similar to literature values~\cite{Gierz:2013aa,Johannsen:2013ab,Johannsen:2015aa}.

For a more detailed analysis, we fit the electronic temperature from the C~$1s$ spectra using the three-temperature model described in Ref.~\cite{Johannsen:2013ab}.
This yields a good fit to $T_e$, along with the time-dependent temperatures of the strongly coupled optical phonons ($T_p$) and the acoustic phonon bath ($T_l$).
The result of this fit is also shown in Fig.~\ref{fig:2}.
The coupling constants we find are $\lambda_1=$\SI{0.06(4)} (coupling to strongly coupled optical phonons) and $\lambda_2=$\SI{0.0029(6)} (coupling to acoustic phonons), similar to what has been reported for lightly doped graphene~\cite{Ulstrup:2012aa,Johannsen:2013ab,Calandra:2007aa}.

The excellent fit to the data, the $T_e$ values obtained from the fit, and the $\mathbf{k}$-dependent C~$1s$ intensity and energy variations discussed later in the paper all support the mechanism of electronic line shape broadening introduced here.
However, one may ask about the importance of the conventional temperature-dependent phonon broadening.
After all, $T_p$ of the strongly coupled optical phonons also reaches \SI{3200}{K} and this happens almost simultaneously with $T_e$ reaching its maximum.
While coupling to phonons might play some role, we expect the effect to be quite small due to the high energies of the optical phonons and due to the fact that only the strongly coupled phonons at $\Gamma$ and $K$ are excited.
As a first simple estimate, we can follow the treatment of Ref.~\cite{Citrin:1977aa} and multiply the intrinsic low-temperature Gaussian contribution to the line width of $\approx$\SI{110}{meV}~\cite{Pozzo:2011aa} with the statistical Bose-Einstein broadening factor for a phonon at $\approx$\SI{200}{meV} at $T_p=$\SI{3200}{K}, giving rise to total line width of less than \SI{190}{meV}.
Alternatively, we can extrapolate the measured temperature-dependent line width from equilibrium XPS data in Ref.~\cite{Pozzo:2011aa} to \SI{3200}{K}, resulting in less than \SI{400}{meV}.
Note that this strongly overestimates the broadening because it corresponds to a situation in which \textit{all} phonons are excited to \SI{3200}{K}.
Still, the resulting broadening is much smaller than the one observed here.

Summarizing at this point, we have demonstrated that selectively increasing the electronic temperature of a material on an ultrafast time scale can lead to a dominant temperature-dependent electronic broadening mechanism in XPS, similar to that predicted in Ref.~\cite{Satpathy:1982aa}.
Moreover, including this broadening mechanism in the description of the XPS line shape gives direct access to the sample's electronic temperature, providing an alternative to measuring the broadening of the Fermi-Dirac distribution \cite{Ulstrup:2014aa,Matveev:2019aa}. 

\begin{figure}[htb!]
    \begin{center}
        \includegraphics[width=0.5\textwidth]{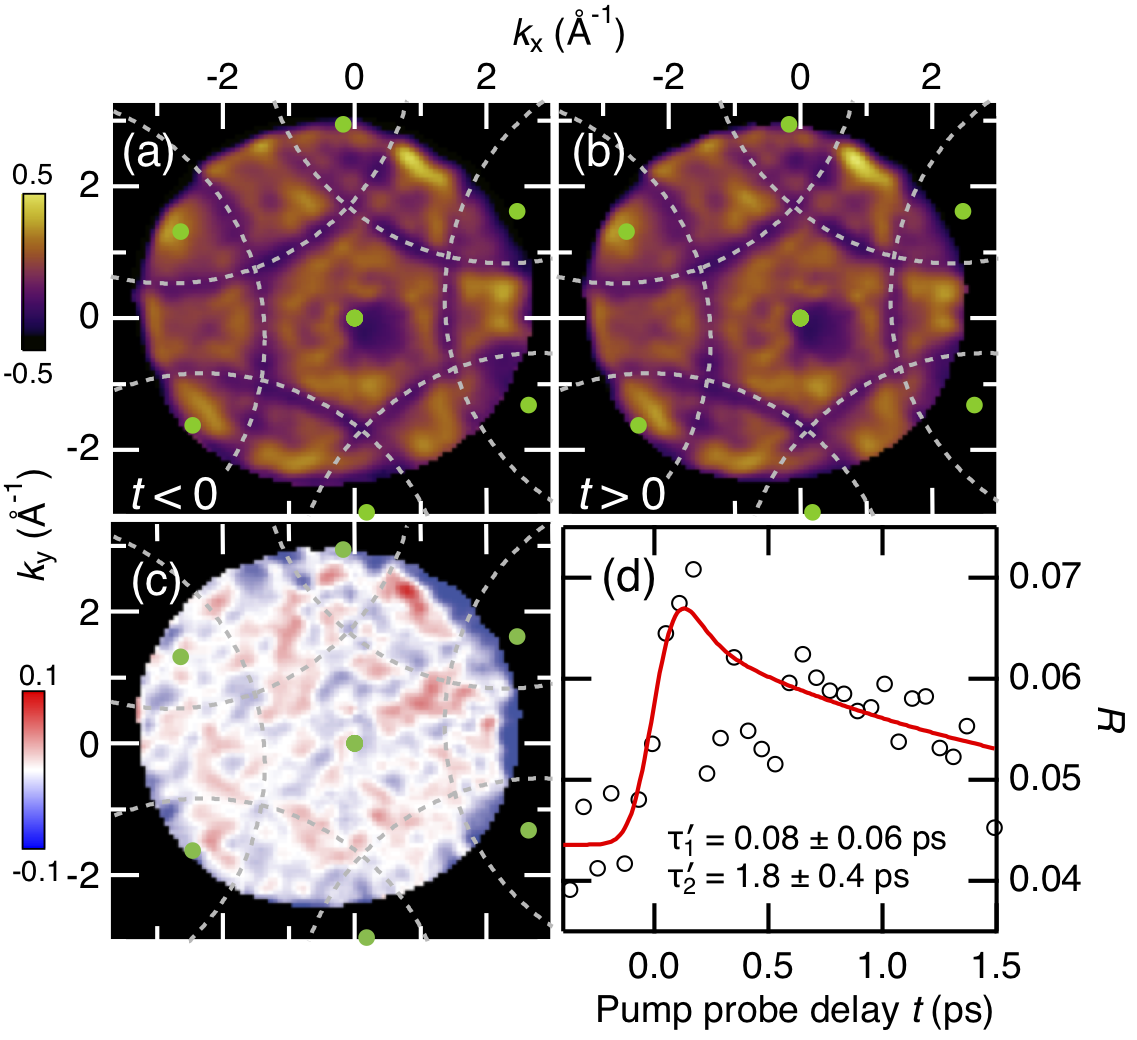}
        \caption{
            (a) C~$1s$ photoemission intensity integrated over negative pump-probe delay times, displayed as a modulation function in momentum space.
            The $\Gamma$ points are shown in green.
            The dashed lines emphasize regions of low photoemission intensity (``dark lines'') arising from a combination of scattering by the graphene reciprocal lattice and reflection at the surface potential.
            (b) The same as in (a), but integrated over positive time delays.
            (c) Difference of the modulation functions in (a) and (b).
            (d) Comparison between the time-dependent modulation function and the equilibrium modulation function, quantified by $R(t)$.
            A fit to the data using a double decaying exponential convoluted with a Gaussian is shown in red.
        }
        \label{fig:3}
    \end{center}
\end{figure}

A momentum microscope simultaneously detects the photoemitted electrons over a large two-dimensional $\mathbf{k}$-range.
We can thus exploit this to further explore the effect of the strong electronic final state scattering on the $\mathbf{k}$-resolved photoemission intensity and core level binding energy.
Figure~\ref{fig:3}(a) shows the C~$1s$ core level intensity, integrated for the time interval before the arrival of the pump pulse.
The data is displayed in terms of a so-called modulation function $\chi$, a quantity commonly used in XPD measurements~\cite{Woodruff:2002aa} (for a definition see Appendix~\ref{appendix:rfactor}).
In XPD, intensity modulations arise from the interference between the part of the electron wave field reaching the detector directly and the parts that are elastically scattered by the atoms surrounding the emitting carbon atom.
A comparison between these intensity modulations and multiple scattering calculations can be used for a local and element-specific structural determination~\cite{Woodruff:2002aa}.
This has been carried out for epitaxial graphene~\cite{Lizzit:2010aa} (see also Appendix~\ref{appendix:EDAC}).
In addition to the scattering-induced XPD features, the photoemission intensity in Fig.~\ref{fig:3}(a) contains a dominant contribution from the curved ``dark lines'' emphasized by dashed lines.
These are due to an interference effect involving scattering by the graphene reciprocal lattice combined with reflection at the surface potential boundary~\cite{Winkelmann:2012aa}.

We can probe the effect of ultrafast electronic final state excitations on the XPD pattern by comparing the XPD patterns after excitation to those at equilibrium~\cite{Greif:2016aa,Ang:2020aa}.
To this end, Fig.~\ref{fig:3}(b) shows the modulation function integrated over positive time delays and Fig.~\ref{fig:3}(c) gives the difference between (a) and (b).
Clearly, the excitation-induced changes are extremely small and the difference appears to be largely noise.
From a mere inspection of these data, we can conclude that the electronic final state line shape broadening does not preclude a time-dependent structural determination by XPD.

A more quantitative analysis of the XPD pattern's time evolution can be performed by introducing $R(t)=\sum_i (\chi_i - \chi_i(t))^2 / (\sum_i (\chi_i^2 + \chi_i(t)^2 )$, where $\chi$ and $\chi(t)$ are the modulation functions at equilibrium and delay time $t$, respectively, and the sum runs over all the points in the diffraction pattern.
The resulting $R(t)$ is shown in Fig.~\ref{fig:3}(d).
As expected, the time-dependent variation of $R(t)$ is small but the pump-induced excitation is still clearly visible.
Moreover, its time dependence is very similar to that of the electronic temperature and we can fit $R(t)$ by a Gaussian pulse with a double exponential decay, using time constants of $\tau'_1 =$\SI{0.08(6)}{ps} and $\tau'_2 =$\SI{1.8(4)}{ps}.

A change of the XPD pattern following the excitation of the system can be expected for purely structural reasons.
After all, the rise of $T_e$ is immediately followed by the excitation of optical phonons and an (eventually) increased temperature of the lattice is certain to lead to a Debye-Waller-like blurring of the diffraction features.
However, given the fact that the behavior of $R(t)$ is extremely similar to that of $T_e$ with even quantitative agreement of the time constants, an electronic origin of the observed change in $R(t)$ appears more likely.
Moreover, only few high-energy optical phonons are populated to reach high values of $T_p$~\cite{Caruso:2020aa} and the lattice temperature $T_l$ does not increase significantly during the time interval observed here (see Fig.~\ref{fig:2}).
Finally, an anharmonic change of the average lattice constant can be ruled out by the fact that the position of the dark lines in the diffraction pattern, and thus the size of the reciprocal lattice vectors, remains fixed.
Indeed, any displacement of the dark lines would stand out in the differential signal shown in Figure~\ref{fig:3}(c) in such a way that an intensity loss (blue) would be seen at the new position position of the dark line and an intensity gain (red) at the original position.
For a small shift, a pair of blue and red lines tracking the dark line position would be expected.

The mechanism for an electronic change of the XPD pattern is essentially the same as that leading to the line shape change in Fig.~\ref{fig:1}: The inelastic electron-hole generation (annihilation) process giving rise to an energy loss (gain) of the photoelectron is necessarily accompanied by a momentum change.
In lightly doped graphene, there are strong constraints on such momentum changes: Electron-hole pair creation or annihilation can proceed within a given Dirac cone (intravalley) or between Dirac cones (intervalley).
The latter process is usually insignificant for carrier scattering in transport because it requires a strongly localized potential scatterer~\cite{Das-Sarma:2011aa}.
In our case it can be important because of the localized character of core hole generation.
The momentum change in the electron-hole pair creation/annihilation process leading to the line broadening in Fig.~\ref{fig:1} is thus approximately either zero for intravalley, or $K - K'$ for intervalley processes.
An additional momentum change corresponding to a reciprocal lattice vector is also possible.
The similar time dependence of $R(t)$ and $T_e$ suggests that the same momentum changes are responsible for the changes of the XPS line shape and the XPD pattern.

Given the large fraction of inelastically deflected electrons leading to the line shape change in Fig.~\ref{fig:1} and the fact that the momentum changes can be large, it appears surprising that the XPD pattern changes so little.
Qualitatively, this can be understood as follows: If a core electron escapes leaving an excited electron-hole pair behind, the momentum change deflects the primary emitted electron and it thereby influences the amplitude of the electron wave field reaching the detector and the surrounding scatterers.
However, it does \textit{not} affect the location of the scatterers relative to the emitter.
Therefore, the phase difference between the part of the wave field reaching the detector directly and the scattered parts remains fixed and so does the location of the XPD features.
Only the relative amplitude of the direct and scattered waves changes, modifying the intensity of the XPD features.
Using simulated diffraction patterns, Appendix~\ref{appendix:trEDAC} shows that the observed change of $R(t)$ is consistent with a situation in which a substantial amount (50 \%) of the photoelectrons have been deflected in momentum because of the creation/annihilation of electron-hole pairs in the photoemission process.

\begin{figure}[htb!]
    \begin{center}
        \includegraphics[width=0.4\textwidth]{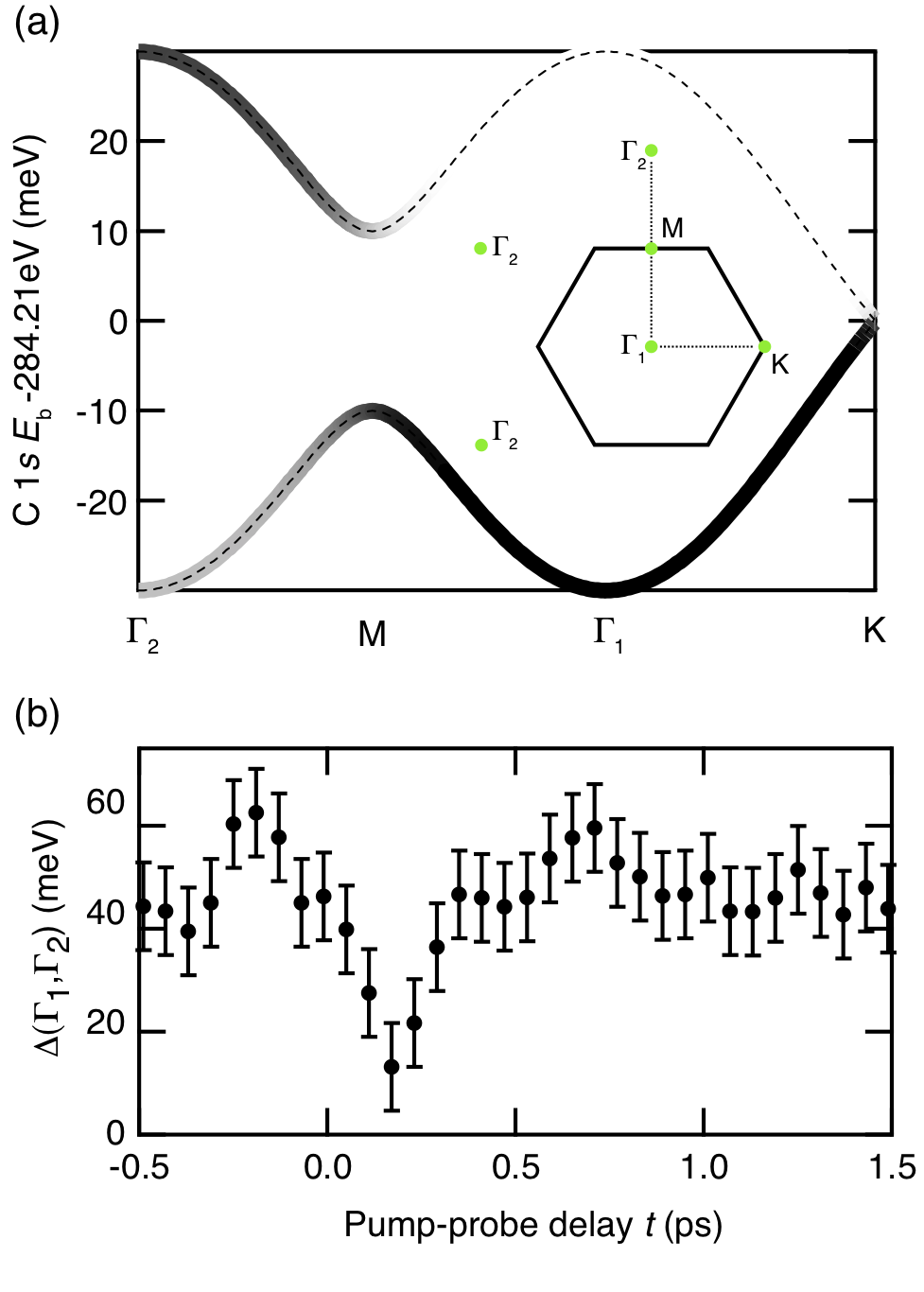}
        \caption{(a) Tight-binding band structure of the C~$1s$ core level states along the dotted line in the inset.
            The dashed lines show the dispersion of the bonding and anti-bonding bands.
            The grayscale shading encodes the expected photoemission intensity due to the sub-lattice interference effects.
            Dark corresponds to high intensity.
            (b) Time-dependent C~$1s$ band width obtained from the energy difference between the C~$1s$ peaks at the $\Gamma_1$ and $\Gamma_2$ points of the first and the neighboring Brillouin zone.
        }
        \label{fig:4}
    \end{center}
\end{figure}

Graphene offers the unique possibility to further test the role of electronic final state scattering in the C~$1s$ spectrum due to the fact that the C~$1s$ state forms a narrow $\sigma$-band with a splitting between bonding and anti-bonding states of 60~meV at the $\Gamma$ point~\cite{Lizzit:2010aa}.
This band dispersion is shown in Fig.~\ref{fig:4}(a).
The splitting is too small to be directly observable given the natural C~$1s$ line width but, due to the selection rules in the bipartite graphene lattice~\cite{Shirley:1995aa,Mucha-Kruczynski:2008aa}, only the bonding band is visible in the first Brillouin zone (BZ), whereas mainly the anti-bonding band is visible in the neighboring zones.
The expected photoemission intensity due to this effect is encoded in the grayscale of the band displayed in Fig.~\ref{fig:4}(a).
This dominance of one particular band makes the dispersion observable as a slight shift of the C~$1s$ peak between the $\Gamma$ points of different BZs, \textit{e.g.}, between $\Gamma_1$ and $\Gamma_2$ in Fig.~\ref{fig:4}(a)~\cite{Lizzit:2010aa}.
It also leads to the curious situation in which the observed shift of the C~$1s$ peak has a periodicity which is \textit{larger} than that of the reciprocal lattice even though the true dispersion is not.

We can exploit this effect for confirming the presence of inelastic momentum deflection in the electronic final state.
When explaining the ultrafast changes of the XPD pattern in Fig.~\ref{fig:3}, we have argued that the inelastic excitation/annihilation of electron-hole pairs during the photoemission process is accompanied by a momentum change of the photoelectron which is either zero or $K - K'$  modulo a reciprocal lattice vector.
Here we immediately notice that momentum changes involving a reciprocal lattice vector could completely eliminate the observed binding energy difference between the $\Gamma_1$ and $\Gamma_2$ points, $\Delta (\Gamma_1, \Gamma_2)$.
Even changes involving a momentum change of $K - K'$, which is the same scattering vector as $\Gamma - K$, could greatly reduce $\Delta (\Gamma_1, \Gamma_2)$ since they mix electrons from the $K$ points, where there is no energy difference between bonding and anti-bonding band, into the $\Gamma$ points.
If electronic scattering effects are indeed important, the dispersion of the $\sigma$-band might no longer be observable.

This is indeed seen when plotting the time-dependent energy difference $\Delta (\Gamma_1, \Gamma_2)$ in Figure~\ref{fig:4}(b).
Upon pumping the system, $\Delta (\Gamma_1, \Gamma_2)$ reduces from around 45~meV to roughly 10~meV on a similar time scale as the electronic temperature change.
This is followed by a recovery to the equilibrium levels that is more swift than the $T_e$ recovery, which can be explained by a non-linear relationship between $T_e$ and the measured $\Delta (\Gamma_1, \Gamma_2)$.
Note that the observed collapse of the binding energy difference should only manifest itself between $\Gamma$ points in neighboring BZs but not at, \textit{e.g.}, the $M$ points of the BZ where the splitting between bonding and anti-bonding band is much smaller and the photoemission matrix element is similar for both bands.
In Appendix~\ref{appendix:Mpnts}, we demonstrate that the binding energy difference between two $M$ points is indeed very small and constant over time.

In Appendix~\ref{appendix:theoCollapse}, we simulate $\mathbf{k}$-dependent C~$1s$ XPS spectra based on the known band structure and photoemission matrix elements and we explore the effect of mixing electrons deflected by a reciprocal lattice vector and/or by $\Gamma-K$ (corresponding to $K - K'$) on the observed $\Delta (\Gamma_1, \Gamma_2)$.
Assuming that half of the emitted electrons are inelastically deflected turns out to be sufficient to reduce $\Delta (\Gamma_1, \Gamma_2)$ from \SI{45}{meV} to \SI{10}{meV}, similar to what is observed in the experiment.
As mentioned above, this scenario is also consistent with the minor change of the XPD pattern reported in Fig.~\ref{fig:3}.

Instead of being due to electronic final state scattering, the ultrafast reduction of $\Delta (\Gamma_1, \Gamma_2)$ could also be caused by a true or apparent collapse of the C~$1s$ bandwidth.
In Appendix ~\ref{appendix:frozenphonon}, we present frozen-phonon calculations that probe the effect of exciting the strongly coupled phonons at $\Gamma$ and $K$ on the electronic structure.
A lattice distortion at $K$, in particular, results in a smaller BZ and back-folding of bands to the original $\Gamma$ points, and this could potentially decrease the observed observed $\Delta (\Gamma_1, \Gamma_2)$.
It turns out, however, that the spectral weight of the back-folded bands would be so small that it could not lead to the effect reported in Figure~\ref{fig:4}.
Moreover, the observable back-folded bands are concentrated in a very small energy window around the original bands, such that we can rule out a significant contribution of the strongly coupled phonons to the C~$1s$ line width broadening.
In order to probe the possible role of temperature-dependent screening effects on the C~$1s$ band width, we have performed temperature-dependent GW calculations of the C~$1s$ band structure, as described in Appendix ~\ref{appendix:gwbands}.
The band structure changes are far too small to play a role for the observed $\Delta (\Gamma_1, \Gamma_2)$ decrease.
Further effects that have been considered as possible causes, but have been excluded, are a suppression of the band formation due to the electric field of the pump pulse~\cite{Crespi:2013aa}, or a similar effect but with the electric field supplied by plasmons, as well as an average interatomic distance increase due to out-of-plane phonons.

\section{conclusions}
In conclusion, we have demonstrated the potential of high-resolution time-resolved XPS to give detailed information on the ultrafast development of electronic many-body effects.
In particular, we have identified an electronic broadening mechanism in which the outgoing core level electron exchanges energy and momentum with the hot electron gas.
This interpretation of the data is supported by an ultrafast suppression of the momentum-dependent binding energy variations associated with the C~$1s$ $\sigma$-band.
Moreover, a quantitative description of the resulting line shape permits the determination of the electronic temperature in agreement with direct measurements in the valence band.
It will be interesting to expand the technique to very short time delays, before the thermalization of the electron gas, where high-resolution XPS will provide an element-specific probe of the electronic excitations and many-body effects.
The ultrafast electronic final state scattering effect observed here could, in principle, impose some severe limitations for techniques such as ultrafast XPD because it could be expected to smear out the diffraction pattern.
We demonstrate that this is not the case and that the XPD pattern shows only very minor changes, even at electronic temperatures of over \SI{4000}{K}.

\appendix
\section{Data preparation}
\label{appendix:datacube}

The raw data obtained for the time-resolved measurements at the PG2 beamline is in the form of tables that require binning into histograms to be visualized and analyzed.
The procedure employs the hextof-processor open source code~\cite{Agustsson:2021aa} and is outlined elsewhere~\cite{Xian:2020aa,Dendzik:2020aa,Kutnyakhov:2020aa}.

In this section, we discuss the calibrations and corrections necessary to convert the binning axes from hardware values (time-of-flight steps, pixels, delay-line position), to physically relevant values (binding energy, $k_x$ and $k_y$, pump-probe delay $t$).

The momentum microscope time-of-flight axis is calibrated to binding energy by measuring the C~$1s$ spectrum while applying different voltages to the sample that shift the spectrum by a known energy ($\pm$~\SI{1}{eV}, \SI{0}{eV}), and subsequently comparing the resulting C~$1s$ spectra to each other and to the ones obtained at the SuperESCA beamline.

The momentum microscope time-of-flight ($tof$) can be converted to binding energy ($E_b$) by the following equation:
\begin{equation}
    E_b = -\frac{1}{2} m_e \left(\frac{l_0}{tof-tof_0}\right)^2 - W + h\nu + e V_s
\end{equation}
Where $m_e$ is the rest electron mass, $l_0$ is the effective microscope drift tube length, $tof_0$ is the time-of-flight offset, $W$ is the effective work function, $h\nu$ is the photon energy, and $V_s$ is the potential applied to the sample.
In particular, note that the conversion is not linear, and the binning needs to be performed directly in binding energy to avoid intensity biases generated by variable bin size across the binning range.
The conversion then depends on 3 unknown parameters: $tof_0$ is found by measuring the so-called ``photon peak'', which is generated by photons reflected off the sample that cause a count peak on the detector.
This is assumed to happen with negligible time-of-flight.
Changing $l_0$ mainly affects the scaling of the binding energy axis, and the parameter can be reliably found by analyzing the voltage-shifted C~$1s$ peaks.
Finally, $W$ can be found ensuring the resulting C~$1s$ peak is found at the same binding energy as the reference spectrum acquired at the SuperESCA beamline.

The binding energy in non-static measurements was found to display a significant $\mathbf{k}$-dependent, space-charge induced shift.
Pump-induced space-charge has, in fact, the effect of shifting and broadening the spectra with a radial Lorentzian-type dependence~\cite{Schonhense:2018aa}.
The apparent binding energy shift can be corrected for by following the algorithm presented in Ref.~\cite{Schonhense:2018aa}, and this was found to significantly improve the quality of the data.
While the central position of the correcting Lorentzian, as well as the Lorentzian width parameter, did not vary significantly with pump-probe time delay, the Lorentzian amplitude was found to display a slow, linear dependence on pump-probe delay over the observed time window.

The $k_x$ and $k_y$ axes have been calibrated by matching the position of the dark lines to the photoemission horizons shifted by 2 reciprocal lattice vectors.
The photoemission horizon corresponds to a circle with a polar angle of \SI{90}{\degree}, which is equivalent to a radius of \SI{3.58}{\angstrom^{-1}} for the graphene C 1s core level in our measurements.

The optical delay line position is converted to pump-probe delay time $t$ after establishing the pump-probe delay $t=0$.
This has been found by analyzing the time evolution of the integrated intensity in a small energy range of \SI{200}{meV} around the graphene C~$1s$ binding energy over acquisition intervals of \SI{400}{s}.
A fit to the data is performed using a double decaying exponential convoluted with a Gaussian, allowing a tracking of any delay drift of the pump or probe over the long acquisition time (\SI{22}{h}).
The delay has also been corrected by the bunch arrival monitor values, removing a large part of the probe jitter caused by the self-amplified spontaneous emission process, and improving the time resolution to \SI{210}{fs}.
This can be compared to the time resolution before correction, which was \SI{275}{fs}.

Gaussian-shaped bins have been employed for the $k_x$ and $k_y$ axes.
The Gaussian $\sigma$ parameters were \SI{0.06}{\angstrom^{-1}}.
The energy and the pump-probe delay axes were binned using rectangular bins with widths of \SI{46}{meV} and \SI{60}{fs}, respectively.

The dataset for the pumped sample with no space-charge correction and with rectangular binning for all axes is openly available~\cite{Curcio:2021aa}.

\section{C 1s spectra}
\label{appendix:spectraC1s}

In this section the C~$1s$ spectra from PG2 and SuperESCA are presented with a binding energy range that allows the small C~$1s$ component from the SiC substrate to be appreciated.

\begin{figure}[htb!]
    \begin{center}
        \includegraphics[width=0.45\textwidth]{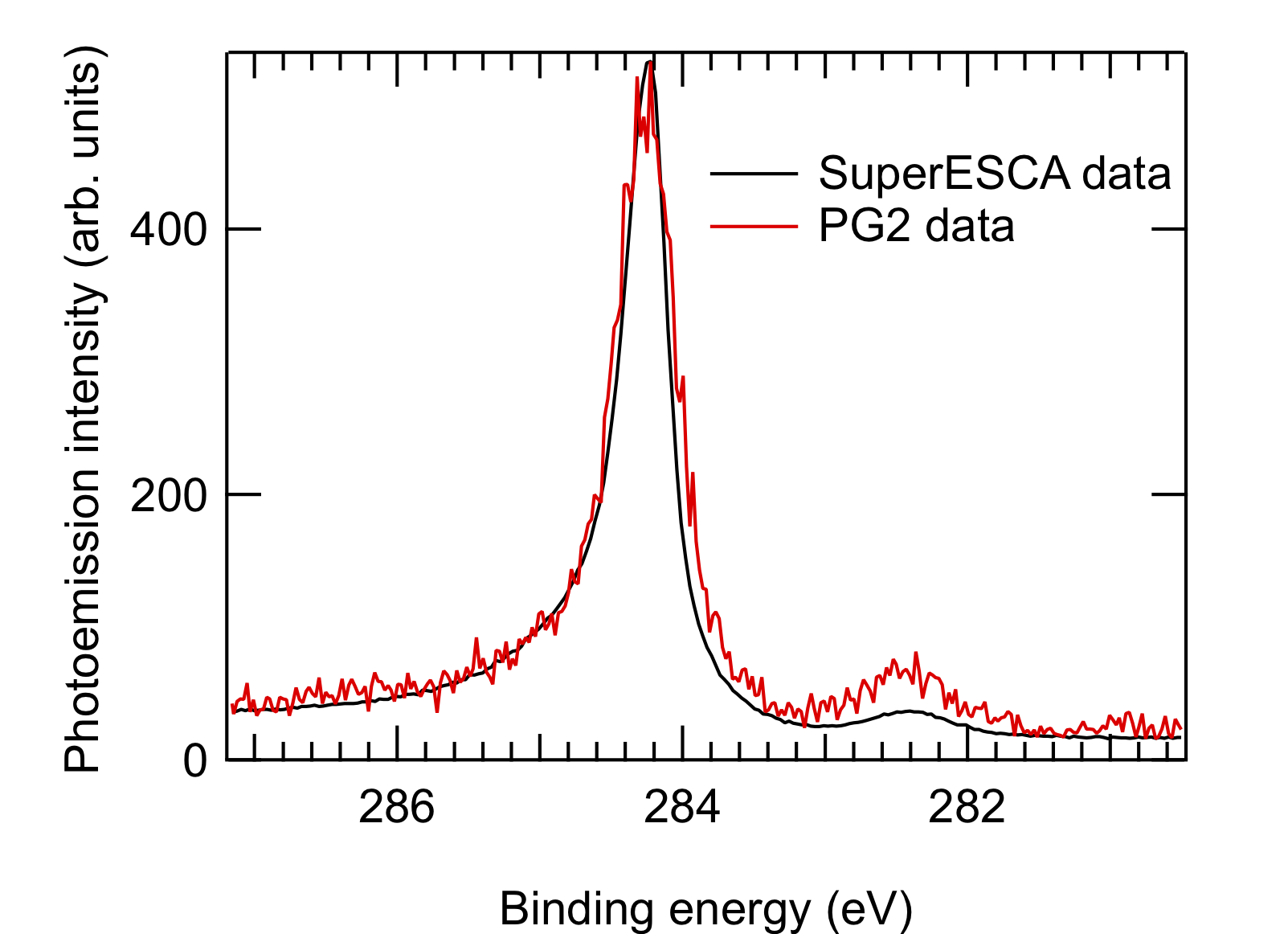}
        \caption{
            C~$1s$ spectra acquired at SuperESCA (black line) and at PG2 (red line).
        }
        \label{fig:S8}
    \end{center}
\end{figure}

The PG2 spectrum is obtained for negative pump probe delays, in the same $k$-range as that of Fig.~\ref{fig:1}.
The SuperESCA spectrum is obtained close to normal emission.
The absolute binding energy was calibrated based on this spectrum combined with a fit of the valence band data using a linear density of states and a Fermi-Dirac distribution.
Note that the intensities of the various components in the spectra shown in Fig.~\ref{fig:S8} are not completely comparable due to the different experimental geometries of the two beamlines.

\section{$D(\epsilon)$ models}
\label{appendix:dos}
Equation~\ref{equ:1} gives a line shape that is strongly influenced by the choice of the density of possible excitations, $J(E)$ and this is directly derived from the sample's density of states $D(\epsilon)$.
\begin{figure}[htb!]
    \begin{center}
        \includegraphics[width=0.4\textwidth]{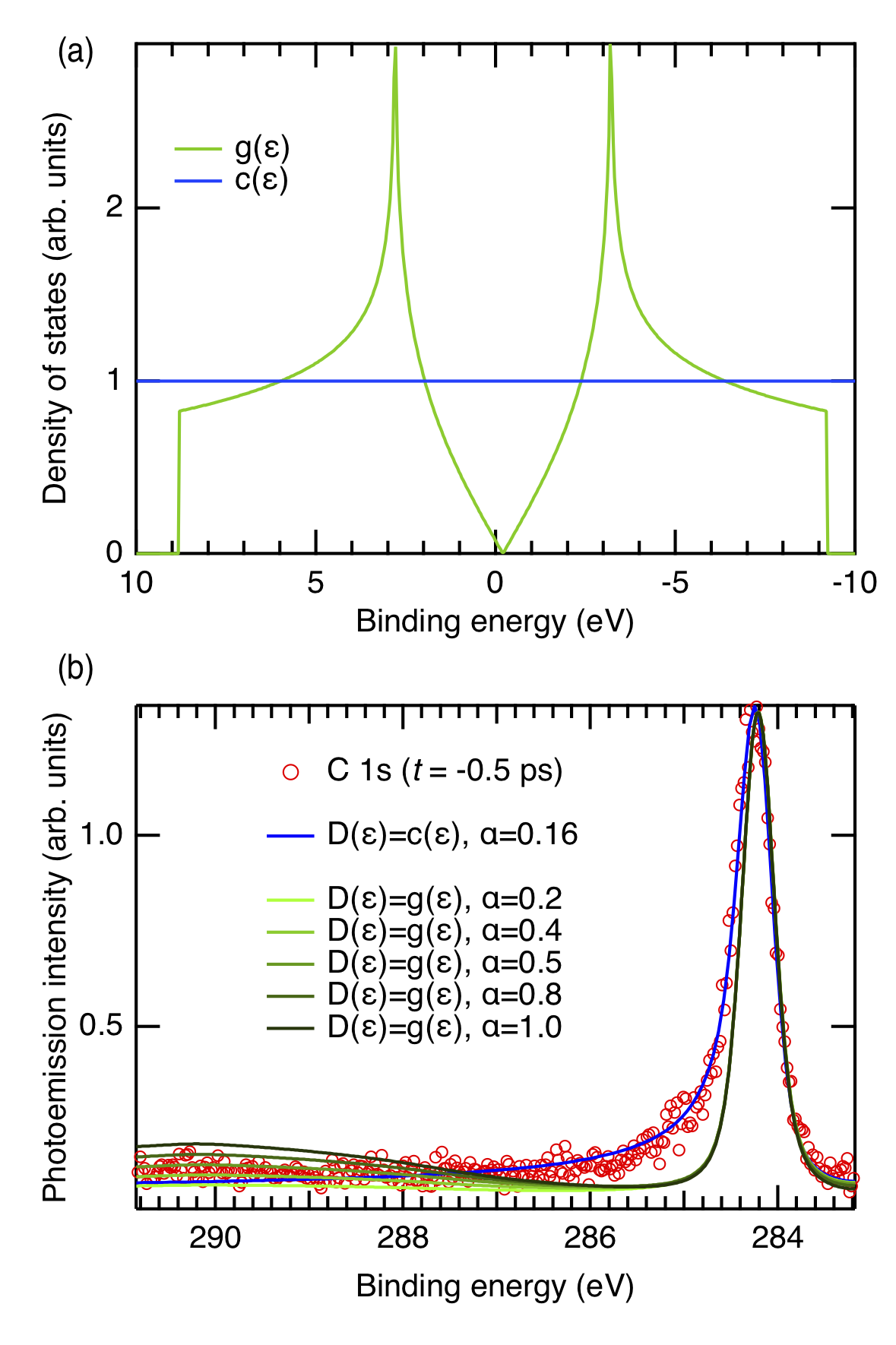}
        \caption{
            (a) Different choices for the density of states $D(E)$ near $E_F$ used for fitting the graphene C~$1s$ spectrum.
            $g{(\epsilon)}$ is a simple model for graphene (nearest neighbor tight-binding calculation for the $\pi$-band.
            $c(\epsilon)$ is a constant density of states.
            (b) C~$1s$ spectrum of graphene from PG2 at negative time delay (markers) shown together with line shapes resulting from equation (\ref{equ:1}) using different choices for $D(\epsilon)$ and the asymmetry parameter $a$.
        }
        \label{fig:S11}
    \end{center}
\end{figure}
The most obvious choice for $D(\epsilon)$ would be the density of states of graphene which, for the purpose of the simple illustration here, we approximate by the nearest neighbor tight-binding density of states for the $\pi$-band $g(\epsilon)$, with the same amount of hole doping as the quasi free-standing graphene used in the experiment (green line in Figure~\ref{fig:S11}(a)).
With this choice for $D(E)$, it is not possible to obtain a satisfactory agreement with the experimental data for any value of the asymmetry parameter $a$, as demonstrated in Figure \ref{fig:S11}(b).
This is because the experimental data shows a continuous asymmetric tail at higher binding energies, while using $g(\epsilon)$ results in a negligible amount of possible excitations at low energies and a very large number of excitations involving transitions between the van Hove singularities (sharp peaks at roughly \SI{2.9}{eV} from the Fermi energy).
Increasing $a$ does therefore not increase the intensity of the low-energy tail close to the main peak while it creates a pronounced intensity increase at a binding energy \SI{5.8}{eV} higher than the peak.
A different choice for $D(\epsilon)$, namely the constant function $c(\epsilon)$ in Figure~\ref{fig:S11}(a), blue line, gives a much better agreement between model and experiment in Figure~\ref{fig:S11}(b).
Indeed, this choice for $D(\epsilon)$ is equivalent to a Doniach-\v Sunji\'c function at \SI{0}{K}, which is a line shape commonly observed to yield good agreement to graphene C~$1s$ spectra \cite{Riedl:2009aa,Lizzit:2010aa}.

\section{R-factor definition}
\label{appendix:rfactor}

The so-called R-factor is a quantity that is used to quantify the agreement between two XPD patterns, $\chi(k_x, k_y)$ and $\chi_0(k_x, k_y)$.
The definition is
\begin{equation}
    R=\frac{\sum_{k_x,k_y}[\chi(k_x,k_y) - \chi_{0}(k_x,k_y)]^2 }{\sum_{k_x,k_y} \chi(k_x,k_y)^2 + \chi_{0}(k_x,k_y)^2 },
\end{equation}
where typically $\chi(k_x, k_y)$ is an experimental modulation function while $\chi_0(k_x, k_y)$ is a simulated modulation function~\cite{Woodruff:1994ab}.
Modulation functions are introduced as
\begin{equation}
    \chi(k_x, k_y)=\frac{I(k_x, k_y)-I_0(k_x, k_y)}{I_0(k_x, k_y)}
\end{equation}
where $I(k_x, k_y)$ is an intensity distribution, and $I_0(k_x, k_y)$ is a smooth function approximating the photoemission intensity in the absence of any scatterers.
The modulation function then represents the photoemission intensity variations that are due to diffraction.

\section{Multiple scattering simulations}
\label{appendix:EDAC}

In order to evaluate the effect of inelastic electron-hole pair generation/annihilation during the photoemission process on the XPD pattern by simulations (see Appendix~\ref{appendix:trEDAC}), it is desirable to obtain a realistic description of the observed diffraction pattern.
To do so, we have performed simulations of XPD patterns using the EDAC code~\cite{Garcia-de-Abajo:2001aa}.
This code can perform multiple scattering simulations of electrons in atomic clusters, and we compared the results to the XPD patterns obtained at the SuperESCA beamline.
The SuperESCA data (see Fig.~\ref{fig:S8}) have the advantage of a much higher signal to noise ratio compared to the PG2 data.
Note that while the same photon energy was used at SuperESCA and PG2, the resulting XPD patterns are not expected to be identical due to differences in experimental geometry/light polarization at the two sources.

Using the known structural parameters of graphene,  optimized values for $V_0$ (muffin tin inner potential), $\lambda_i$ (inelastic mean free path), $R_{max}$ (radius of the atomic cluster), $l_{max}$ (maximum angular momentum quantum number used in the calculation), and number of scattering orders, have been found by direct comparison with the SuperESCA experimental XPD pattern, as well as by convergence tests for the calculation.
The optimization of $V_0$ and $\lambda_i$ has been accomplished by minimizing the R-factor, calculated by excluding the area affected by the dark lines, since they are not captured by our multiple scattering simulations that use an abrupt step in the muffin tin surface potential.
Multiple scattering simulations using the parameters found here can then be used to simulate XPD patterns in the different experimental geometry of the PG2 beamline.

\begin{figure}[htb!]
    \begin{center}
        \includegraphics[width=0.4\textwidth]{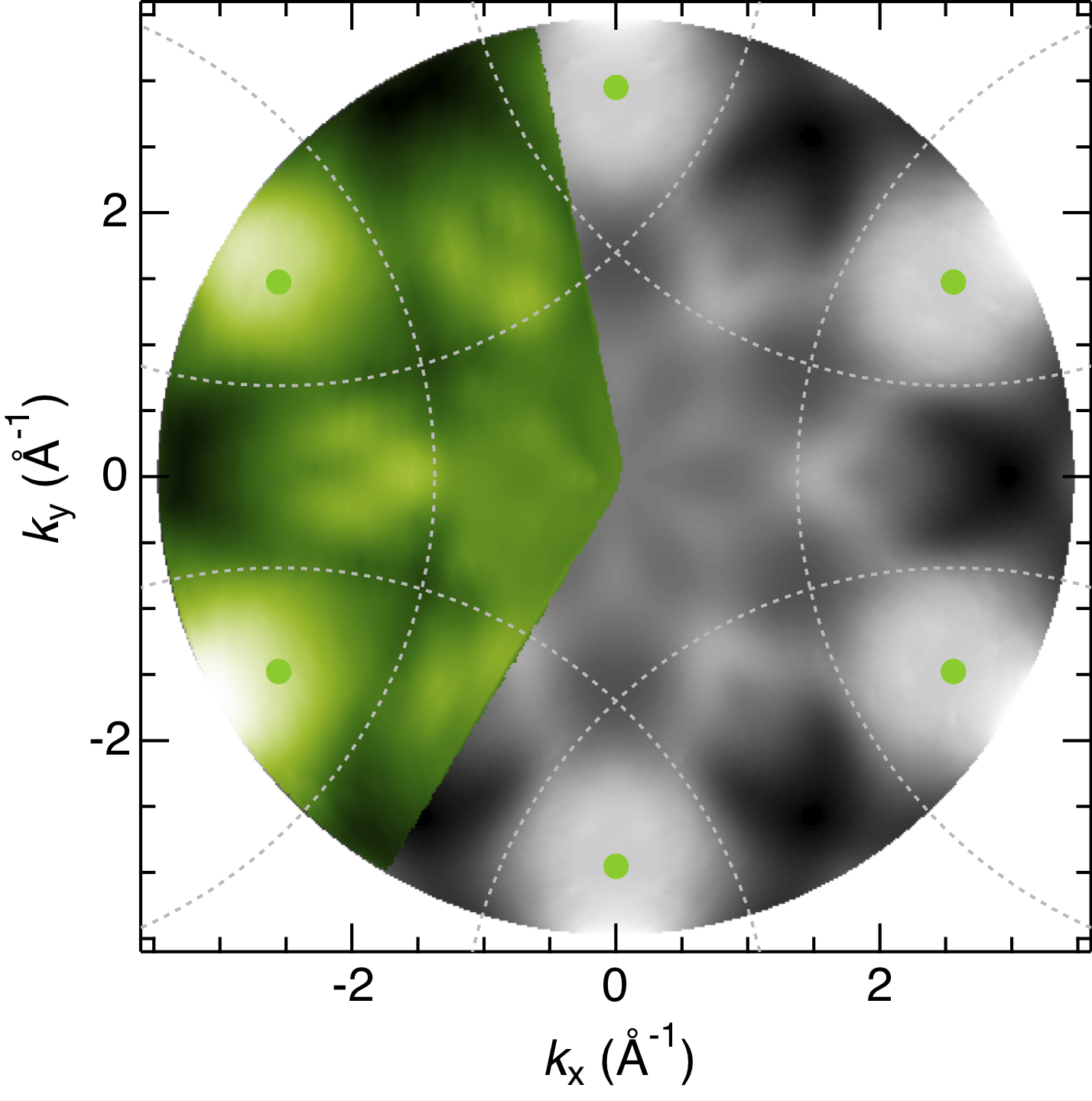}
        \caption{Analysis of the XPD data collected at the SuperESCA beamline.
            Experimental (green) and simulated (gray) XPD pattern for the graphene C~$1s$ component at $h\nu=$\SI{337.5}{eV}.
            The gray dashed lines are the photoemission horizon location (polar angle $\theta = \SI{90}{\degree}$) shifted by 2 BZs.
            $\Gamma$ points are indicated by green markers.
        }
        \label{fig:S8a}
    \end{center}
\end{figure}

Fig.~\ref{fig:S8a} shows the resulting comparison between the experimental pattern and the optimized multiple scattering simulation, obtained with the following simulation parameters: $V_0=\SI{17.2}{eV}$, $\lambda_i=\SI{5}{\angstrom}$, $l_{max}=8$, $R_{max}=\SI{15}{\angstrom}$, number of scattering orders~$= 17$.
The kinetic energy was set at \SI{49.1}{eV}, which is the kinetic energy measured by the electron analyzer for the graphene C~$1s$ component.
The simulated structure only includes the graphene layer, with no substrate, since the lattices have a different unit cell, and the effect of the substrate would thus be averaged out in a local cluster calculation.
Neglecting the substrate is known to be an excellent approximation~\cite{Lizzit:2010aa}.

\section{Effect of inelastic electron-hole pair generation/annihilation on R(t)}
\label{appendix:trEDAC}

\begin{figure*}[htb!]
    \begin{center}
        \includegraphics[width=\textwidth]{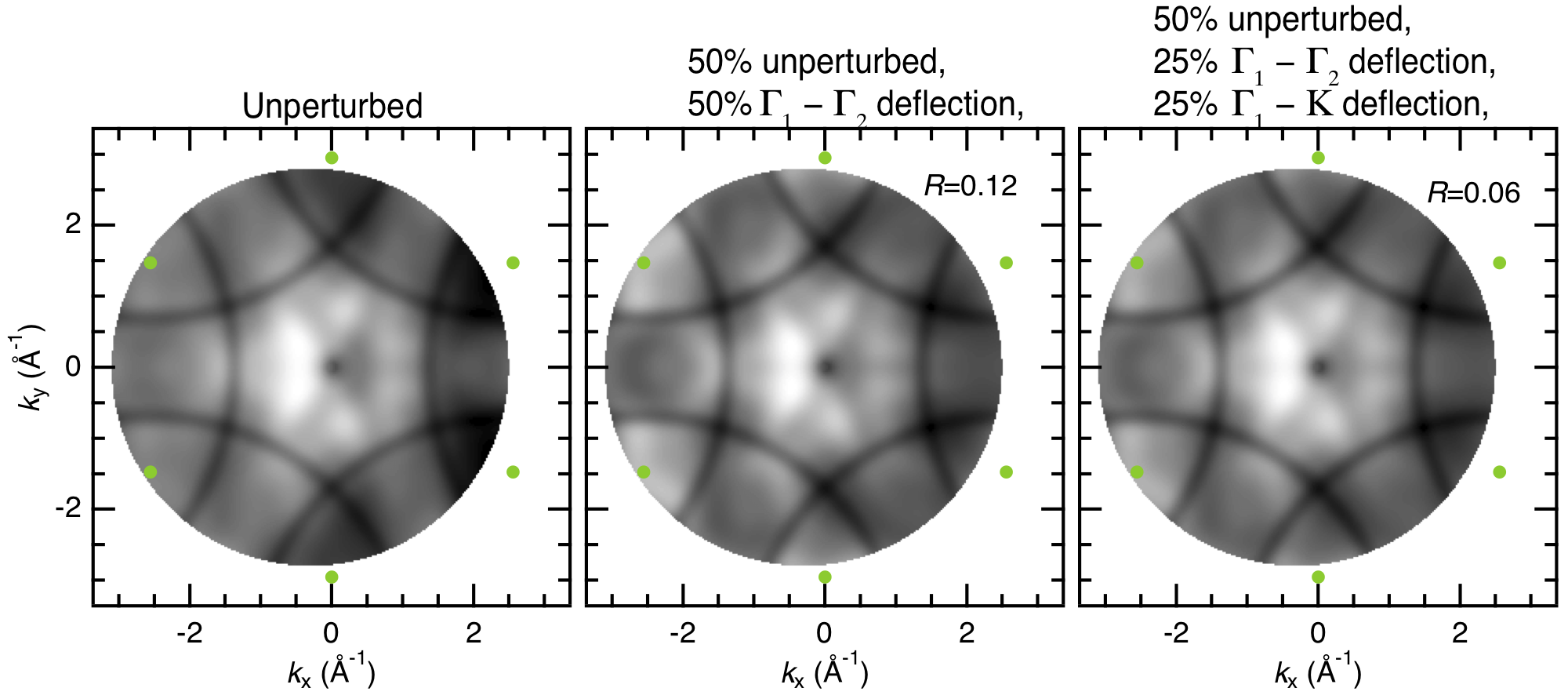}
        \caption{Multiple scattering simulations for the C~$1s$ XPD pattern from the PG2 beamline at FLASH.
            The simulation parameters are the same as what has been found to optimize agreement with the SuperESCA experimental data in Fig.~\ref{fig:S4}.
            The simulated experimental geometry, however, is the same as PG2 geometry.
            Patterns are displayed with artificially added dark lines.}
        \label{fig:S4}
    \end{center}
\end{figure*}

Being able to simulate XPD patterns, we can evaluate the effect of core electron deflection by inelastic electron-hole generation/annihilation at high electronic temperatures.
Specifically, we can test if such effects would be consistent with the very small change of $R(t)$ shown in Fig.~3(d).
We have argued that, at high electronic temperature, the C~$1s$ photoelectrons have a high probability of leaving the solid with one or several electron-hole pairs created or annihilated.
The energy loss (gain) is accompanied by a momentum change that, due to the strong constraints by the electronic structure of graphene is either nearly zero or $K-K'$ for intravalley and intervalley excitations, respectively (modulo a reciprocal lattice vector).
We can now simulate how the presence of such excitations would affect the XPD pattern.

The excitation (annihilation) of electron-hole pairs is part of the photoemission process and can be considered to be instantaneous for our purposes.
We thus have to simulate the situation of a photoelectron escaping that is deflected by a reciprocal lattice vector, by $K-K'$ (or the identical vector $\Gamma-K$) or by a combination of both.
We use the following way to implement this in the multiple scattering simulation: Since the initial state for C~$1s$ has angular momentum $l=0$, due to the photoemission selection rules, the final state angular momentum quantum number is $l=1$.
Therefore the directionality of the final state is determined by a single vector, the electric field direction of the probe pulse.
Deflecting the momentum of the photoelectron can thus be implemented by changing the electric field direction in the multiple scattering simulations accordingly.
The contribution of such deflected photoelectrons to the XPD pattern can then be estimated by summing the photoemission intensity from photoelectrons deflected by $\Gamma-K$ or a reciprocal lattice vector and those without a momentum change.

In Fig.~\ref{fig:S4} the resulting modulation functions are shown.
Panel (a) displays the simulated pattern in the PG2 experimental conditions without the inclusion of inelastically deflected photoelectrons.
Note that this XPD pattern is slightly different from that in Fig.~\ref{fig:S8} due to the fact the impinging photon beam axis has a fixed angle with respect to the surface of the sample, reducing the symmetry of the pattern.
Note also that the pattern is quite similar to the experimental pattern from PG2 as shown in Fig.~3(a), with both showing an intensity depletion around $\Gamma_1$ and a minimum in the azimuthal modulation around the $M$ points at the edges of the observed pattern.

Fig.~\ref{fig:S4}(b) and (c) show the XPD pattern with a substantial contribution of inelastically deflected photoelectrons.
The deflection is included either only by the sum of six equivalent $\Gamma_1 - \Gamma_2$ reciprocal lattice vectors or by these and the six equivalent $\Gamma_1-K$ vectors.
We shall see in Appendix~\ref{appendix:theoCollapse} that such an admixture of inelastically scattered intensity is sufficient to cause the observed reduction of the C~$1s$ binding energy variation $\Delta (\Gamma_1, \Gamma_2)$.

The change of the XPD patterns upon mixing deflected photoelectrons can be tracked by an R-factor defined the same way as $R(t)$.
The corresponding values are given in the figure.
Interestingly, even substantial inelastic contributions only lead to a small change of the XPD pattern with $R$ values that are comparable to the maximum $R(t)$ in Fig.~3(d).
As already pointed out in connection with Fig.~3, the main reason for this is that the direction of the primary photoelectron wave in the XPD process mainly influences the intensity of the diffraction features and not so much their locations which, in turn, are governed by the atomic structure.

We stress that the dark lines have been added artificially in Fig.~\ref{fig:S4}, by multiplying the modulation function of the pattern by a value smaller than 1 with a Gaussian profile as a function of distance from the dark line.
This has been done to give a better estimate of the R-factor increase.

\section{Quantitative theoretical estimation of $\Delta (\Gamma_1, \Gamma_2)$}
\label{appendix:theoCollapse}

\begin{figure}[htb!]
    \begin{center}
        \includegraphics[width=0.5\textwidth]{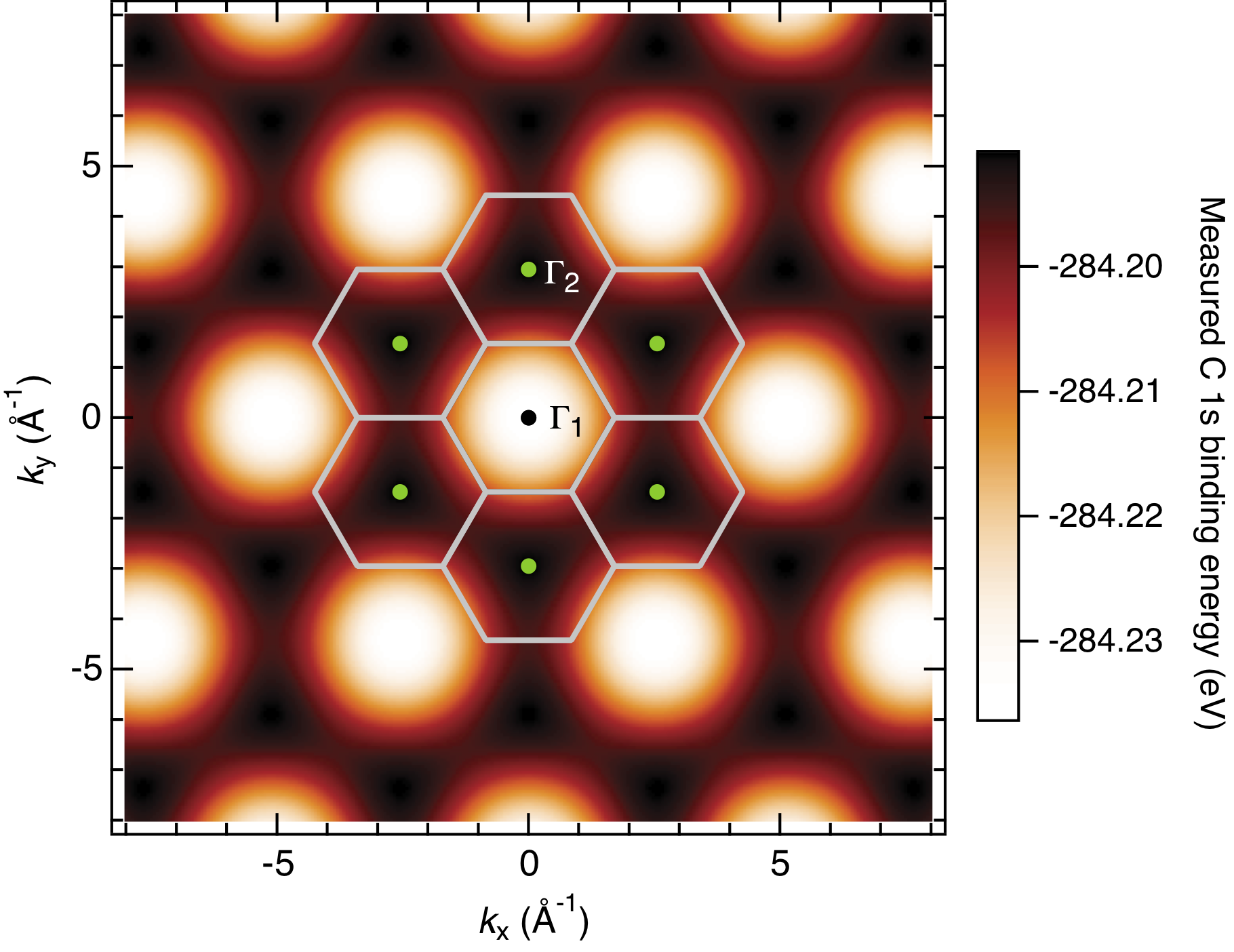}
        \caption{Map of the expected $\mathbf{k}$-dependent C~$1s$ binding energy for simulated data.
            The reciprocal lattice and the position of the $\Gamma$ points are indicated.
            Note that, due to the sub-lattice interference effect, the observed periodicity of the C~$1s$ binding energy modulation is twice as large as the reciprocal lattice.}
        \label{fig:S6}
    \end{center}
\end{figure}

As for the XPD pattern, we can simulate the effect of some photoelectrons being deflected by a reciprocal lattice vector or by $K-K'$ (i.e. $\Gamma-K$) on $\Delta (\Gamma_1, \Gamma_2)$, the observed energy difference between the $\Gamma_1$ and $\Gamma_2$ points.
We first simulate how the C~$1s$ dispersion shown in Fig.~4(a) is expected to be observed as a $\mathbf{k}$-dependent shift of the core level binding energy.
For each $\mathbf{k}$ point, a simulated C~$1s$ spectrum is generated by summing two Doniach-\v Sunji\'c functions with line shape parameters that are given by fitting the experimental spectra.
The peak energies are taken from the tight-binding model in Fig.~4(a), i.e., from a $\sigma$-band with a bonding anti-bonding splitting of \SI{60}{meV}.
The peak intensities are calculated based on the expected sub-lattice interference in graphene~\cite{Lizzit:2010aa}.
While each spectrum is then the sum of two C~$1s$ peaks, these individual contributions cannot be resolved because their separation is much smaller than the width of each peak.
In fact, the entire spectrum can be well-approximated by a single Doniach-\v Sunji\'c function.
The binding energy resulting from fitting a single Doniach-\v Sunji\' c function to the synthetic data is plotted in Fig.~\ref{fig:S6}.
The \textit{observed} bonding anti-bonding splitting, i.e., $\Delta (\Gamma_1, \Gamma_2)$, is \SI{45}{meV} and not \SI{60}{meV}.
This is caused by the incomplete suppression of the bonding band close to $\Gamma_2$ which is in contrast to the complete extinction of the anti-bonding band at $\Gamma_1$ (see Fig.~4(a)).

\begin{figure}[htb!]
    \begin{center}
        \includegraphics[width=0.5\textwidth]{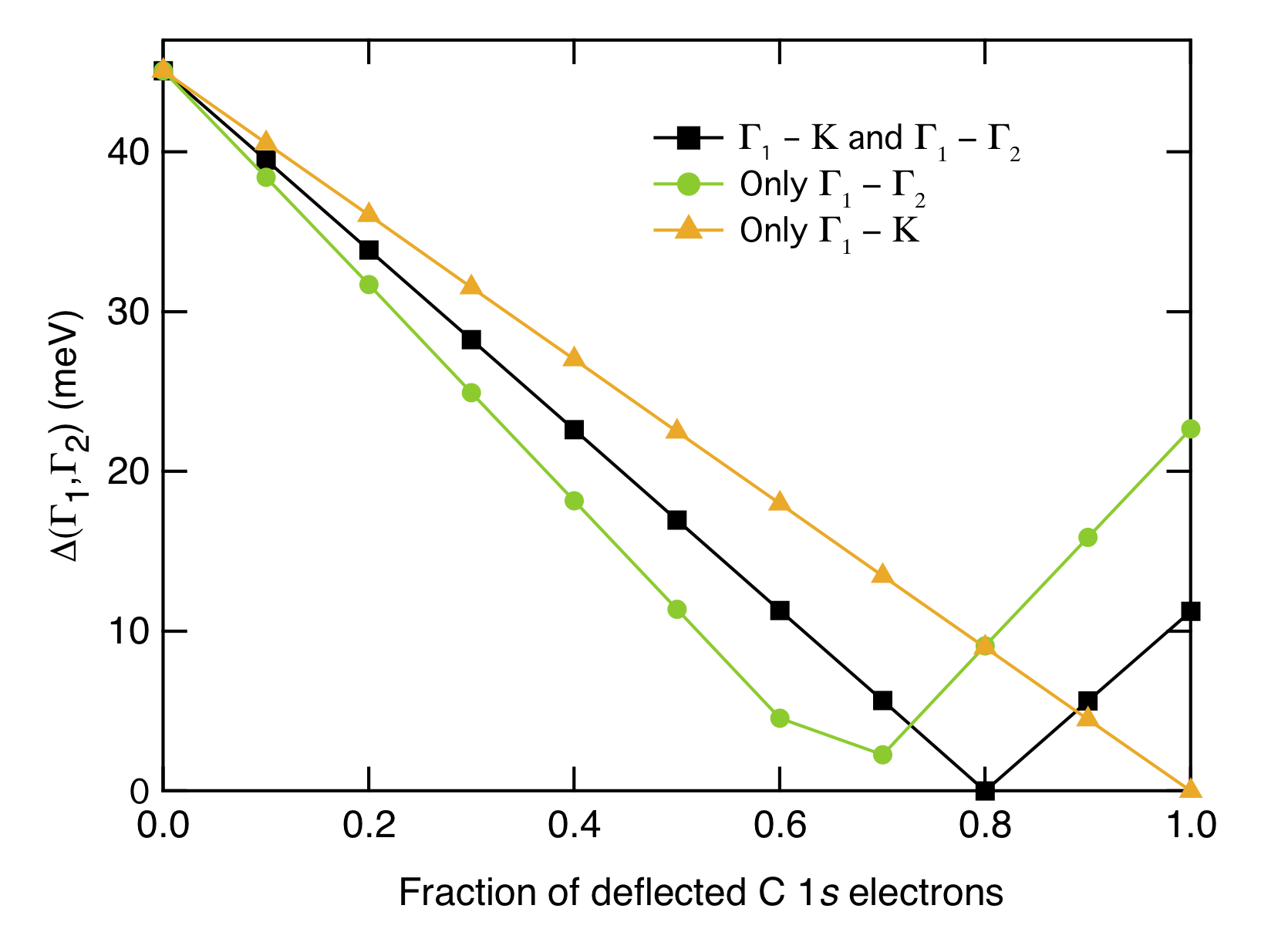}
        \caption{Observed C~$1s$ band width, expressed in terms of  $\Delta (\Gamma_1, \Gamma_2)$,  for simulated data when considering deflections of the outgoing core electrons by different momentum changes.
        }
        \label{fig:S7}
    \end{center}
\end{figure}

To investigate the change of $\Delta (\Gamma_1, \Gamma_2)$ as a result of inelastic electron deflection, we create the same artificial C~$1s$ spectra as described before but add additional spectra that have been shifted by some $\mathbf{k}$ vector before performing the analysis leading to the $\mathbf{k}$-resolved apparent binding energy shown in Fig.~\ref{fig:S6}.
The results are summarized in Fig.~\ref{fig:S7}.
We consider momentum deflections corresponding to $\Gamma_1-K$ for the six closest $K$-points, as well as $\Gamma_1-\Gamma_2$ deflections for the six closest reciprocal lattice vectors.
We study how mixing these into the photoemission intensity would affect $\Delta (\Gamma_1, \Gamma_2)$.
In order to observe a change of $\Delta (\Gamma_1, \Gamma_2)$ from \SI{45}{meV} to \SI{10}{meV}, \SI{60}{\percent} of the observed signal needs to have inelastic origin in the case that both $\Gamma_1-\Gamma_2$ and $\Gamma_1-K$ deflection is taken into account, or \SI{50}{\percent} of the observed signal needs to have been deflected in the case that only $\Gamma_1-\Gamma_2$ deflection is taken into account.
Such relatively high inelastic contributions appear plausible in view of the substantial line broadening at the highest electronic temperatures in Fig.~1 and we stress again that this scenario would not be inconsistent with the only minor change of the XPD pattern in Fig.~3.

\section{C~$1s$ binding energy at the M points}
\label{appendix:Mpnts}

\begin{figure}[htb!]
    \begin{center}
        \includegraphics[width=0.5\textwidth]{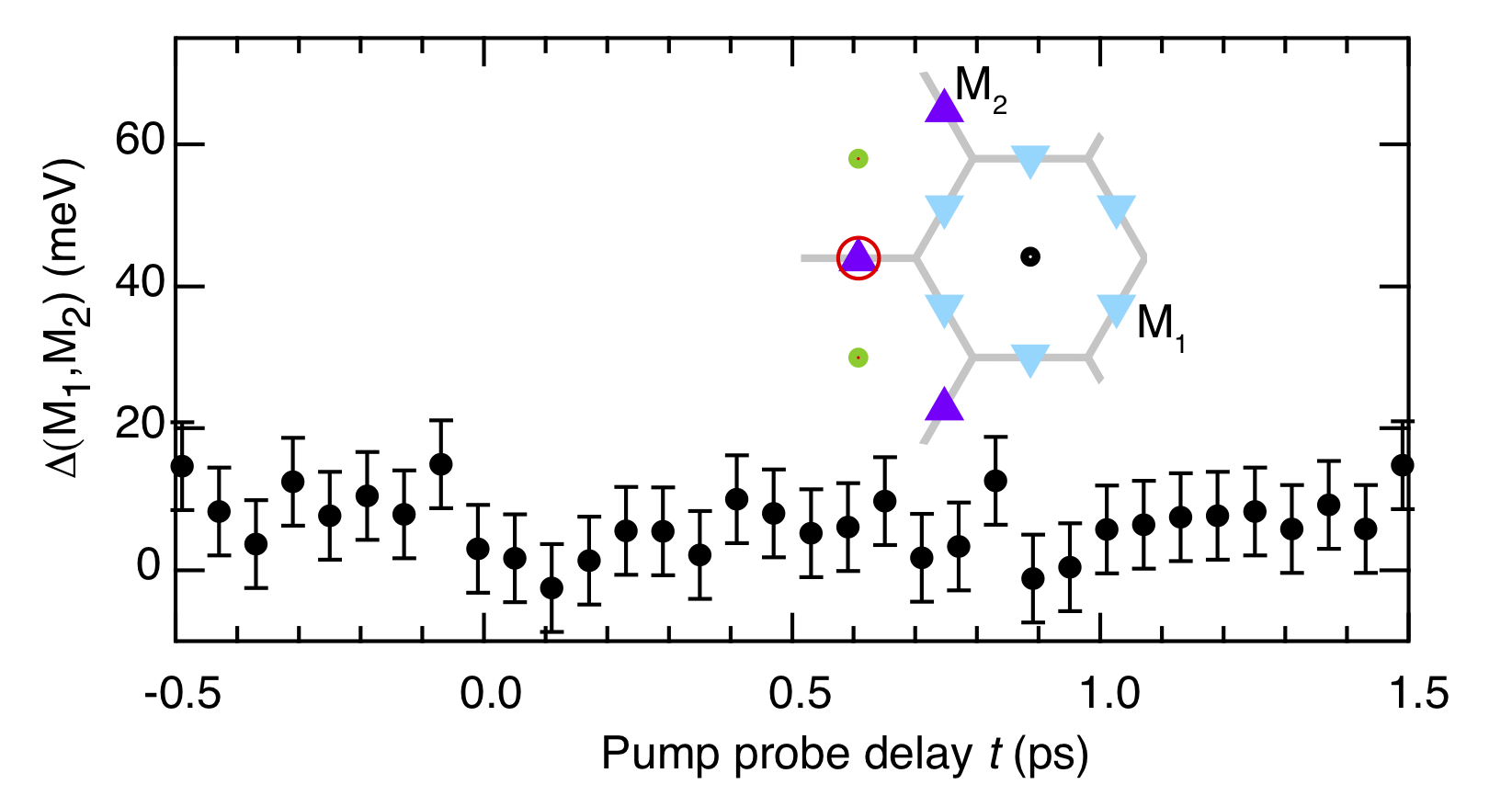}
        \caption{Binding energy changes measured at $M$ points in the first and neighboring Brillouin zones.
            The vertical energy scale corresponds to that shown in Fig.~4.
            In the inset, BZ boundaries are displayed in gray, $\Gamma$ points are displayed as circles (black in the first BZ, green in the neighboring BZs), first BZ $M$ points ($M_1$) as downward pointing blue triangles, neighboring BZ $M$ points ($M_2$) as upwards pointing purple triangles.
            The red circle around the leftmost M point marks the area used for the analysis presented in Fig.~\ref{fig:1}.
        }
        \label{fig:S5}
    \end{center}
\end{figure}

In order to verify that the observed C~$1s$ binding energy difference $\Delta (\Gamma_1, \Gamma_2)$ is not an experimental artefact, the C~$1s$ band was also measured at non-equivalent $M$ points placed at different distances from the normal emission axis (see inset in Fig.~\ref{fig:S5}).
In the area visible on the detector, it is possible to find six $M_{1}$ points between the first BZ and neighboring zones (downward pointing blue triangles in Fig.~\ref{fig:S5}), and three $M_{2}$ points further away from normal emission (upward pointing purple triangles in Fig.~\ref{fig:S5}).

The time-resolved binding energy difference $\Delta(M_1, M_2)$ for the average of the two inequivalent $M$ point families is shown in Fig.~\ref{fig:S5}.
The binding energy difference shows an unequivocally reduced change compared to $\Delta (\Gamma_1, \Gamma_2)$ (or none at all), confirming the reciprocal-lattice-aware origin for the C~$1s$ binding energy.
Note that the theoretical difference in observed binding energy for the different types of $M$ points used here is \SI{15}{meV}, and not strictly \SI{0}{meV}.
This is again caused by the different photoemission intensities of the bonding and anti-bonding bands, caused by the selection rules of the bipartite lattice.

\section{Frozen phonon band structure}
\label{appendix:frozenphonon}

\begin{figure} 
    \includegraphics[width=0.5\textwidth]{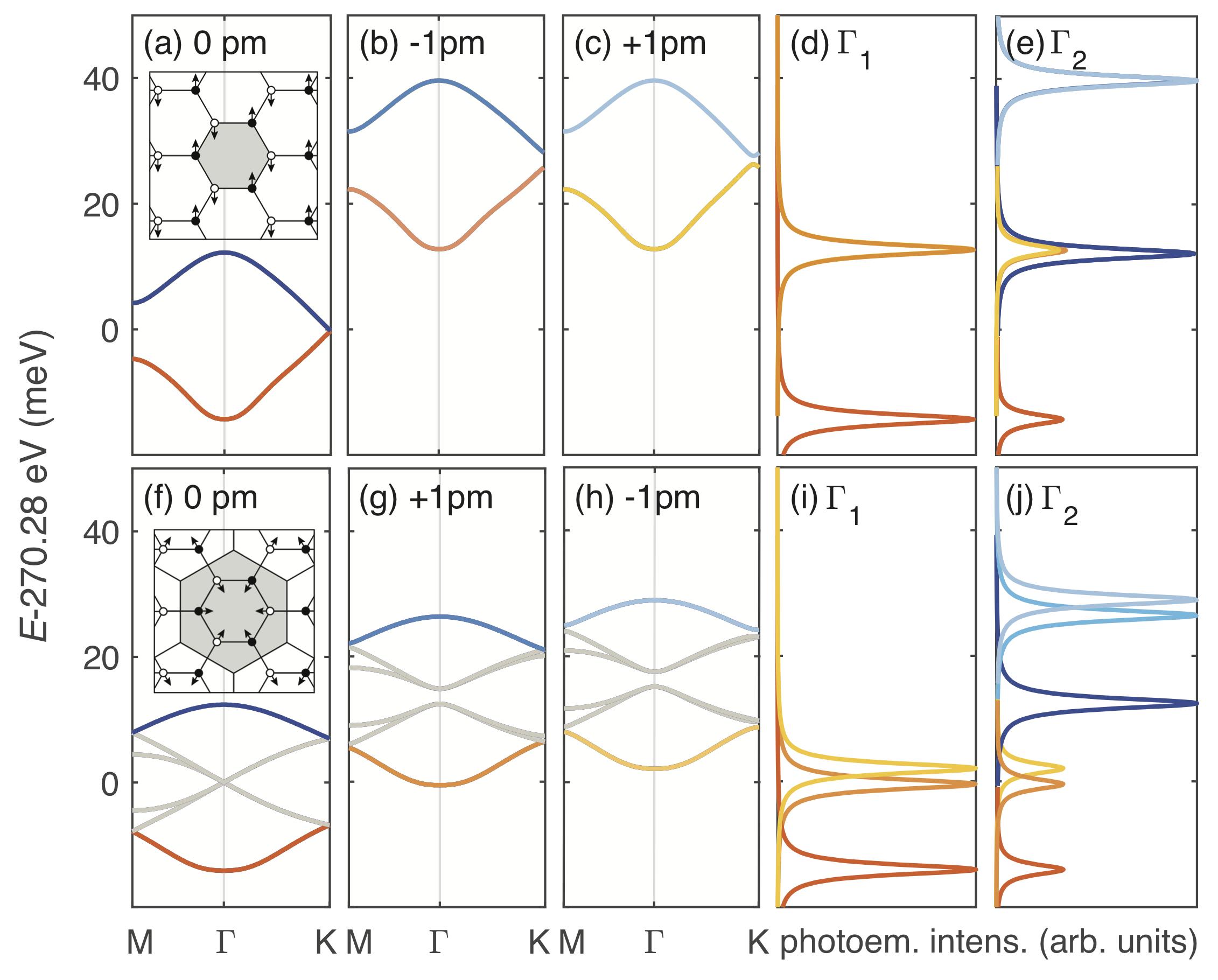}
    \caption{
    (a)-(c) Graphene 1$s$ DFT band structures corresponding to the $E_{2g} (LO)$ phonon mode at the equilibrium (a), and for atomic displacements of \SI{1}{pm} (b) and \SI{-1}{pm} (c), in the directions indicated by the arrows in the inset of panel (a).
    (d),(e) Photoemission intensity at $\Gamma_1$ and $\Gamma_2$, respectively, estimated from a tight-binding model as explained in the text. The colors correspond to photoemission from the bands in (a)-(c).
    (f)-(h) Graphene 1$s$ DFT band structures corresponding to the $A_1' (TO)$ mode, folded within the supercell represented in the inset of panel (f), at the equilibrium (f) and for atomic displacements of 1pm (g) and \SI{1}{-pm} (h), in the directions indicated in the inset of panel (f).
    (i),(j) Photoemission intensity at $\Gamma_1$ and $\Gamma_2$, estimated from a tight-binding model.
    The symmetry points, K and M, refer to the primitive and supercell (reduced) BZ in panels (a)-(c) and (f)-(h), respectively. All energies are given with respect to the vacuum level.
    }
    \label{Fig:S12}
\end{figure}

Almost immediately after the electronic excitation, energy is transferred to the strongly coupled $E_{2g} (LO)$ and $A_1' (TO)$ optical phonon modes in the vicinity of the  $\Gamma$ and $K$ points, respectively~\cite{Caruso:2020aa}, increasing their temperature $T_p$ (see Figure \ref{fig:2}).
Given the high energies of these modes, a temperature of around $T_p \approx$\SI{3000}{K} is reached without a significant increase in population.
Nevertheless, we explore the effect of such excited lattice vibrations on the core level dispersion, considering it as a possible cause of the line width increase and the ultrafast change of $\Delta (\Gamma_1, \Gamma_2)$.
For this purpose,  we adopt the so-called ``frozen phonon'' approach. Specifically, we calculate the C~$1s$ bands for \SI{1}{pm} atomic displacements corresponding to the lattice deformations due to $E_{2g} (LO)$ and $A_1' (TO)$ phonon modes.

The C~$1s$ dispersions were obtained by all-electron Density Functional Theory (DFT) calculations using the Perdew-Burke-Ernzerhof exchange-correlation functional, as implemented in the code CRYSTAL14~
\cite{Dirac:1930uy,Dovesi:2014tt,Dovesi:wq,crystal}.
An atomic natural orbital Gaussian basis set has been used and our calculations were done using a $16\times16$ ($8\times8$) Monkhorst-Pack $k$-grid for the $E_{2g} (LO)$ ($A_1' (TO)$) phonon, respectively. 
For convergence, we applied a Fock (Kohn-Sham) matrix mixing of 97\% between subsequent self-consistent field cycles and the convergence on total energy was set to $10^{-8}$ Hartrees.

In order to evaluate the effect of the band structure distortions on the C~$1s$ spectra at  $\Gamma_1$ and $\Gamma_2$, we have employed a tight-binding model and followed the procedure of  Ref.~\onlinecite{MOSER201729} to calculate the modulus-squares of the Bloch-states' Fourier components at $\Gamma_1$ and $\Gamma_2$ ---which are directly related to the photoemission intensities~\cite{Shirley:1995aa,Mucha-Kruczynski:2008aa}.  The tight-binding parameters where obtained from the  DFT band structures. In order to incorporate the lattice deformations, we assumed that the hopping between the $s$-orbitals decreases exponentially as a function of the atomic distances. Our model consists of three parameters: the hopping between $s$-orbitals at the equilibrium distance, the exponential decay rate and the on-site energy (i.e., the energy distance with respect to the vacuum level).
Note that the relative photoemission intensities between  $\Gamma_1$ and $\Gamma_2$ are not directly comparable. 

The effect of a static lattice distortion following the displacement pattern of the $E_{2g} (LO)$ mode at $\Gamma$ is shown in Figure ~\ref{Fig:S12}(b) and (c), starting from the equilibrium dispersion in Figure \ref{Fig:S12}(a). Displacements of $\pm$~\SI{1}{pm} are applied along the arrows in the inset of   Figure ~\ref{Fig:S12}(a). The lattice distortion results in a small overall shift of the band structure to lower binding energy and hardly any effect on $\Delta (\Gamma_1, \Gamma_2)$.
A noticeable effect on the $\Delta (\Gamma_1, \Gamma_2)$ can only be obtained by going to larger displacements but these can be ruled out because they would result in an  \textit{increase} of $\Delta (\Gamma_1, \Gamma_2)$ ---which is opposite to the observed behavior.
Moreover, larger displacements would induce major changes of the band structure near the Fermi level (not shown), which are not observed experimentally \cite{Gierz:2013aa,Johannsen:2013ab,Johannsen:2015aa,Aeschlimann:2017aa,Rohde:2018ab}. Figure \ref{Fig:S12}(d) and (e) show the simulated photoemission spectra at  $\Gamma_1$ and $\Gamma_2$. Each panel shows the photoemission intensity for the equilibrium structure and the two distorted structures, such that the peak colors corresponds to the band colors in Figure \ref{Fig:S12}(a)-(c). In the first BZ zone, at $\Gamma_1$, photoemission from the anti-bonding bands is completely suppressed whereas near $\Gamma_2$ the bonding bands still have some spectral weight. As expected from the band structure calculations, the core levels spectra merely show a small shift. 

Fig.~\ref{Fig:S12}(f)-(j) shows the corresponding results for the  $A_1'$ phonon at K. In contrast to the  $E_{2g}$ mode, the static lattice distortion results in a larger unit cell in real space and, therefore, a smaller BZ (see the unit cell indicated in gray in Figure~\ref{Fig:S12}(f)). The DFT band structures are shown within this supercell BZ. The smaller size of the BZ causes a back-folding of the band structure from the original BZ and the resulting new bands at $\Gamma$ are shown in gray. In principle, these bands could be relevant for both the broadening of the C $1s$ peak and the reduction of $\Delta (\Gamma_1, \Gamma_2)$.
In order to simulate the photoemission intensities at $\Gamma_1$ and  $\Gamma_2$, we used our tight-binding model for the supercell shown in the inset of Fig. \ref{Fig:S12}(f), adapated the tight-binding bands to be similar to the DFT bands in  Figure \ref{Fig:S12}(f)-(h) and then unfolded those bands into the BZ of undeformed graphene. The resulting intensities are shown in Figure~\ref{Fig:S12}(i) and (j). As a result of the photoemission matrix elements (essentially due to the same sub-lattice interference effects that suppress the bonding band at $\Gamma_2$~\cite{Shirley:1995aa,Mucha-Kruczynski:2008aa}), the intensity from the gray back-folded bands is very small compared to the other bands. The presence of these bands can thus not explain the observed reduction of $\Delta (\Gamma_1, \Gamma_2)$ or the broadening of the peak. 
We also note that, as for the $E_{2g}$ mode, larger unphysical static distortions would lead to an apparent increase of $\Delta(\Gamma_1, \Gamma_2)$, rather than to the observed decrease.

Overall, our calculations suggest that neither the broadening nor the $\Delta (\Gamma_1, \Gamma_2)$ reduction can be explained in terms of a band structure change caused by the excitation of the $E_{2g} (LO)$ and $A_1' (TO)$ optical phonons.

\section{GW corrected band structure}
\label{appendix:gwbands}
In principle, the ultrafast reduction of $\Delta (\Gamma_1, \Gamma_2)$ may be caused by temperature-dependent effects on the electronic structure.
In order to exclude this, we have performed finite-temperature GW calculations of the C~$1s$ band structure. 
For this purpose we calculated the electronic structure using the FlapwMBPT software package \cite{flapwmbpt} with the generalized gradient approximation as parameterized in  Perdew-Burke-Ernzerhof \cite{perdew1996generalized}. We employed an all-electron basis set and a $21\times21\times2$ $k$-grid. 15 steps of quasi-self consistent GW were sufficient to converge $\Delta (\Gamma_1, \Gamma_2)$ to better than \SI{1.0}{meV}.
Interestingly, at  
zero-temperature we find that $\Delta_{T=0} (\Gamma_1, \Gamma_2)= $\SI{46}{meV}
---which is in better agreement with the experimental result than the previously reported DFT value of \SI{25}{meV}~\cite{Lizzit:2010aa}.
However, no appreciable temperature dependence of the band structure is observed.
In fact, we also find 
$\Delta_{T=3500 \mathrm{K}} (\Gamma_1, \Gamma_2)=$ \SI{46}{meV}.
This suggests that the time-dependent variation of the C~$1s$ line shape cannot be ascribed to a temperature dependence of electronic many-body effects.

\begin{acknowledgments}
This work was supported by VILLUM FONDEN via the Centre of Excellence for Dirac Materials (Grant No. 11744). 
    S. U. acknowledges financial support from VILLUM FONDEN under the Young Investigator Program (grant no. 15375).
     J. A. M. and S. U. acknowledge financial support from the Danish Council for Independent Research, Natural Sciences under the Sapere Aude program (Grant Nos. DFF-6108-00409 and DFF-9064-00057B).
    The DESY team acknowledges financial support from the Deutsche Forschungsgemeisnchaft (DFG, German Research Foundation) -- SFB 925 -- Project ID 170620586.
    We thank Holger Meyer and Sven Gieschen from the University of Hamburg for HEXTOF instrumentation support, as well as Fabio Caruso for helpful discussions.
    This research was carried out at FLASH at DESY, a member of the Helmholtz Association.
    The research leading to this result has been supported by the project ``CALIPSOplus'' under the Grant Agreement 730872 from the EU Framework Programme for Research and Innovation HORIZON 2020".
    Funding by the BMBF (Grant No. 05K19PGA) is gratefully acknowledged by C. T. and Y.-J. C.
    S. P. acknowledges supports from Spanish MINECO for the computational resources provided through Grant No. PID2019-109539GB-C43.

\end{acknowledgments}

\begin{thebibliography}{53}%
\makeatletter
\providecommand \@ifxundefined [1]{%
 \@ifx{#1\undefined}
}%
\providecommand \@ifnum [1]{%
 \ifnum #1\expandafter \@firstoftwo
 \else \expandafter \@secondoftwo
 \fi
}%
\providecommand \@ifx [1]{%
 \ifx #1\expandafter \@firstoftwo
 \else \expandafter \@secondoftwo
 \fi
}%
\providecommand \natexlab [1]{#1}%
\providecommand \enquote  [1]{``#1''}%
\providecommand \bibnamefont  [1]{#1}%
\providecommand \bibfnamefont [1]{#1}%
\providecommand \citenamefont [1]{#1}%
\providecommand \href@noop [0]{\@secondoftwo}%
\providecommand \href [0]{\begingroup \@sanitize@url \@href}%
\providecommand \@href[1]{\@@startlink{#1}\@@href}%
\providecommand \@@href[1]{\endgroup#1\@@endlink}%
\providecommand \@sanitize@url [0]{\catcode `\\12\catcode `\$12\catcode
  `\&12\catcode `\#12\catcode `\^12\catcode `\_12\catcode `\%12\relax}%
\providecommand \@@startlink[1]{}%
\providecommand \@@endlink[0]{}%
\providecommand \url  [0]{\begingroup\@sanitize@url \@url }%
\providecommand \@url [1]{\endgroup\@href {#1}{\urlprefix }}%
\providecommand \urlprefix  [0]{URL }%
\providecommand \Eprint [0]{\href }%
\providecommand \doibase [0]{https://doi.org/}%
\providecommand \selectlanguage [0]{\@gobble}%
\providecommand \bibinfo  [0]{\@secondoftwo}%
\providecommand \bibfield  [0]{\@secondoftwo}%
\providecommand \translation [1]{[#1]}%
\providecommand \BibitemOpen [0]{}%
\providecommand \bibitemStop [0]{}%
\providecommand \bibitemNoStop [0]{.\EOS\space}%
\providecommand \EOS [0]{\spacefactor3000\relax}%
\providecommand \BibitemShut  [1]{\csname bibitem#1\endcsname}%
\let\auto@bib@innerbib\@empty
\bibitem [{\citenamefont {Nordling}\ \emph {et~al.}(1957)\citenamefont
  {Nordling}, \citenamefont {Sokolowski},\ and\ \citenamefont
  {Siegbahn}}]{Nordling:1957aa}%
  \BibitemOpen
  \bibfield  {author} {\bibinfo {author} {\bibfnamefont {C.}~\bibnamefont
  {Nordling}}, \bibinfo {author} {\bibfnamefont {E.}~\bibnamefont
  {Sokolowski}},\ and\ \bibinfo {author} {\bibfnamefont {K.}~\bibnamefont
  {Siegbahn}},\ }\bibfield  {title} {\bibinfo {title} {Precision method for
  obtaining absolute values of atomic binding energies},\ }\href
  {https://doi.org/10.1103/PhysRev.105.1676} {\bibfield  {journal} {\bibinfo
  {journal} {Phys. Rev.}\ }\textbf {\bibinfo {volume} {105}},\ \bibinfo {pages}
  {1676} (\bibinfo {year} {1957})}\BibitemShut {NoStop}%
\bibitem [{\citenamefont {Citrin}\ \emph {et~al.}(1977)\citenamefont {Citrin},
  \citenamefont {Wertheim},\ and\ \citenamefont {Baer}}]{Citrin:1977aa}%
  \BibitemOpen
  \bibfield  {author} {\bibinfo {author} {\bibfnamefont {P.~H.}\ \bibnamefont
  {Citrin}}, \bibinfo {author} {\bibfnamefont {G.~K.}\ \bibnamefont
  {Wertheim}},\ and\ \bibinfo {author} {\bibfnamefont {Y.}~\bibnamefont
  {Baer}},\ }\bibfield  {title} {\bibinfo {title} {Many-body processes in x-ray
  photoemission line shapes from {Li, Na, Mg, and Al} metals},\ }\href@noop {}
  {\bibfield  {journal} {\bibinfo  {journal} {Physical Review B}\ }\textbf
  {\bibinfo {volume} {16}},\ \bibinfo {pages} {4256} (\bibinfo {year}
  {1977})}\BibitemShut {NoStop}%
\bibitem [{\citenamefont {Doniach}\ and\ \citenamefont
  {Sunjic}(1970)}]{Doniach:1970aa}%
  \BibitemOpen
  \bibfield  {author} {\bibinfo {author} {\bibfnamefont {S.}~\bibnamefont
  {Doniach}}\ and\ \bibinfo {author} {\bibfnamefont {M.}~\bibnamefont
  {Sunjic}},\ }\bibfield  {title} {\bibinfo {title} {Many-electron singularity
  in x-ray photoemission and x-ray line spectra from metals},\ }\href@noop {}
  {\bibfield  {journal} {\bibinfo  {journal} {Journal of Physics C: Solid State
  Physics}\ }\textbf {\bibinfo {volume} {3}},\ \bibinfo {pages} {285 }
  (\bibinfo {year} {1970})}\BibitemShut {NoStop}%
\bibitem [{\citenamefont {Satpathy}\ and\ \citenamefont
  {Dow}(1982)}]{Satpathy:1982aa}%
  \BibitemOpen
  \bibfield  {author} {\bibinfo {author} {\bibfnamefont {S.}~\bibnamefont
  {Satpathy}}\ and\ \bibinfo {author} {\bibfnamefont {J.~D.}\ \bibnamefont
  {Dow}},\ }\bibfield  {title} {\bibinfo {title} {Temperature dependence of
  x-ray photoemission spectra: Fermi-sea recoil effects},\ }\href
  {https://doi.org/https://doi.org/10.1016/0038-1098(82)90808-0} {\bibfield
  {journal} {\bibinfo  {journal} {Solid State Communications}\ }\textbf
  {\bibinfo {volume} {42}},\ \bibinfo {pages} {637 } (\bibinfo {year}
  {1982})}\BibitemShut {NoStop}%
\bibitem [{\citenamefont {de~la Torre}\ \emph {et~al.}(2021)\citenamefont
  {de~la Torre}, \citenamefont {Kennes}, \citenamefont {Claassen},
  \citenamefont {Gerber}, \citenamefont {McIver},\ and\ \citenamefont
  {Sentef}}]{Torre:2021wg}%
  \BibitemOpen
  \bibfield  {author} {\bibinfo {author} {\bibfnamefont {A.}~\bibnamefont
  {de~la Torre}}, \bibinfo {author} {\bibfnamefont {D.~M.}\ \bibnamefont
  {Kennes}}, \bibinfo {author} {\bibfnamefont {M.}~\bibnamefont {Claassen}},
  \bibinfo {author} {\bibfnamefont {S.}~\bibnamefont {Gerber}}, \bibinfo
  {author} {\bibfnamefont {J.~W.}\ \bibnamefont {McIver}},\ and\ \bibinfo
  {author} {\bibfnamefont {M.~A.}\ \bibnamefont {Sentef}},\ }\href@noop {}
  {\bibinfo {title} {Nonthermal pathways to ultrafast control in quantum
  materials}} (\bibinfo {year} {2021}),\ \Eprint
  {https://arxiv.org/abs/2103.14888} {arXiv:2103.14888 [cond-mat.str-el]}
  \BibitemShut {NoStop}%
\bibitem [{\citenamefont {Zong}\ \emph {et~al.}(2021)\citenamefont {Zong},
  \citenamefont {Kogar},\ and\ \citenamefont {Gedik}}]{Zong:2021ui}%
  \BibitemOpen
  \bibfield  {author} {\bibinfo {author} {\bibfnamefont {A.}~\bibnamefont
  {Zong}}, \bibinfo {author} {\bibfnamefont {A.}~\bibnamefont {Kogar}},\ and\
  \bibinfo {author} {\bibfnamefont {N.}~\bibnamefont {Gedik}},\ }\href@noop {}
  {\bibinfo {title} {Unconventional light-induced states visualized by
  ultrafast electron diffraction and microscopy}} (\bibinfo {year} {2021}),\
  \Eprint {https://arxiv.org/abs/2104.10373} {arXiv:2104.10373
  [cond-mat.mtrl-sci]} \BibitemShut {NoStop}%
\bibitem [{\citenamefont {Passlack}\ \emph {et~al.}(2006)\citenamefont
  {Passlack}, \citenamefont {Mathias}, \citenamefont {Andreyev}, \citenamefont
  {Mittnacht}, \citenamefont {Aeschlimann},\ and\ \citenamefont
  {Bauer}}]{Passlack:2006aa}%
  \BibitemOpen
  \bibfield  {author} {\bibinfo {author} {\bibfnamefont {S.}~\bibnamefont
  {Passlack}}, \bibinfo {author} {\bibfnamefont {S.}~\bibnamefont {Mathias}},
  \bibinfo {author} {\bibfnamefont {O.}~\bibnamefont {Andreyev}}, \bibinfo
  {author} {\bibfnamefont {D.}~\bibnamefont {Mittnacht}}, \bibinfo {author}
  {\bibfnamefont {M.}~\bibnamefont {Aeschlimann}},\ and\ \bibinfo {author}
  {\bibfnamefont {M.}~\bibnamefont {Bauer}},\ }\bibfield  {title} {\bibinfo
  {title} {Space charge effects in photoemission with a low repetition, high
  intensity femtosecond laser source},\ }\href
  {https://doi.org/10.1063/1.2217985} {\bibfield  {journal} {\bibinfo
  {journal} {Journal of Applied Physics}\ }\textbf {\bibinfo {volume} {100}},\
  \bibinfo {pages} {024912} (\bibinfo {year} {2006})}\BibitemShut {NoStop}%
\bibitem [{\citenamefont {Pietzsch}\ \emph {et~al.}(2008)\citenamefont
  {Pietzsch}, \citenamefont {F{\"o}hlisch}, \citenamefont {Beye}, \citenamefont
  {Deppe}, \citenamefont {Hennies}, \citenamefont {Nagasono}, \citenamefont
  {Suljoti}, \citenamefont {Wurth}, \citenamefont {Gahl}, \citenamefont
  {D{\"o}brich},\ and\ \citenamefont {Melnikov}}]{Pietzsch:2008aa}%
  \BibitemOpen
  \bibfield  {author} {\bibinfo {author} {\bibfnamefont {A.}~\bibnamefont
  {Pietzsch}}, \bibinfo {author} {\bibfnamefont {A.}~\bibnamefont
  {F{\"o}hlisch}}, \bibinfo {author} {\bibfnamefont {M.}~\bibnamefont {Beye}},
  \bibinfo {author} {\bibfnamefont {M.}~\bibnamefont {Deppe}}, \bibinfo
  {author} {\bibfnamefont {F.}~\bibnamefont {Hennies}}, \bibinfo {author}
  {\bibfnamefont {M.}~\bibnamefont {Nagasono}}, \bibinfo {author}
  {\bibfnamefont {E.}~\bibnamefont {Suljoti}}, \bibinfo {author} {\bibfnamefont
  {W.}~\bibnamefont {Wurth}}, \bibinfo {author} {\bibfnamefont
  {C.}~\bibnamefont {Gahl}}, \bibinfo {author} {\bibfnamefont {K.}~\bibnamefont
  {D{\"o}brich}},\ and\ \bibinfo {author} {\bibfnamefont {A.}~\bibnamefont
  {Melnikov}},\ }\bibfield  {title} {\bibinfo {title} {Towards time resolved
  core level photoelectron spectroscopy with femtosecond x-ray free-electron
  lasers},\ }\href {https://doi.org/10.1088/1367-2630/10/3/033004} {\bibfield
  {journal} {\bibinfo  {journal} {New Journal of Physics}\ }\textbf {\bibinfo
  {volume} {10}},\ \bibinfo {pages} {033004} (\bibinfo {year}
  {2008})}\BibitemShut {NoStop}%
\bibitem [{\citenamefont {Hellmann}\ \emph {et~al.}(2009)\citenamefont
  {Hellmann}, \citenamefont {Rossnagel}, \citenamefont {Marczynski-B\"uhlow},\
  and\ \citenamefont {Kipp}}]{Hellmann:2009aa}%
  \BibitemOpen
  \bibfield  {author} {\bibinfo {author} {\bibfnamefont {S.}~\bibnamefont
  {Hellmann}}, \bibinfo {author} {\bibfnamefont {K.}~\bibnamefont {Rossnagel}},
  \bibinfo {author} {\bibfnamefont {M.}~\bibnamefont {Marczynski-B\"uhlow}},\
  and\ \bibinfo {author} {\bibfnamefont {L.}~\bibnamefont {Kipp}},\ }\bibfield
  {title} {\bibinfo {title} {Vacuum space-charge effects in solid-state
  photoemission},\ }\href {https://doi.org/10.1103/PhysRevB.79.035402}
  {\bibfield  {journal} {\bibinfo  {journal} {Phys. Rev. B}\ }\textbf {\bibinfo
  {volume} {79}},\ \bibinfo {pages} {035402} (\bibinfo {year}
  {2009})}\BibitemShut {NoStop}%
\bibitem [{\citenamefont {Hellmann}\ \emph {et~al.}(2012)\citenamefont
  {Hellmann}, \citenamefont {Sohrt}, \citenamefont {Beye}, \citenamefont
  {Rohwer}, \citenamefont {Sorgenfrei}, \citenamefont {Marczynski-B{\"u}hlow},
  \citenamefont {Kall{\"a}ne}, \citenamefont {Redlin}, \citenamefont {Hennies},
  \citenamefont {Bauer},\ and\ \citenamefont {Rossnagel}}]{Hellmann:2012ab}%
  \BibitemOpen
  \bibfield  {author} {\bibinfo {author} {\bibfnamefont {S.}~\bibnamefont
  {Hellmann}}, \bibinfo {author} {\bibfnamefont {C.}~\bibnamefont {Sohrt}},
  \bibinfo {author} {\bibfnamefont {M.}~\bibnamefont {Beye}}, \bibinfo {author}
  {\bibfnamefont {T.}~\bibnamefont {Rohwer}}, \bibinfo {author} {\bibfnamefont
  {F.}~\bibnamefont {Sorgenfrei}}, \bibinfo {author} {\bibfnamefont
  {M.}~\bibnamefont {Marczynski-B{\"u}hlow}}, \bibinfo {author} {\bibfnamefont
  {M.}~\bibnamefont {Kall{\"a}ne}}, \bibinfo {author} {\bibfnamefont
  {H.}~\bibnamefont {Redlin}}, \bibinfo {author} {\bibfnamefont
  {F.}~\bibnamefont {Hennies}}, \bibinfo {author} {\bibfnamefont
  {M.}~\bibnamefont {Bauer}},\ and\ \bibinfo {author} {\bibfnamefont
  {K.}~\bibnamefont {Rossnagel}},\ }\bibfield  {title} {\bibinfo {title}
  {Time-resolved x-ray photoelectron spectroscopy at {FLASH}},\ }\href
  {https://doi.org/10.1088/1367-2630/14/1/013062} {\bibfield  {journal}
  {\bibinfo  {journal} {New Journal of Physics}\ }\textbf {\bibinfo {volume}
  {14}},\ \bibinfo {pages} {013062} (\bibinfo {year} {2012})}\BibitemShut
  {NoStop}%
\bibitem [{\citenamefont {Sch{\"o}nhense}\ \emph {et~al.}(2021)\citenamefont
  {Sch{\"o}nhense}, \citenamefont {Kutnyakhov}, \citenamefont {Pressacco},
  \citenamefont {Heber}, \citenamefont {Wind}, \citenamefont {Agustsson},
  \citenamefont {Babenkov}, \citenamefont {Vasilyev}, \citenamefont
  {Fedchenko}, \citenamefont {Chernov}, \citenamefont {Rettig}, \citenamefont
  {Sch{\"o}nhense}, \citenamefont {Wenthaus}, \citenamefont {Brenner},
  \citenamefont {Dziarzhytski}, \citenamefont {Paluttke}, \citenamefont
  {Mahatha}, \citenamefont {Schirmel}, \citenamefont {Redlin}, \citenamefont
  {Manschwetus}, \citenamefont {Hartl}, \citenamefont {Matveyev}, \citenamefont
  {Gloskovskii}, \citenamefont {Schlueter}, \citenamefont {Shokeen},
  \citenamefont {Duerr}, \citenamefont {Allison}, \citenamefont {Beye},
  \citenamefont {Rossnagel}, \citenamefont {Elmers},\ and\ \citenamefont
  {Medjanik}}]{Schonhense:2021aa}%
  \BibitemOpen
  \bibfield  {author} {\bibinfo {author} {\bibfnamefont {G.}~\bibnamefont
  {Sch{\"o}nhense}}, \bibinfo {author} {\bibfnamefont {D.}~\bibnamefont
  {Kutnyakhov}}, \bibinfo {author} {\bibfnamefont {F.}~\bibnamefont
  {Pressacco}}, \bibinfo {author} {\bibfnamefont {M.}~\bibnamefont {Heber}},
  \bibinfo {author} {\bibfnamefont {N.}~\bibnamefont {Wind}}, \bibinfo {author}
  {\bibfnamefont {S.~Y.}\ \bibnamefont {Agustsson}}, \bibinfo {author}
  {\bibfnamefont {S.}~\bibnamefont {Babenkov}}, \bibinfo {author}
  {\bibfnamefont {D.}~\bibnamefont {Vasilyev}}, \bibinfo {author}
  {\bibfnamefont {O.}~\bibnamefont {Fedchenko}}, \bibinfo {author}
  {\bibfnamefont {S.}~\bibnamefont {Chernov}}, \bibinfo {author} {\bibfnamefont
  {L.}~\bibnamefont {Rettig}}, \bibinfo {author} {\bibfnamefont
  {B.}~\bibnamefont {Sch{\"o}nhense}}, \bibinfo {author} {\bibfnamefont
  {L.}~\bibnamefont {Wenthaus}}, \bibinfo {author} {\bibfnamefont
  {G.}~\bibnamefont {Brenner}}, \bibinfo {author} {\bibfnamefont
  {S.}~\bibnamefont {Dziarzhytski}}, \bibinfo {author} {\bibfnamefont
  {S.}~\bibnamefont {Paluttke}}, \bibinfo {author} {\bibfnamefont {S.~K.}\
  \bibnamefont {Mahatha}}, \bibinfo {author} {\bibfnamefont {N.}~\bibnamefont
  {Schirmel}}, \bibinfo {author} {\bibfnamefont {H.}~\bibnamefont {Redlin}},
  \bibinfo {author} {\bibfnamefont {B.}~\bibnamefont {Manschwetus}}, \bibinfo
  {author} {\bibfnamefont {I.}~\bibnamefont {Hartl}}, \bibinfo {author}
  {\bibfnamefont {Y.}~\bibnamefont {Matveyev}}, \bibinfo {author}
  {\bibfnamefont {A.}~\bibnamefont {Gloskovskii}}, \bibinfo {author}
  {\bibfnamefont {C.}~\bibnamefont {Schlueter}}, \bibinfo {author}
  {\bibfnamefont {V.}~\bibnamefont {Shokeen}}, \bibinfo {author} {\bibfnamefont
  {H.}~\bibnamefont {Duerr}}, \bibinfo {author} {\bibfnamefont {T.~K.}\
  \bibnamefont {Allison}}, \bibinfo {author} {\bibfnamefont {M.}~\bibnamefont
  {Beye}}, \bibinfo {author} {\bibfnamefont {K.}~\bibnamefont {Rossnagel}},
  \bibinfo {author} {\bibfnamefont {H.~J.}\ \bibnamefont {Elmers}},\ and\
  \bibinfo {author} {\bibfnamefont {K.}~\bibnamefont {Medjanik}},\ }\bibfield
  {title} {\bibinfo {title} {Suppression of the vacuum space-charge effect in
  fs-photoemission by a retarding electrostatic front lens},\ }\href
  {https://doi.org/10.1063/5.0046567} {\bibfield  {journal} {\bibinfo
  {journal} {Review of Scientific Instruments}\ }\textbf {\bibinfo {volume}
  {92}},\ \bibinfo {pages} {000000} (\bibinfo {year} {2021})}\BibitemShut
  {NoStop}%
\bibitem [{\citenamefont {Dendzik}\ \emph {et~al.}(2020)\citenamefont
  {Dendzik}, \citenamefont {Xian}, \citenamefont {Perfetto}, \citenamefont
  {Sangalli}, \citenamefont {Kutnyakhov}, \citenamefont {Dong}, \citenamefont
  {Beaulieu}, \citenamefont {Pincelli}, \citenamefont {Pressacco},
  \citenamefont {Curcio}, \citenamefont {Agustsson}, \citenamefont {Heber},
  \citenamefont {Hauer}, \citenamefont {Wurth}, \citenamefont {Brenner},
  \citenamefont {Acremann}, \citenamefont {Hofmann}, \citenamefont {Wolf},
  \citenamefont {Marini}, \citenamefont {Stefanucci}, \citenamefont {Rettig},\
  and\ \citenamefont {Ernstorfer}}]{Dendzik:2020aa}%
  \BibitemOpen
  \bibfield  {author} {\bibinfo {author} {\bibfnamefont {M.}~\bibnamefont
  {Dendzik}}, \bibinfo {author} {\bibfnamefont {R.~P.}\ \bibnamefont {Xian}},
  \bibinfo {author} {\bibfnamefont {E.}~\bibnamefont {Perfetto}}, \bibinfo
  {author} {\bibfnamefont {D.}~\bibnamefont {Sangalli}}, \bibinfo {author}
  {\bibfnamefont {D.}~\bibnamefont {Kutnyakhov}}, \bibinfo {author}
  {\bibfnamefont {S.}~\bibnamefont {Dong}}, \bibinfo {author} {\bibfnamefont
  {S.}~\bibnamefont {Beaulieu}}, \bibinfo {author} {\bibfnamefont
  {T.}~\bibnamefont {Pincelli}}, \bibinfo {author} {\bibfnamefont
  {F.}~\bibnamefont {Pressacco}}, \bibinfo {author} {\bibfnamefont
  {D.}~\bibnamefont {Curcio}}, \bibinfo {author} {\bibfnamefont {S.~Y.}\
  \bibnamefont {Agustsson}}, \bibinfo {author} {\bibfnamefont {M.}~\bibnamefont
  {Heber}}, \bibinfo {author} {\bibfnamefont {J.}~\bibnamefont {Hauer}},
  \bibinfo {author} {\bibfnamefont {W.}~\bibnamefont {Wurth}}, \bibinfo
  {author} {\bibfnamefont {G.}~\bibnamefont {Brenner}}, \bibinfo {author}
  {\bibfnamefont {Y.}~\bibnamefont {Acremann}}, \bibinfo {author}
  {\bibfnamefont {P.}~\bibnamefont {Hofmann}}, \bibinfo {author} {\bibfnamefont
  {M.}~\bibnamefont {Wolf}}, \bibinfo {author} {\bibfnamefont {A.}~\bibnamefont
  {Marini}}, \bibinfo {author} {\bibfnamefont {G.}~\bibnamefont {Stefanucci}},
  \bibinfo {author} {\bibfnamefont {L.}~\bibnamefont {Rettig}},\ and\ \bibinfo
  {author} {\bibfnamefont {R.}~\bibnamefont {Ernstorfer}},\ }\bibfield  {title}
  {\bibinfo {title} {Observation of an excitonic {M}ott transition through
  ultrafast core-cum-conduction photoemission spectroscopy},\ }\href
  {https://doi.org/10.1103/PhysRevLett.125.096401} {\bibfield  {journal}
  {\bibinfo  {journal} {Phys. Rev. Lett.}\ }\textbf {\bibinfo {volume} {125}},\
  \bibinfo {pages} {096401} (\bibinfo {year} {2020})}\BibitemShut {NoStop}%
\bibitem [{\citenamefont {Gierz}\ \emph {et~al.}(2013)\citenamefont {Gierz},
  \citenamefont {Petersen}, \citenamefont {Mitrano}, \citenamefont {Cacho},
  \citenamefont {Turcu}, \citenamefont {Springate}, \citenamefont {St{\"o}hr},
  \citenamefont {K{\"o}hler}, \citenamefont {Starke},\ and\ \citenamefont
  {Cavalleri}}]{Gierz:2013aa}%
  \BibitemOpen
  \bibfield  {author} {\bibinfo {author} {\bibfnamefont {I.}~\bibnamefont
  {Gierz}}, \bibinfo {author} {\bibfnamefont {J.~C.}\ \bibnamefont {Petersen}},
  \bibinfo {author} {\bibfnamefont {M.}~\bibnamefont {Mitrano}}, \bibinfo
  {author} {\bibfnamefont {C.}~\bibnamefont {Cacho}}, \bibinfo {author}
  {\bibfnamefont {I.~C.~E.}\ \bibnamefont {Turcu}}, \bibinfo {author}
  {\bibfnamefont {E.}~\bibnamefont {Springate}}, \bibinfo {author}
  {\bibfnamefont {A.}~\bibnamefont {St{\"o}hr}}, \bibinfo {author}
  {\bibfnamefont {A.}~\bibnamefont {K{\"o}hler}}, \bibinfo {author}
  {\bibfnamefont {U.}~\bibnamefont {Starke}},\ and\ \bibinfo {author}
  {\bibfnamefont {A.}~\bibnamefont {Cavalleri}},\ }\bibfield  {title} {\bibinfo
  {title} {Snapshots of non-equilibrium {Dirac} carrier distributions in
  graphene},\ }\href {http://dx.doi.org/10.1038/nmat3757} {\bibfield  {journal}
  {\bibinfo  {journal} {Nature Materials}\ }\textbf {\bibinfo {volume} {12}},\
  \bibinfo {pages} {1119} (\bibinfo {year} {2013})}\BibitemShut {NoStop}%
\bibitem [{\citenamefont {Johannsen}\ \emph {et~al.}(2013)\citenamefont
  {Johannsen}, \citenamefont {Ulstrup}, \citenamefont {Cilento}, \citenamefont
  {Crepaldi}, \citenamefont {Zacchigna}, \citenamefont {Cacho}, \citenamefont
  {Turcu}, \citenamefont {Springate}, \citenamefont {Fromm}, \citenamefont
  {Raidel}, \citenamefont {Seyller}, \citenamefont {Parmigiani}, \citenamefont
  {Grioni},\ and\ \citenamefont {Hofmann}}]{Johannsen:2013ab}%
  \BibitemOpen
  \bibfield  {author} {\bibinfo {author} {\bibfnamefont {J.~C.}\ \bibnamefont
  {Johannsen}}, \bibinfo {author} {\bibfnamefont {S.}~\bibnamefont {Ulstrup}},
  \bibinfo {author} {\bibfnamefont {F.}~\bibnamefont {Cilento}}, \bibinfo
  {author} {\bibfnamefont {A.}~\bibnamefont {Crepaldi}}, \bibinfo {author}
  {\bibfnamefont {M.}~\bibnamefont {Zacchigna}}, \bibinfo {author}
  {\bibfnamefont {C.}~\bibnamefont {Cacho}}, \bibinfo {author} {\bibfnamefont
  {I.~C.~E.}\ \bibnamefont {Turcu}}, \bibinfo {author} {\bibfnamefont
  {E.}~\bibnamefont {Springate}}, \bibinfo {author} {\bibfnamefont
  {F.}~\bibnamefont {Fromm}}, \bibinfo {author} {\bibfnamefont
  {C.}~\bibnamefont {Raidel}}, \bibinfo {author} {\bibfnamefont
  {T.}~\bibnamefont {Seyller}}, \bibinfo {author} {\bibfnamefont
  {F.}~\bibnamefont {Parmigiani}}, \bibinfo {author} {\bibfnamefont
  {M.}~\bibnamefont {Grioni}},\ and\ \bibinfo {author} {\bibfnamefont
  {P.}~\bibnamefont {Hofmann}},\ }\bibfield  {title} {\bibinfo {title} {Direct
  view of hot carrier dynamics in graphene},\ }\href
  {https://doi.org/10.1103/PhysRevLett.111.027403} {\bibfield  {journal}
  {\bibinfo  {journal} {Phys. Rev. Lett.}\ }\textbf {\bibinfo {volume} {111}},\
  \bibinfo {pages} {027403} (\bibinfo {year} {2013})}\BibitemShut {NoStop}%
\bibitem [{\citenamefont {Johannsen}\ \emph {et~al.}(2015)\citenamefont
  {Johannsen}, \citenamefont {Ulstrup}, \citenamefont {Crepaldi}, \citenamefont
  {Cilento}, \citenamefont {Zacchigna}, \citenamefont {Miwa}, \citenamefont
  {Cacho}, \citenamefont {Chapman}, \citenamefont {Springate}, \citenamefont
  {Fromm}, \citenamefont {Raidel}, \citenamefont {Seyller}, \citenamefont
  {King}, \citenamefont {Parmigiani}, \citenamefont {Grioni},\ and\
  \citenamefont {Hofmann}}]{Johannsen:2015aa}%
  \BibitemOpen
  \bibfield  {author} {\bibinfo {author} {\bibfnamefont {J.~C.}\ \bibnamefont
  {Johannsen}}, \bibinfo {author} {\bibfnamefont {S.}~\bibnamefont {Ulstrup}},
  \bibinfo {author} {\bibfnamefont {A.}~\bibnamefont {Crepaldi}}, \bibinfo
  {author} {\bibfnamefont {F.}~\bibnamefont {Cilento}}, \bibinfo {author}
  {\bibfnamefont {M.}~\bibnamefont {Zacchigna}}, \bibinfo {author}
  {\bibfnamefont {J.~A.}\ \bibnamefont {Miwa}}, \bibinfo {author}
  {\bibfnamefont {C.}~\bibnamefont {Cacho}}, \bibinfo {author} {\bibfnamefont
  {R.~T.}\ \bibnamefont {Chapman}}, \bibinfo {author} {\bibfnamefont
  {E.}~\bibnamefont {Springate}}, \bibinfo {author} {\bibfnamefont
  {F.}~\bibnamefont {Fromm}}, \bibinfo {author} {\bibfnamefont
  {C.}~\bibnamefont {Raidel}}, \bibinfo {author} {\bibfnamefont
  {T.}~\bibnamefont {Seyller}}, \bibinfo {author} {\bibfnamefont {P.~D.}\
  \bibnamefont {King}}, \bibinfo {author} {\bibfnamefont {F.}~\bibnamefont
  {Parmigiani}}, \bibinfo {author} {\bibfnamefont {M.}~\bibnamefont {Grioni}},\
  and\ \bibinfo {author} {\bibfnamefont {P.}~\bibnamefont {Hofmann}},\
  }\bibfield  {title} {\bibinfo {title} {Tunable carrier multiplication and
  cooling in graphene},\ }\href {https://doi.org/10.1021/nl503614v} {\bibfield
  {journal} {\bibinfo  {journal} {Nano Letters}\ }\textbf {\bibinfo {volume}
  {15}},\ \bibinfo {pages} {326} (\bibinfo {year} {2015})}\BibitemShut
  {NoStop}%
\bibitem [{\citenamefont {Aeschlimann}\ \emph {et~al.}(2017)\citenamefont
  {Aeschlimann}, \citenamefont {Krause}, \citenamefont {Ch{\'a}vez-Cervantes},
  \citenamefont {Bromberger}, \citenamefont {Jago}, \citenamefont {Mali{\'c}},
  \citenamefont {Al-Temimy}, \citenamefont {Coletti}, \citenamefont
  {Cavalleri},\ and\ \citenamefont {Gierz}}]{Aeschlimann:2017aa}%
  \BibitemOpen
  \bibfield  {author} {\bibinfo {author} {\bibfnamefont {S.}~\bibnamefont
  {Aeschlimann}}, \bibinfo {author} {\bibfnamefont {R.}~\bibnamefont {Krause}},
  \bibinfo {author} {\bibfnamefont {M.}~\bibnamefont {Ch{\'a}vez-Cervantes}},
  \bibinfo {author} {\bibfnamefont {H.}~\bibnamefont {Bromberger}}, \bibinfo
  {author} {\bibfnamefont {R.}~\bibnamefont {Jago}}, \bibinfo {author}
  {\bibfnamefont {E.}~\bibnamefont {Mali{\'c}}}, \bibinfo {author}
  {\bibfnamefont {A.}~\bibnamefont {Al-Temimy}}, \bibinfo {author}
  {\bibfnamefont {C.}~\bibnamefont {Coletti}}, \bibinfo {author} {\bibfnamefont
  {A.}~\bibnamefont {Cavalleri}},\ and\ \bibinfo {author} {\bibfnamefont
  {I.}~\bibnamefont {Gierz}},\ }\bibfield  {title} {\bibinfo {title} {Ultrafast
  momentum imaging of pseudospin-flip excitations in graphene},\ }\href
  {http://dx.doi.org/10.1103/PhysRevB.96.020301} {\bibfield  {journal}
  {\bibinfo  {journal} {Physical Review B}\ }\textbf {\bibinfo {volume} {96}}
  (\bibinfo {year} {2017})}\BibitemShut {NoStop}%
\bibitem [{\citenamefont {Rohde}\ \emph {et~al.}(2018)\citenamefont {Rohde},
  \citenamefont {Stange}, \citenamefont {M{\"u}ller}, \citenamefont {Behrendt},
  \citenamefont {Oloff}, \citenamefont {Hanff}, \citenamefont {Albert},
  \citenamefont {Hein}, \citenamefont {Rossnagel},\ and\ \citenamefont
  {Bauer}}]{Rohde:2018ab}%
  \BibitemOpen
  \bibfield  {author} {\bibinfo {author} {\bibfnamefont {G.}~\bibnamefont
  {Rohde}}, \bibinfo {author} {\bibfnamefont {A.}~\bibnamefont {Stange}},
  \bibinfo {author} {\bibfnamefont {A.}~\bibnamefont {M{\"u}ller}}, \bibinfo
  {author} {\bibfnamefont {M.}~\bibnamefont {Behrendt}}, \bibinfo {author}
  {\bibfnamefont {L.-P.}\ \bibnamefont {Oloff}}, \bibinfo {author}
  {\bibfnamefont {K.}~\bibnamefont {Hanff}}, \bibinfo {author} {\bibfnamefont
  {T.}~\bibnamefont {Albert}}, \bibinfo {author} {\bibfnamefont
  {P.}~\bibnamefont {Hein}}, \bibinfo {author} {\bibfnamefont {K.}~\bibnamefont
  {Rossnagel}},\ and\ \bibinfo {author} {\bibfnamefont {M.}~\bibnamefont
  {Bauer}},\ }\bibfield  {title} {\bibinfo {title} {Ultrafast formation of a
  fermi-dirac distributed electron gas},\ }\href
  {http://dx.doi.org/10.1103/PhysRevLett.121.256401} {\bibfield  {journal}
  {\bibinfo  {journal} {Physical Review Letters}\ }\textbf {\bibinfo {volume}
  {121}} (\bibinfo {year} {2018})}\BibitemShut {NoStop}%
\bibitem [{\citenamefont {Caruso}\ \emph {et~al.}(2020)\citenamefont {Caruso},
  \citenamefont {Novko},\ and\ \citenamefont {Draxl}}]{Caruso:2020aa}%
  \BibitemOpen
  \bibfield  {author} {\bibinfo {author} {\bibfnamefont {F.}~\bibnamefont
  {Caruso}}, \bibinfo {author} {\bibfnamefont {D.}~\bibnamefont {Novko}},\ and\
  \bibinfo {author} {\bibfnamefont {C.}~\bibnamefont {Draxl}},\ }\bibfield
  {title} {\bibinfo {title} {Photoemission signatures of nonequilibrium carrier
  dynamics from first principles},\ }\href
  {https://doi.org/10.1103/PhysRevB.101.035128} {\bibfield  {journal} {\bibinfo
   {journal} {Phys. Rev. B}\ }\textbf {\bibinfo {volume} {101}},\ \bibinfo
  {pages} {035128} (\bibinfo {year} {2020})}\BibitemShut {NoStop}%
\bibitem [{\citenamefont {Pozzo}\ \emph {et~al.}(2011)\citenamefont {Pozzo},
  \citenamefont {Alf\`e}, \citenamefont {Lacovig}, \citenamefont {Hofmann},
  \citenamefont {Lizzit},\ and\ \citenamefont {Baraldi}}]{Pozzo:2011aa}%
  \BibitemOpen
  \bibfield  {author} {\bibinfo {author} {\bibfnamefont {M.}~\bibnamefont
  {Pozzo}}, \bibinfo {author} {\bibfnamefont {D.}~\bibnamefont {Alf\`e}},
  \bibinfo {author} {\bibfnamefont {P.}~\bibnamefont {Lacovig}}, \bibinfo
  {author} {\bibfnamefont {P.}~\bibnamefont {Hofmann}}, \bibinfo {author}
  {\bibfnamefont {S.}~\bibnamefont {Lizzit}},\ and\ \bibinfo {author}
  {\bibfnamefont {A.}~\bibnamefont {Baraldi}},\ }\bibfield  {title} {\bibinfo
  {title} {Thermal expansion of supported and freestanding graphene: Lattice
  constant versus interatomic distance},\ }\href
  {https://doi.org/10.1103/PhysRevLett.106.135501} {\bibfield  {journal}
  {\bibinfo  {journal} {Physical Review Letters}\ }\textbf {\bibinfo {volume}
  {106}},\ \bibinfo {pages} {135501} (\bibinfo {year} {2011})}\BibitemShut
  {NoStop}%
\bibitem [{\citenamefont {Lizzit}\ \emph {et~al.}(2010)\citenamefont {Lizzit},
  \citenamefont {Zampieri}, \citenamefont {Petaccia}, \citenamefont
  {Larciprete}, \citenamefont {Lacovig}, \citenamefont {Rienks}, \citenamefont
  {Bihlmayer}, \citenamefont {Baraldi},\ and\ \citenamefont
  {Hofmann}}]{Lizzit:2010aa}%
  \BibitemOpen
  \bibfield  {author} {\bibinfo {author} {\bibfnamefont {S.}~\bibnamefont
  {Lizzit}}, \bibinfo {author} {\bibfnamefont {G.}~\bibnamefont {Zampieri}},
  \bibinfo {author} {\bibfnamefont {L.}~\bibnamefont {Petaccia}}, \bibinfo
  {author} {\bibfnamefont {R.}~\bibnamefont {Larciprete}}, \bibinfo {author}
  {\bibfnamefont {P.}~\bibnamefont {Lacovig}}, \bibinfo {author} {\bibfnamefont
  {E.~D.~L.}\ \bibnamefont {Rienks}}, \bibinfo {author} {\bibfnamefont
  {G.}~\bibnamefont {Bihlmayer}}, \bibinfo {author} {\bibfnamefont
  {A.}~\bibnamefont {Baraldi}},\ and\ \bibinfo {author} {\bibfnamefont
  {P.}~\bibnamefont {Hofmann}},\ }\bibfield  {title} {\bibinfo {title} {Band
  dispersion in the deep {C} $1s$ core level of graphene},\ }\href@noop {}
  {\bibfield  {journal} {\bibinfo  {journal} {Nature Physics}\ }\textbf
  {\bibinfo {volume} {6}},\ \bibinfo {pages} {345} (\bibinfo {year}
  {2010})}\BibitemShut {NoStop}%
\bibitem [{\citenamefont {Kutnyakhov}\ \emph {et~al.}(2020)\citenamefont
  {Kutnyakhov}, \citenamefont {Xian}, \citenamefont {Dendzik}, \citenamefont
  {Heber}, \citenamefont {Pressacco}, \citenamefont {Agustsson}, \citenamefont
  {Wenthaus}, \citenamefont {Meyer}, \citenamefont {Gieschen}, \citenamefont
  {Mercurio}, \citenamefont {Benz}, \citenamefont {B{\"u}hlman}, \citenamefont
  {D{\"a}ster}, \citenamefont {Gort}, \citenamefont {Curcio}, \citenamefont
  {Volckaert}, \citenamefont {Bianchi}, \citenamefont {Sanders}, \citenamefont
  {Miwa}, \citenamefont {Ulstrup}, \citenamefont {Oelsner}, \citenamefont
  {Tusche}, \citenamefont {Chen}, \citenamefont {Vasilyev}, \citenamefont
  {Medjanik}, \citenamefont {Brenner}, \citenamefont {Dziarzhytski},
  \citenamefont {Redlin}, \citenamefont {Manschwetus}, \citenamefont {Dong},
  \citenamefont {Hauer}, \citenamefont {Rettig}, \citenamefont {Diekmann},
  \citenamefont {Rossnagel}, \citenamefont {Demsar}, \citenamefont {Elmers},
  \citenamefont {Hofmann}, \citenamefont {Ernstorfer}, \citenamefont
  {Sch{\"o}nhense}, \citenamefont {Acremann},\ and\ \citenamefont
  {Wurth}}]{Kutnyakhov:2020aa}%
  \BibitemOpen
  \bibfield  {author} {\bibinfo {author} {\bibfnamefont {D.}~\bibnamefont
  {Kutnyakhov}}, \bibinfo {author} {\bibfnamefont {R.~P.}\ \bibnamefont
  {Xian}}, \bibinfo {author} {\bibfnamefont {M.}~\bibnamefont {Dendzik}},
  \bibinfo {author} {\bibfnamefont {M.}~\bibnamefont {Heber}}, \bibinfo
  {author} {\bibfnamefont {F.}~\bibnamefont {Pressacco}}, \bibinfo {author}
  {\bibfnamefont {S.~Y.}\ \bibnamefont {Agustsson}}, \bibinfo {author}
  {\bibfnamefont {L.}~\bibnamefont {Wenthaus}}, \bibinfo {author}
  {\bibfnamefont {H.}~\bibnamefont {Meyer}}, \bibinfo {author} {\bibfnamefont
  {S.}~\bibnamefont {Gieschen}}, \bibinfo {author} {\bibfnamefont
  {G.}~\bibnamefont {Mercurio}}, \bibinfo {author} {\bibfnamefont
  {A.}~\bibnamefont {Benz}}, \bibinfo {author} {\bibfnamefont {K.}~\bibnamefont
  {B{\"u}hlman}}, \bibinfo {author} {\bibfnamefont {S.}~\bibnamefont
  {D{\"a}ster}}, \bibinfo {author} {\bibfnamefont {R.}~\bibnamefont {Gort}},
  \bibinfo {author} {\bibfnamefont {D.}~\bibnamefont {Curcio}}, \bibinfo
  {author} {\bibfnamefont {K.}~\bibnamefont {Volckaert}}, \bibinfo {author}
  {\bibfnamefont {M.}~\bibnamefont {Bianchi}}, \bibinfo {author} {\bibfnamefont
  {C.}~\bibnamefont {Sanders}}, \bibinfo {author} {\bibfnamefont {J.~A.}\
  \bibnamefont {Miwa}}, \bibinfo {author} {\bibfnamefont {S.}~\bibnamefont
  {Ulstrup}}, \bibinfo {author} {\bibfnamefont {A.}~\bibnamefont {Oelsner}},
  \bibinfo {author} {\bibfnamefont {C.}~\bibnamefont {Tusche}}, \bibinfo
  {author} {\bibfnamefont {Y.-J.}\ \bibnamefont {Chen}}, \bibinfo {author}
  {\bibfnamefont {D.}~\bibnamefont {Vasilyev}}, \bibinfo {author}
  {\bibfnamefont {K.}~\bibnamefont {Medjanik}}, \bibinfo {author}
  {\bibfnamefont {G.}~\bibnamefont {Brenner}}, \bibinfo {author} {\bibfnamefont
  {S.}~\bibnamefont {Dziarzhytski}}, \bibinfo {author} {\bibfnamefont
  {H.}~\bibnamefont {Redlin}}, \bibinfo {author} {\bibfnamefont
  {B.}~\bibnamefont {Manschwetus}}, \bibinfo {author} {\bibfnamefont
  {S.}~\bibnamefont {Dong}}, \bibinfo {author} {\bibfnamefont {J.}~\bibnamefont
  {Hauer}}, \bibinfo {author} {\bibfnamefont {L.}~\bibnamefont {Rettig}},
  \bibinfo {author} {\bibfnamefont {F.}~\bibnamefont {Diekmann}}, \bibinfo
  {author} {\bibfnamefont {K.}~\bibnamefont {Rossnagel}}, \bibinfo {author}
  {\bibfnamefont {J.}~\bibnamefont {Demsar}}, \bibinfo {author} {\bibfnamefont
  {H.-J.}\ \bibnamefont {Elmers}}, \bibinfo {author} {\bibfnamefont
  {P.}~\bibnamefont {Hofmann}}, \bibinfo {author} {\bibfnamefont
  {R.}~\bibnamefont {Ernstorfer}}, \bibinfo {author} {\bibfnamefont
  {G.}~\bibnamefont {Sch{\"o}nhense}}, \bibinfo {author} {\bibfnamefont
  {Y.}~\bibnamefont {Acremann}},\ and\ \bibinfo {author} {\bibfnamefont
  {W.}~\bibnamefont {Wurth}},\ }\bibfield  {title} {\bibinfo {title} {Time- and
  momentum-resolved photoemission studies using time-of-flight momentum
  microscopy at a free-electron laser},\ }\href
  {https://doi.org/10.1063/1.5118777} {\bibfield  {journal} {\bibinfo
  {journal} {Review of Scientific Instruments}\ }\textbf {\bibinfo {volume}
  {91}},\ \bibinfo {pages} {013109} (\bibinfo {year} {2020})},\ \Eprint
  {https://arxiv.org/abs/https://doi.org/10.1063/1.5118777}
  {https://doi.org/10.1063/1.5118777} \BibitemShut {NoStop}%
\bibitem [{\citenamefont {Riedl}\ \emph {et~al.}(2009)\citenamefont {Riedl},
  \citenamefont {Coletti}, \citenamefont {Iwasaki}, \citenamefont {Zakharov},\
  and\ \citenamefont {Starke}}]{Riedl:2009aa}%
  \BibitemOpen
  \bibfield  {author} {\bibinfo {author} {\bibfnamefont {C.}~\bibnamefont
  {Riedl}}, \bibinfo {author} {\bibfnamefont {C.}~\bibnamefont {Coletti}},
  \bibinfo {author} {\bibfnamefont {T.}~\bibnamefont {Iwasaki}}, \bibinfo
  {author} {\bibfnamefont {A.~A.}\ \bibnamefont {Zakharov}},\ and\ \bibinfo
  {author} {\bibfnamefont {U.}~\bibnamefont {Starke}},\ }\bibfield  {title}
  {\bibinfo {title} {Quasi-free-standing epitaxial graphene on {SiC} obtained
  by hydrogen intercalation},\ }\href
  {https://doi.org/10.1103/PhysRevLett.103.246804} {\bibfield  {journal}
  {\bibinfo  {journal} {Physical Review Letters}\ }\textbf {\bibinfo {volume}
  {103}},\ \bibinfo {eid} {246804} (\bibinfo {year} {2009})}\BibitemShut
  {NoStop}%
\bibitem [{\citenamefont {Speck}\ \emph {et~al.}(2011)\citenamefont {Speck},
  \citenamefont {Jobst}, \citenamefont {Fromm}, \citenamefont {Ostler},
  \citenamefont {Waldmann}, \citenamefont {Hundhausen}, \citenamefont {Weber},\
  and\ \citenamefont {Seyller}}]{Speck:2011aa}%
  \BibitemOpen
  \bibfield  {author} {\bibinfo {author} {\bibfnamefont {F.}~\bibnamefont
  {Speck}}, \bibinfo {author} {\bibfnamefont {J.}~\bibnamefont {Jobst}},
  \bibinfo {author} {\bibfnamefont {F.}~\bibnamefont {Fromm}}, \bibinfo
  {author} {\bibfnamefont {M.}~\bibnamefont {Ostler}}, \bibinfo {author}
  {\bibfnamefont {D.}~\bibnamefont {Waldmann}}, \bibinfo {author}
  {\bibfnamefont {M.}~\bibnamefont {Hundhausen}}, \bibinfo {author}
  {\bibfnamefont {H.~B.}\ \bibnamefont {Weber}},\ and\ \bibinfo {author}
  {\bibfnamefont {T.}~\bibnamefont {Seyller}},\ }\bibfield  {title} {\bibinfo
  {title} {The quasi-free-standing nature of graphene on h-saturated
  {SiC}(0001)},\ }\href {https://doi.org/10.1063/1.3643034} {\bibfield
  {journal} {\bibinfo  {journal} {Applied Physics Letters}\ }\textbf {\bibinfo
  {volume} {99}},\ \bibinfo {eid} {122106} (\bibinfo {year}
  {2011})}\BibitemShut {NoStop}%
\bibitem [{\citenamefont {Xian}\ \emph {et~al.}(2020)\citenamefont {Xian},
  \citenamefont {Acremann}, \citenamefont {Agustsson}, \citenamefont {Dendzik},
  \citenamefont {B{\"u}hlmann}, \citenamefont {Curcio}, \citenamefont
  {Kutnyakhov}, \citenamefont {Pressacco}, \citenamefont {Heber}, \citenamefont
  {Dong}, \citenamefont {Pincelli}, \citenamefont {Demsar}, \citenamefont
  {Wurth}, \citenamefont {Hofmann}, \citenamefont {Wolf}, \citenamefont
  {Scheidgen}, \citenamefont {Rettig},\ and\ \citenamefont
  {Ernstorfer}}]{Xian:2020aa}%
  \BibitemOpen
  \bibfield  {author} {\bibinfo {author} {\bibfnamefont {R.~P.}\ \bibnamefont
  {Xian}}, \bibinfo {author} {\bibfnamefont {Y.}~\bibnamefont {Acremann}},
  \bibinfo {author} {\bibfnamefont {S.~Y.}\ \bibnamefont {Agustsson}}, \bibinfo
  {author} {\bibfnamefont {M.}~\bibnamefont {Dendzik}}, \bibinfo {author}
  {\bibfnamefont {K.}~\bibnamefont {B{\"u}hlmann}}, \bibinfo {author}
  {\bibfnamefont {D.}~\bibnamefont {Curcio}}, \bibinfo {author} {\bibfnamefont
  {D.}~\bibnamefont {Kutnyakhov}}, \bibinfo {author} {\bibfnamefont
  {F.}~\bibnamefont {Pressacco}}, \bibinfo {author} {\bibfnamefont
  {M.}~\bibnamefont {Heber}}, \bibinfo {author} {\bibfnamefont
  {S.}~\bibnamefont {Dong}}, \bibinfo {author} {\bibfnamefont {T.}~\bibnamefont
  {Pincelli}}, \bibinfo {author} {\bibfnamefont {J.}~\bibnamefont {Demsar}},
  \bibinfo {author} {\bibfnamefont {W.}~\bibnamefont {Wurth}}, \bibinfo
  {author} {\bibfnamefont {P.}~\bibnamefont {Hofmann}}, \bibinfo {author}
  {\bibfnamefont {M.}~\bibnamefont {Wolf}}, \bibinfo {author} {\bibfnamefont
  {M.}~\bibnamefont {Scheidgen}}, \bibinfo {author} {\bibfnamefont
  {L.}~\bibnamefont {Rettig}},\ and\ \bibinfo {author} {\bibfnamefont
  {R.}~\bibnamefont {Ernstorfer}},\ }\bibfield  {title} {\bibinfo {title} {An
  open-source, end-to-end workflow for multidimensional photoemission
  spectroscopy},\ }\href {https://doi.org/10.1038/s41597-020-00769-8}
  {\bibfield  {journal} {\bibinfo  {journal} {Scientific Data}\ }\textbf
  {\bibinfo {volume} {7}},\ \bibinfo {pages} {442} (\bibinfo {year}
  {2020})}\BibitemShut {NoStop}%
\bibitem [{\citenamefont {Keunecke}\ \emph {et~al.}(2020)\citenamefont
  {Keunecke}, \citenamefont {Reutzel}, \citenamefont {Schmitt}, \citenamefont
  {Osterkorn}, \citenamefont {Mishra}, \citenamefont {M\"oller}, \citenamefont
  {Bennecke}, \citenamefont {Jansen}, \citenamefont {Steil}, \citenamefont
  {Manmana}, \citenamefont {Steil}, \citenamefont {Kehrein},\ and\
  \citenamefont {Mathias}}]{Keunecke:2020aa}%
  \BibitemOpen
  \bibfield  {author} {\bibinfo {author} {\bibfnamefont {M.}~\bibnamefont
  {Keunecke}}, \bibinfo {author} {\bibfnamefont {M.}~\bibnamefont {Reutzel}},
  \bibinfo {author} {\bibfnamefont {D.}~\bibnamefont {Schmitt}}, \bibinfo
  {author} {\bibfnamefont {A.}~\bibnamefont {Osterkorn}}, \bibinfo {author}
  {\bibfnamefont {T.~A.}\ \bibnamefont {Mishra}}, \bibinfo {author}
  {\bibfnamefont {C.}~\bibnamefont {M\"oller}}, \bibinfo {author}
  {\bibfnamefont {W.}~\bibnamefont {Bennecke}}, \bibinfo {author}
  {\bibfnamefont {G.~S.~M.}\ \bibnamefont {Jansen}}, \bibinfo {author}
  {\bibfnamefont {D.}~\bibnamefont {Steil}}, \bibinfo {author} {\bibfnamefont
  {S.~R.}\ \bibnamefont {Manmana}}, \bibinfo {author} {\bibfnamefont
  {S.}~\bibnamefont {Steil}}, \bibinfo {author} {\bibfnamefont
  {S.}~\bibnamefont {Kehrein}},\ and\ \bibinfo {author} {\bibfnamefont
  {S.}~\bibnamefont {Mathias}},\ }\bibfield  {title} {\bibinfo {title}
  {Electromagnetic dressing of the electron energy spectrum of {A}u(111) at
  high momenta},\ }\href {https://doi.org/10.1103/PhysRevB.102.161403}
  {\bibfield  {journal} {\bibinfo  {journal} {Phys. Rev. B}\ }\textbf {\bibinfo
  {volume} {102}},\ \bibinfo {pages} {161403} (\bibinfo {year}
  {2020})}\BibitemShut {NoStop}%
\bibitem [{\citenamefont {Abrami}\ \emph {et~al.}(1995)\citenamefont {Abrami},
  \citenamefont {Barnaba}, \citenamefont {Battistello}, \citenamefont {Bianco},
  \citenamefont {Brena}, \citenamefont {Cautero}, \citenamefont {Chen},
  \citenamefont {Cocco}, \citenamefont {Comelli}, \citenamefont {Contrino},
  \citenamefont {DeBona}, \citenamefont {Di~Fonzo}, \citenamefont {Fava},
  \citenamefont {Finetti}, \citenamefont {Furlan}, \citenamefont {Galimberti},
  \citenamefont {Gambitta}, \citenamefont {Giuressi}, \citenamefont {Godnig},
  \citenamefont {Jark}, \citenamefont {Lizzit}, \citenamefont {Mazzolini},
  \citenamefont {Melpignano}, \citenamefont {Olivi}, \citenamefont {Paolucci},
  \citenamefont {Pugliese}, \citenamefont {Qian}, \citenamefont {Rosei},
  \citenamefont {Sandrin}, \citenamefont {Savoia}, \citenamefont {Sergo},
  \citenamefont {Sostero}, \citenamefont {Tommasini}, \citenamefont {Tudor},
  \citenamefont {Vivoda}, \citenamefont {Wei},\ and\ \citenamefont
  {Zanini}}]{Abrami:1995aa}%
  \BibitemOpen
  \bibfield  {author} {\bibinfo {author} {\bibfnamefont {A.}~\bibnamefont
  {Abrami}}, \bibinfo {author} {\bibfnamefont {M.}~\bibnamefont {Barnaba}},
  \bibinfo {author} {\bibfnamefont {L.}~\bibnamefont {Battistello}}, \bibinfo
  {author} {\bibfnamefont {A.}~\bibnamefont {Bianco}}, \bibinfo {author}
  {\bibfnamefont {B.}~\bibnamefont {Brena}}, \bibinfo {author} {\bibfnamefont
  {G.}~\bibnamefont {Cautero}}, \bibinfo {author} {\bibfnamefont
  {H.}~\bibnamefont {Chen}, \bibfnamefont {Q}}, \bibinfo {author}
  {\bibfnamefont {D.}~\bibnamefont {Cocco}}, \bibinfo {author} {\bibfnamefont
  {G.}~\bibnamefont {Comelli}}, \bibinfo {author} {\bibfnamefont
  {S.}~\bibnamefont {Contrino}}, \bibinfo {author} {\bibfnamefont
  {F.}~\bibnamefont {DeBona}}, \bibinfo {author} {\bibfnamefont
  {S.}~\bibnamefont {Di~Fonzo}}, \bibinfo {author} {\bibfnamefont
  {C.}~\bibnamefont {Fava}}, \bibinfo {author} {\bibfnamefont {P.}~\bibnamefont
  {Finetti}}, \bibinfo {author} {\bibfnamefont {P.}~\bibnamefont {Furlan}},
  \bibinfo {author} {\bibfnamefont {A.}~\bibnamefont {Galimberti}}, \bibinfo
  {author} {\bibfnamefont {A.}~\bibnamefont {Gambitta}}, \bibinfo {author}
  {\bibfnamefont {D.}~\bibnamefont {Giuressi}}, \bibinfo {author}
  {\bibfnamefont {R.}~\bibnamefont {Godnig}}, \bibinfo {author} {\bibfnamefont
  {W.}~\bibnamefont {Jark}}, \bibinfo {author} {\bibfnamefont {S.}~\bibnamefont
  {Lizzit}}, \bibinfo {author} {\bibfnamefont {F.}~\bibnamefont {Mazzolini}},
  \bibinfo {author} {\bibfnamefont {P.}~\bibnamefont {Melpignano}}, \bibinfo
  {author} {\bibfnamefont {L.}~\bibnamefont {Olivi}}, \bibinfo {author}
  {\bibfnamefont {G.}~\bibnamefont {Paolucci}}, \bibinfo {author}
  {\bibfnamefont {R.}~\bibnamefont {Pugliese}}, \bibinfo {author}
  {\bibfnamefont {S.~N.}\ \bibnamefont {Qian}}, \bibinfo {author}
  {\bibfnamefont {R.}~\bibnamefont {Rosei}}, \bibinfo {author} {\bibfnamefont
  {G.}~\bibnamefont {Sandrin}}, \bibinfo {author} {\bibfnamefont
  {A.}~\bibnamefont {Savoia}}, \bibinfo {author} {\bibfnamefont
  {R.}~\bibnamefont {Sergo}}, \bibinfo {author} {\bibfnamefont
  {G.}~\bibnamefont {Sostero}}, \bibinfo {author} {\bibfnamefont
  {R.}~\bibnamefont {Tommasini}}, \bibinfo {author} {\bibfnamefont
  {M.}~\bibnamefont {Tudor}}, \bibinfo {author} {\bibfnamefont
  {D.}~\bibnamefont {Vivoda}}, \bibinfo {author} {\bibfnamefont {F.-Q.}\
  \bibnamefont {Wei}},\ and\ \bibinfo {author} {\bibfnamefont {F.}~\bibnamefont
  {Zanini}},\ }\bibfield  {title} {\bibinfo {title} {Super {ESCA: F}irst
  beamline operating at {ELETTRA}},\ }\href@noop {} {\bibfield  {journal}
  {\bibinfo  {journal} {Review of Scientific Instruments}\ }\textbf {\bibinfo
  {volume} {66}},\ \bibinfo {pages} {1618} (\bibinfo {year}
  {1995})}\BibitemShut {NoStop}%
\bibitem [{\citenamefont {Sch{\"o}nhense}\ \emph {et~al.}(2018)\citenamefont
  {Sch{\"o}nhense}, \citenamefont {Medjanik}, \citenamefont {Fedchenko},
  \citenamefont {Chernov}, \citenamefont {Ellguth}, \citenamefont {Vasilyev},
  \citenamefont {Oelsner}, \citenamefont {Viefhaus}, \citenamefont
  {Kutnyakhov}, \citenamefont {Wurth}, \citenamefont {Elmers},\ and\
  \citenamefont {Sch{\"o}nhense}}]{Schonhense:2018aa}%
  \BibitemOpen
  \bibfield  {author} {\bibinfo {author} {\bibfnamefont {B.}~\bibnamefont
  {Sch{\"o}nhense}}, \bibinfo {author} {\bibfnamefont {K.}~\bibnamefont
  {Medjanik}}, \bibinfo {author} {\bibfnamefont {O.}~\bibnamefont {Fedchenko}},
  \bibinfo {author} {\bibfnamefont {S.}~\bibnamefont {Chernov}}, \bibinfo
  {author} {\bibfnamefont {M.}~\bibnamefont {Ellguth}}, \bibinfo {author}
  {\bibfnamefont {D.}~\bibnamefont {Vasilyev}}, \bibinfo {author}
  {\bibfnamefont {A.}~\bibnamefont {Oelsner}}, \bibinfo {author} {\bibfnamefont
  {J.}~\bibnamefont {Viefhaus}}, \bibinfo {author} {\bibfnamefont
  {D.}~\bibnamefont {Kutnyakhov}}, \bibinfo {author} {\bibfnamefont
  {W.}~\bibnamefont {Wurth}}, \bibinfo {author} {\bibfnamefont {H.~J.}\
  \bibnamefont {Elmers}},\ and\ \bibinfo {author} {\bibfnamefont
  {G.}~\bibnamefont {Sch{\"o}nhense}},\ }\bibfield  {title} {\bibinfo {title}
  {Multidimensional photoemission spectroscopy---the space-charge limit},\
  }\href {https://doi.org/10.1088/1367-2630/aaa262} {\bibfield  {journal}
  {\bibinfo  {journal} {New Journal of Physics}\ }\textbf {\bibinfo {volume}
  {20}},\ \bibinfo {pages} {033004} (\bibinfo {year} {2018})}\BibitemShut
  {NoStop}%
\bibitem [{\citenamefont {Coletti}\ \emph {et~al.}(2011)\citenamefont
  {Coletti}, \citenamefont {Emtsev}, \citenamefont {Zakharov}, \citenamefont
  {Ouisse}, \citenamefont {Chaussende},\ and\ \citenamefont
  {Starke}}]{Coletti:2011aa}%
  \BibitemOpen
  \bibfield  {author} {\bibinfo {author} {\bibfnamefont {C.}~\bibnamefont
  {Coletti}}, \bibinfo {author} {\bibfnamefont {K.~V.}\ \bibnamefont {Emtsev}},
  \bibinfo {author} {\bibfnamefont {A.~A.}\ \bibnamefont {Zakharov}}, \bibinfo
  {author} {\bibfnamefont {T.}~\bibnamefont {Ouisse}}, \bibinfo {author}
  {\bibfnamefont {D.}~\bibnamefont {Chaussende}},\ and\ \bibinfo {author}
  {\bibfnamefont {U.}~\bibnamefont {Starke}},\ }\bibfield  {title} {\bibinfo
  {title} {Large area quasi-free standing monolayer graphene on
  3c-{SiC}(111)},\ }\href {https://doi.org/10.1063/1.3618674} {\bibfield
  {journal} {\bibinfo  {journal} {Applied Physics Letters}\ }\textbf {\bibinfo
  {volume} {99}},\ \bibinfo {pages} {081904} (\bibinfo {year}
  {2011})}\BibitemShut {NoStop}%
\bibitem [{\citenamefont {Hughes}\ and\ \citenamefont
  {Scarfe}(1996)}]{Hughes:1996ab}%
  \BibitemOpen
  \bibfield  {author} {\bibinfo {author} {\bibfnamefont {H.~P.}\ \bibnamefont
  {Hughes}}\ and\ \bibinfo {author} {\bibfnamefont {J.~A.}\ \bibnamefont
  {Scarfe}},\ }\bibfield  {title} {\bibinfo {title} {Lineshapes in core-level
  photoemission from metals: I. theory and computational analysis},\ }\href
  {http://stacks.iop.org/0953-8984/8/i=10/a=014} {\bibfield  {journal}
  {\bibinfo  {journal} {Journal of Physics: Condensed Matter}\ }\textbf
  {\bibinfo {volume} {8}},\ \bibinfo {pages} {1421} (\bibinfo {year}
  {1996})}\BibitemShut {NoStop}%
\bibitem [{\citenamefont {Sernelius}(2015)}]{Sernelius:2015ab}%
  \BibitemOpen
  \bibfield  {author} {\bibinfo {author} {\bibfnamefont {B.~E.}\ \bibnamefont
  {Sernelius}},\ }\bibfield  {title} {\bibinfo {title} {Core-level spectra from
  graphene},\ }\href {https://doi.org/10.1103/PhysRevB.91.045402} {\bibfield
  {journal} {\bibinfo  {journal} {Phys. Rev. B}\ }\textbf {\bibinfo {volume}
  {91}},\ \bibinfo {pages} {045402} (\bibinfo {year} {2015})}\BibitemShut
  {NoStop}%
\bibitem [{\citenamefont {Ulstrup}\ \emph {et~al.}(2014)\citenamefont
  {Ulstrup}, \citenamefont {Johannsen}, \citenamefont {Grioni},\ and\
  \citenamefont {Hofmann}}]{Ulstrup:2014aa}%
  \BibitemOpen
  \bibfield  {author} {\bibinfo {author} {\bibfnamefont {S.}~\bibnamefont
  {Ulstrup}}, \bibinfo {author} {\bibfnamefont {J.~C.}\ \bibnamefont
  {Johannsen}}, \bibinfo {author} {\bibfnamefont {M.}~\bibnamefont {Grioni}},\
  and\ \bibinfo {author} {\bibfnamefont {P.}~\bibnamefont {Hofmann}},\
  }\bibfield  {title} {\bibinfo {title} {Extracting the temperature of hot
  carriers in time- and angle-resolved photoemission},\ }\href
  {https://doi.org/http://dx.doi.org/10.1063/1.4863322} {\bibfield  {journal}
  {\bibinfo  {journal} {Review of Scientific Instruments}\ }\textbf {\bibinfo
  {volume} {85}},\ \bibinfo {eid} {013907} (\bibinfo {year}
  {2014})}\BibitemShut {NoStop}%
\bibitem [{\citenamefont {Ulstrup}\ \emph {et~al.}(2012)\citenamefont
  {Ulstrup}, \citenamefont {Bianchi}, \citenamefont {Hatch}, \citenamefont
  {Guan}, \citenamefont {Baraldi}, \citenamefont {Alf\`e}, \citenamefont
  {Hornek\ae{}r},\ and\ \citenamefont {Hofmann}}]{Ulstrup:2012aa}%
  \BibitemOpen
  \bibfield  {author} {\bibinfo {author} {\bibfnamefont {S.}~\bibnamefont
  {Ulstrup}}, \bibinfo {author} {\bibfnamefont {M.}~\bibnamefont {Bianchi}},
  \bibinfo {author} {\bibfnamefont {R.}~\bibnamefont {Hatch}}, \bibinfo
  {author} {\bibfnamefont {D.}~\bibnamefont {Guan}}, \bibinfo {author}
  {\bibfnamefont {A.}~\bibnamefont {Baraldi}}, \bibinfo {author} {\bibfnamefont
  {D.}~\bibnamefont {Alf\`e}}, \bibinfo {author} {\bibfnamefont
  {L.}~\bibnamefont {Hornek\ae{}r}},\ and\ \bibinfo {author} {\bibfnamefont
  {P.}~\bibnamefont {Hofmann}},\ }\bibfield  {title} {\bibinfo {title}
  {High-temperature behavior of supported graphene: Electron-phonon coupling
  and substrate-induced doping},\ }\href
  {https://doi.org/10.1103/PhysRevB.86.161402} {\bibfield  {journal} {\bibinfo
  {journal} {Phys. Rev. B}\ }\textbf {\bibinfo {volume} {86}},\ \bibinfo
  {pages} {161402} (\bibinfo {year} {2012})}\BibitemShut {NoStop}%
\bibitem [{\citenamefont {Calandra}\ and\ \citenamefont
  {Mauri}(2007)}]{Calandra:2007aa}%
  \BibitemOpen
  \bibfield  {author} {\bibinfo {author} {\bibfnamefont {M.}~\bibnamefont
  {Calandra}}\ and\ \bibinfo {author} {\bibfnamefont {F.}~\bibnamefont
  {Mauri}},\ }\bibfield  {title} {\bibinfo {title} {Electron-phonon coupling
  and electron self-energy in electron-doped graphene: Calculation of
  angular-resolved photoemission spectra},\ }\href
  {https://doi.org/10.1103/PhysRevB.76.205411} {\bibfield  {journal} {\bibinfo
  {journal} {Physical Review B}\ }\textbf {\bibinfo {volume} {76}},\ \bibinfo
  {eid} {205411} (\bibinfo {year} {2007})}\BibitemShut {NoStop}%
\bibitem [{\citenamefont {Matveev}\ \emph {et~al.}(2019)\citenamefont
  {Matveev}, \citenamefont {Shvaika}, \citenamefont {Devereaux},\ and\
  \citenamefont {Freericks}}]{Matveev:2019aa}%
  \BibitemOpen
  \bibfield  {author} {\bibinfo {author} {\bibfnamefont {O.}~\bibnamefont
  {Matveev}}, \bibinfo {author} {\bibfnamefont {A.}~\bibnamefont {Shvaika}},
  \bibinfo {author} {\bibfnamefont {T.}~\bibnamefont {Devereaux}},\ and\
  \bibinfo {author} {\bibfnamefont {J.}~\bibnamefont {Freericks}},\ }\bibfield
  {title} {\bibinfo {title} {Stroboscopic tests for thermalization of electrons
  in pump-probe experiments},\ }\bibfield  {journal} {\bibinfo  {journal}
  {Physical Review Letters}\ }\textbf {\bibinfo {volume} {122}},\ \href
  {https://doi.org/10.1103/physrevlett.122.247402}
  {10.1103/physrevlett.122.247402} (\bibinfo {year} {2019})\BibitemShut
  {NoStop}%
\bibitem [{\citenamefont {Woodruff}(2002)}]{Woodruff:2002aa}%
  \BibitemOpen
  \bibfield  {author} {\bibinfo {author} {\bibfnamefont {D.}~\bibnamefont
  {Woodruff}},\ }\bibfield  {title} {\bibinfo {title} {Photoelectron
  diffraction: Past, present and future},\ }\href@noop {} {\bibfield  {journal}
  {\bibinfo  {journal} {Journal of Electron Spectroscopy and Related
  Phenomena}\ }\textbf {\bibinfo {volume} {126}},\ \bibinfo {pages} {55}
  (\bibinfo {year} {2002})}\BibitemShut {NoStop}%
\bibitem [{\citenamefont {Winkelmann}\ \emph {et~al.}(2012)\citenamefont
  {Winkelmann}, \citenamefont {Ellguth}, \citenamefont {Tusche}, \citenamefont
  {{\"U}nal}, \citenamefont {Henk},\ and\ \citenamefont
  {Kirschner}}]{Winkelmann:2012aa}%
  \BibitemOpen
  \bibfield  {author} {\bibinfo {author} {\bibfnamefont {A.}~\bibnamefont
  {Winkelmann}}, \bibinfo {author} {\bibfnamefont {M.}~\bibnamefont {Ellguth}},
  \bibinfo {author} {\bibfnamefont {C.}~\bibnamefont {Tusche}}, \bibinfo
  {author} {\bibfnamefont {A.~A.}\ \bibnamefont {{\"U}nal}}, \bibinfo {author}
  {\bibfnamefont {J.}~\bibnamefont {Henk}},\ and\ \bibinfo {author}
  {\bibfnamefont {J.}~\bibnamefont {Kirschner}},\ }\bibfield  {title} {\bibinfo
  {title} {Momentum-resolved photoelectron interference in crystal surface
  barrier scattering},\ }\href {http://dx.doi.org/10.1103/PhysRevB.86.085427}
  {\bibfield  {journal} {\bibinfo  {journal} {Physical Review B}\ }\textbf
  {\bibinfo {volume} {86}} (\bibinfo {year} {2012})}\BibitemShut {NoStop}%
\bibitem [{\citenamefont {Greif}\ \emph {et~al.}(2016)\citenamefont {Greif},
  \citenamefont {Kasmi}, \citenamefont {Castiglioni}, \citenamefont {Lucchini},
  \citenamefont {Gallmann}, \citenamefont {Keller}, \citenamefont
  {Osterwalder},\ and\ \citenamefont {Hengsberger}}]{Greif:2016aa}%
  \BibitemOpen
  \bibfield  {author} {\bibinfo {author} {\bibfnamefont {M.}~\bibnamefont
  {Greif}}, \bibinfo {author} {\bibfnamefont {L.}~\bibnamefont {Kasmi}},
  \bibinfo {author} {\bibfnamefont {L.}~\bibnamefont {Castiglioni}}, \bibinfo
  {author} {\bibfnamefont {M.}~\bibnamefont {Lucchini}}, \bibinfo {author}
  {\bibfnamefont {L.}~\bibnamefont {Gallmann}}, \bibinfo {author}
  {\bibfnamefont {U.}~\bibnamefont {Keller}}, \bibinfo {author} {\bibfnamefont
  {J.}~\bibnamefont {Osterwalder}},\ and\ \bibinfo {author} {\bibfnamefont
  {M.}~\bibnamefont {Hengsberger}},\ }\bibfield  {title} {\bibinfo {title}
  {Access to phases of coherent phonon excitations by femtosecond ultraviolet
  photoelectron diffraction},\ }\href
  {https://doi.org/10.1103/PhysRevB.94.054309} {\bibfield  {journal} {\bibinfo
  {journal} {Phys. Rev. B}\ }\textbf {\bibinfo {volume} {94}},\ \bibinfo
  {pages} {054309} (\bibinfo {year} {2016})}\BibitemShut {NoStop}%
\bibitem [{\citenamefont {Ang}\ \emph {et~al.}(2020)\citenamefont {Ang},
  \citenamefont {Fukatsu}, \citenamefont {Kimura}, \citenamefont {Yamamoto},
  \citenamefont {Yonezawa}, \citenamefont {Nitta}, \citenamefont {Fleurence},
  \citenamefont {Yamamoto}, \citenamefont {Matsuda}, \citenamefont
  {Yamada-Takamura},\ and\ \citenamefont {Hayashi}}]{Ang:2020aa}%
  \BibitemOpen
  \bibfield  {author} {\bibinfo {author} {\bibfnamefont {A.~K.~R.}\
  \bibnamefont {Ang}}, \bibinfo {author} {\bibfnamefont {Y.}~\bibnamefont
  {Fukatsu}}, \bibinfo {author} {\bibfnamefont {K.}~\bibnamefont {Kimura}},
  \bibinfo {author} {\bibfnamefont {Y.}~\bibnamefont {Yamamoto}}, \bibinfo
  {author} {\bibfnamefont {T.}~\bibnamefont {Yonezawa}}, \bibinfo {author}
  {\bibfnamefont {H.}~\bibnamefont {Nitta}}, \bibinfo {author} {\bibfnamefont
  {A.}~\bibnamefont {Fleurence}}, \bibinfo {author} {\bibfnamefont
  {S.}~\bibnamefont {Yamamoto}}, \bibinfo {author} {\bibfnamefont
  {I.}~\bibnamefont {Matsuda}}, \bibinfo {author} {\bibfnamefont
  {Y.}~\bibnamefont {Yamada-Takamura}},\ and\ \bibinfo {author} {\bibfnamefont
  {K.}~\bibnamefont {Hayashi}},\ }\bibfield  {title} {\bibinfo {title}
  {Time-resolved x-ray photoelectron diffraction using an angle-resolved
  time-of-flight electron analyzer},\ }\href
  {https://doi.org/10.35848/1347-4065/abb57e} {\bibfield  {journal} {\bibinfo
  {journal} {Japanese Journal of Applied Physics}\ }\textbf {\bibinfo {volume}
  {59}},\ \bibinfo {pages} {100902} (\bibinfo {year} {2020})}\BibitemShut
  {NoStop}%
\bibitem [{\citenamefont {Das~Sarma}\ \emph {et~al.}(2011)\citenamefont
  {Das~Sarma}, \citenamefont {Adam}, \citenamefont {Hwang},\ and\ \citenamefont
  {Rossi}}]{Das-Sarma:2011aa}%
  \BibitemOpen
  \bibfield  {author} {\bibinfo {author} {\bibfnamefont {S.}~\bibnamefont
  {Das~Sarma}}, \bibinfo {author} {\bibfnamefont {S.}~\bibnamefont {Adam}},
  \bibinfo {author} {\bibfnamefont {E.~H.}\ \bibnamefont {Hwang}},\ and\
  \bibinfo {author} {\bibfnamefont {E.}~\bibnamefont {Rossi}},\ }\bibfield
  {title} {\bibinfo {title} {Electronic transport in two-dimensional
  graphene},\ }\href {https://doi.org/10.1103/RevModPhys.83.407} {\bibfield
  {journal} {\bibinfo  {journal} {Rev. Mod. Phys.}\ }\textbf {\bibinfo {volume}
  {83}},\ \bibinfo {pages} {407} (\bibinfo {year} {2011})}\BibitemShut
  {NoStop}%
\bibitem [{\citenamefont {Shirley}\ \emph {et~al.}(1995)\citenamefont
  {Shirley}, \citenamefont {Terminello}, \citenamefont {Santoni},\ and\
  \citenamefont {Himpsel}}]{Shirley:1995aa}%
  \BibitemOpen
  \bibfield  {author} {\bibinfo {author} {\bibfnamefont {E.~L.}\ \bibnamefont
  {Shirley}}, \bibinfo {author} {\bibfnamefont {L.~J.}\ \bibnamefont
  {Terminello}}, \bibinfo {author} {\bibfnamefont {A.}~\bibnamefont
  {Santoni}},\ and\ \bibinfo {author} {\bibfnamefont {F.~J.}\ \bibnamefont
  {Himpsel}},\ }\bibfield  {title} {\bibinfo {title} {Brillouin-zone-selection
  effects in graphite photoelectron angular distributions},\ }\href
  {https://doi.org/10.1103/PhysRevB.51.13614} {\bibfield  {journal} {\bibinfo
  {journal} {Phys. Rev. B}\ }\textbf {\bibinfo {volume} {51}},\ \bibinfo
  {pages} {13614} (\bibinfo {year} {1995})}\BibitemShut {NoStop}%
\bibitem [{\citenamefont {Mucha-Kruczynski}\ \emph {et~al.}(2008)\citenamefont
  {Mucha-Kruczynski}, \citenamefont {Tsyplyatyev}, \citenamefont {Grishin},
  \citenamefont {McCann}, \citenamefont {Fal'ko}, \citenamefont {Bostwick},\
  and\ \citenamefont {Rotenberg}}]{Mucha-Kruczynski:2008aa}%
  \BibitemOpen
  \bibfield  {author} {\bibinfo {author} {\bibfnamefont {M.}~\bibnamefont
  {Mucha-Kruczynski}}, \bibinfo {author} {\bibfnamefont {O.}~\bibnamefont
  {Tsyplyatyev}}, \bibinfo {author} {\bibfnamefont {A.}~\bibnamefont
  {Grishin}}, \bibinfo {author} {\bibfnamefont {E.}~\bibnamefont {McCann}},
  \bibinfo {author} {\bibfnamefont {V.~I.}\ \bibnamefont {Fal'ko}}, \bibinfo
  {author} {\bibfnamefont {A.}~\bibnamefont {Bostwick}},\ and\ \bibinfo
  {author} {\bibfnamefont {E.}~\bibnamefont {Rotenberg}},\ }\bibfield  {title}
  {\bibinfo {title} {Characterization of graphene through anisotropy of
  constant-energy maps in angle-resolved photoemission},\ }\href
  {https://doi.org/10.1103/PhysRevB.77.195403} {\bibfield  {journal} {\bibinfo
  {journal} {Physical Review B}\ }\textbf {\bibinfo {volume} {77}},\ \bibinfo
  {eid} {195403} (\bibinfo {year} {2008})}\BibitemShut {NoStop}%
\bibitem [{\citenamefont {Crespi}\ \emph {et~al.}(2013)\citenamefont {Crespi},
  \citenamefont {Corrielli}, \citenamefont {Valle}, \citenamefont {Osellame},\
  and\ \citenamefont {Longhi}}]{Crespi:2013aa}%
  \BibitemOpen
  \bibfield  {author} {\bibinfo {author} {\bibfnamefont {A.}~\bibnamefont
  {Crespi}}, \bibinfo {author} {\bibfnamefont {G.}~\bibnamefont {Corrielli}},
  \bibinfo {author} {\bibfnamefont {G.~D.}\ \bibnamefont {Valle}}, \bibinfo
  {author} {\bibfnamefont {R.}~\bibnamefont {Osellame}},\ and\ \bibinfo
  {author} {\bibfnamefont {S.}~\bibnamefont {Longhi}},\ }\bibfield  {title}
  {\bibinfo {title} {Dynamic band collapse in photonic graphene},\ }\href
  {https://doi.org/10.1088/1367-2630/15/1/013012} {\bibfield  {journal}
  {\bibinfo  {journal} {New Journal of Physics}\ }\textbf {\bibinfo {volume}
  {15}},\ \bibinfo {pages} {013012} (\bibinfo {year} {2013})}\BibitemShut
  {NoStop}%
\bibitem [{\citenamefont {Agustsson}\ \emph {et~al.}(2021)\citenamefont
  {Agustsson}, \citenamefont {Curcio}, \citenamefont {Xian}, \citenamefont
  {Sohail}, \citenamefont {Heber}, \citenamefont {Scholz},\ and\ \citenamefont
  {Acremann}}]{Agustsson:2021aa}%
  \BibitemOpen
  \bibfield  {author} {\bibinfo {author} {\bibfnamefont {S.~Y.}\ \bibnamefont
  {Agustsson}}, \bibinfo {author} {\bibfnamefont {D.}~\bibnamefont {Curcio}},
  \bibinfo {author} {\bibfnamefont {R.~P.}\ \bibnamefont {Xian}}, \bibinfo
  {author} {\bibfnamefont {M.~Z.}\ \bibnamefont {Sohail}}, \bibinfo {author}
  {\bibfnamefont {M.}~\bibnamefont {Heber}}, \bibinfo {author} {\bibfnamefont
  {M.}~\bibnamefont {Scholz}},\ and\ \bibinfo {author} {\bibfnamefont
  {Y.}~\bibnamefont {Acremann}},\ }\href
  {https://doi.org/10.5281/zenodo.4651325} {\bibinfo {title}
  {{momentoscope/hextof-processor:hextofprocessor-v1.0.4}}} (\bibinfo {year}
  {2021})\BibitemShut {NoStop}%
\bibitem [{\citenamefont {Curcio}\ \emph {et~al.}(2021)\citenamefont {Curcio},
  \citenamefont {Pakdel}, \citenamefont {Volckaert}, \citenamefont {Miwa},
  \citenamefont {Ulstrup}, \citenamefont {Lanat\`a}, \citenamefont {Bianchi},
  \citenamefont {Kutnyakhov}, \citenamefont {Pressacco}, \citenamefont
  {Brenner}, \citenamefont {Dziarzhytski}, \citenamefont {Redlin},
  \citenamefont {Agustsson}, \citenamefont {Medjanik}, \citenamefont
  {Vasilyev}, \citenamefont {Elmers}, \citenamefont {Sch{\"o}nhense},
  \citenamefont {Tusche}, \citenamefont {Chen}, \citenamefont {Speck},
  \citenamefont {Seyller}, \citenamefont {B{\"u}hlmann}, \citenamefont {Gort},
  \citenamefont {Diekmann}, \citenamefont {Rossnagel}, \citenamefont
  {Acremann}, \citenamefont {Demsar}, \citenamefont {Wurth}, \citenamefont
  {Lizzit}, \citenamefont {Bignardi}, \citenamefont {Lacovig}, \citenamefont
  {Lizzit}, \citenamefont {Sanders},\ and\ \citenamefont
  {Hofmann}}]{Curcio:2021aa}%
  \BibitemOpen
  \bibfield  {author} {\bibinfo {author} {\bibfnamefont {D.}~\bibnamefont
  {Curcio}}, \bibinfo {author} {\bibfnamefont {S.}~\bibnamefont {Pakdel}},
  \bibinfo {author} {\bibfnamefont {K.}~\bibnamefont {Volckaert}}, \bibinfo
  {author} {\bibfnamefont {J.~A.}\ \bibnamefont {Miwa}}, \bibinfo {author}
  {\bibfnamefont {S.}~\bibnamefont {Ulstrup}}, \bibinfo {author} {\bibfnamefont
  {N.}~\bibnamefont {Lanat\`a}}, \bibinfo {author} {\bibfnamefont
  {M.}~\bibnamefont {Bianchi}}, \bibinfo {author} {\bibfnamefont
  {D.}~\bibnamefont {Kutnyakhov}}, \bibinfo {author} {\bibfnamefont
  {F.}~\bibnamefont {Pressacco}}, \bibinfo {author} {\bibfnamefont
  {G.}~\bibnamefont {Brenner}}, \bibinfo {author} {\bibfnamefont
  {S.}~\bibnamefont {Dziarzhytski}}, \bibinfo {author} {\bibfnamefont
  {H.}~\bibnamefont {Redlin}}, \bibinfo {author} {\bibfnamefont
  {S.}~\bibnamefont {Agustsson}}, \bibinfo {author} {\bibfnamefont
  {K.}~\bibnamefont {Medjanik}}, \bibinfo {author} {\bibfnamefont
  {D.}~\bibnamefont {Vasilyev}}, \bibinfo {author} {\bibfnamefont {H.-J.}\
  \bibnamefont {Elmers}}, \bibinfo {author} {\bibfnamefont {G.}~\bibnamefont
  {Sch{\"o}nhense}}, \bibinfo {author} {\bibfnamefont {C.}~\bibnamefont
  {Tusche}}, \bibinfo {author} {\bibfnamefont {Y.-J.}\ \bibnamefont {Chen}},
  \bibinfo {author} {\bibfnamefont {F.}~\bibnamefont {Speck}}, \bibinfo
  {author} {\bibfnamefont {T.}~\bibnamefont {Seyller}}, \bibinfo {author}
  {\bibfnamefont {K.}~\bibnamefont {B{\"u}hlmann}}, \bibinfo {author}
  {\bibfnamefont {R.}~\bibnamefont {Gort}}, \bibinfo {author} {\bibfnamefont
  {F.}~\bibnamefont {Diekmann}}, \bibinfo {author} {\bibfnamefont
  {K.}~\bibnamefont {Rossnagel}}, \bibinfo {author} {\bibfnamefont
  {Y.}~\bibnamefont {Acremann}}, \bibinfo {author} {\bibfnamefont
  {J.}~\bibnamefont {Demsar}}, \bibinfo {author} {\bibfnamefont
  {W.}~\bibnamefont {Wurth}}, \bibinfo {author} {\bibfnamefont
  {D.}~\bibnamefont {Lizzit}}, \bibinfo {author} {\bibfnamefont
  {L.}~\bibnamefont {Bignardi}}, \bibinfo {author} {\bibfnamefont
  {P.}~\bibnamefont {Lacovig}}, \bibinfo {author} {\bibfnamefont
  {S.}~\bibnamefont {Lizzit}}, \bibinfo {author} {\bibfnamefont {C.~E.}\
  \bibnamefont {Sanders}},\ and\ \bibinfo {author} {\bibfnamefont
  {P.}~\bibnamefont {Hofmann}},\ }\bibfield  {title} {\bibinfo {title}
  {{Ultrafast electronic line width broadening in the C 1s core level of
  graphene}},\ }\href {https://doi.org/10.5281/zenodo.4773850}
  {10.5281/zenodo.4773850} (\bibinfo {year} {2021})\BibitemShut {NoStop}%
\bibitem [{\citenamefont {Woodruff}\ and\ \citenamefont
  {Bradshaw}(1994)}]{Woodruff:1994ab}%
  \BibitemOpen
  \bibfield  {author} {\bibinfo {author} {\bibfnamefont {D.~P.}\ \bibnamefont
  {Woodruff}}\ and\ \bibinfo {author} {\bibfnamefont {A.~M.}\ \bibnamefont
  {Bradshaw}},\ }\bibfield  {title} {\bibinfo {title} {Adsorbate structure
  determination on surfaces using photoelectron diffraction},\ }\href@noop {}
  {\bibfield  {journal} {\bibinfo  {journal} {Reports on Progress in Physics}\
  }\textbf {\bibinfo {volume} {57}},\ \bibinfo {pages} {1029} (\bibinfo {year}
  {1994})}\BibitemShut {NoStop}%
\bibitem [{\citenamefont {Garc\'ia~de Abajo}\ \emph {et~al.}(2001)\citenamefont
  {Garc\'ia~de Abajo}, \citenamefont {Van~Hove},\ and\ \citenamefont
  {Fadley}}]{Garcia-de-Abajo:2001aa}%
  \BibitemOpen
  \bibfield  {author} {\bibinfo {author} {\bibfnamefont {F.~J.}\ \bibnamefont
  {Garc\'ia~de Abajo}}, \bibinfo {author} {\bibfnamefont {M.~A.}\ \bibnamefont
  {Van~Hove}},\ and\ \bibinfo {author} {\bibfnamefont {C.~S.}\ \bibnamefont
  {Fadley}},\ }\bibfield  {title} {\bibinfo {title} {Multiple scattering of
  electrons in solids and molecules: A cluster-model approach},\ }\href
  {https://doi.org/10.1103/PhysRevB.63.075404} {\bibfield  {journal} {\bibinfo
  {journal} {Phys. Rev. B}\ }\textbf {\bibinfo {volume} {63}},\ \bibinfo
  {pages} {075404} (\bibinfo {year} {2001})}\BibitemShut {NoStop}%
\bibitem [{\citenamefont {Dirac}(1930)}]{Dirac:1930uy}%
  \BibitemOpen
  \bibfield  {author} {\bibinfo {author} {\bibfnamefont {P.~A.}\ \bibnamefont
  {Dirac}},\ }\bibfield  {title} {\bibinfo {title} {Note on exchange phenomena
  in the thomas atom},\ }in\ \href@noop {} {\emph {\bibinfo {booktitle} {Math.
  Proc. Cambridge Philos. Soc.}}},\ Vol.~\bibinfo {volume} {26}\ (\bibinfo
  {organization} {Cambridge University Press},\ \bibinfo {year} {1930})\ pp.\
  \bibinfo {pages} {376--385}\BibitemShut {NoStop}%
\bibitem [{\citenamefont {Dovesi}\ \emph {et~al.}(2014)\citenamefont {Dovesi},
  \citenamefont {Orlando}, \citenamefont {Erba}, \citenamefont
  {Zicovich-Wilson}, \citenamefont {Civalleri}, \citenamefont {Casassa},
  \citenamefont {Maschio}, \citenamefont {Ferrabone}, \citenamefont
  {De~La~Pierre}, \citenamefont {D'Arco} \emph {et~al.}}]{Dovesi:2014tt}%
  \BibitemOpen
  \bibfield  {author} {\bibinfo {author} {\bibfnamefont {R.}~\bibnamefont
  {Dovesi}}, \bibinfo {author} {\bibfnamefont {R.}~\bibnamefont {Orlando}},
  \bibinfo {author} {\bibfnamefont {A.}~\bibnamefont {Erba}}, \bibinfo {author}
  {\bibfnamefont {C.~M.}\ \bibnamefont {Zicovich-Wilson}}, \bibinfo {author}
  {\bibfnamefont {B.}~\bibnamefont {Civalleri}}, \bibinfo {author}
  {\bibfnamefont {S.}~\bibnamefont {Casassa}}, \bibinfo {author} {\bibfnamefont
  {L.}~\bibnamefont {Maschio}}, \bibinfo {author} {\bibfnamefont
  {M.}~\bibnamefont {Ferrabone}}, \bibinfo {author} {\bibfnamefont
  {M.}~\bibnamefont {De~La~Pierre}}, \bibinfo {author} {\bibfnamefont
  {P.}~\bibnamefont {D'Arco}}, \emph {et~al.},\ }\bibfield  {title} {\bibinfo
  {title} {Crystal14: A program for the ab initio investigation of crystalline
  solids},\ }\href@noop {} {\bibfield  {journal} {\bibinfo  {journal} {Int. J.
  Quantum Chem.}\ }\textbf {\bibinfo {volume} {114}},\ \bibinfo {pages} {1287}
  (\bibinfo {year} {2014})}\BibitemShut {NoStop}%
\bibitem [{\citenamefont {Dovesi}\ \emph {et~al.}()\citenamefont {Dovesi},
  \citenamefont {Saunders}, \citenamefont {Roetti}, \citenamefont {Orlando},
  \citenamefont {Zicovich-Wilson}, \citenamefont {Pascale}, \citenamefont
  {Civalleri}, \citenamefont {Doll}, \citenamefont {Harrison}, \citenamefont
  {Bush} \emph {et~al.}}]{Dovesi:wq}%
  \BibitemOpen
  \bibfield  {author} {\bibinfo {author} {\bibfnamefont {R.}~\bibnamefont
  {Dovesi}}, \bibinfo {author} {\bibfnamefont {V.}~\bibnamefont {Saunders}},
  \bibinfo {author} {\bibfnamefont {C.}~\bibnamefont {Roetti}}, \bibinfo
  {author} {\bibfnamefont {R.}~\bibnamefont {Orlando}}, \bibinfo {author}
  {\bibfnamefont {C.}~\bibnamefont {Zicovich-Wilson}}, \bibinfo {author}
  {\bibfnamefont {F.}~\bibnamefont {Pascale}}, \bibinfo {author} {\bibfnamefont
  {B.}~\bibnamefont {Civalleri}}, \bibinfo {author} {\bibfnamefont
  {K.}~\bibnamefont {Doll}}, \bibinfo {author} {\bibfnamefont {N.}~\bibnamefont
  {Harrison}}, \bibinfo {author} {\bibfnamefont {I.}~\bibnamefont {Bush}},
  \emph {et~al.},\ }\href {http://www.crystal.unito.it/Manuals/crystal14.pdf}
  {\bibinfo {title} {Crystal14 user's manual; university of torino: Torino,
  italy, 2014}}\BibitemShut {NoStop}%
\bibitem [{cry()}]{crystal}%
  \BibitemOpen
  \href {http://www.crystal.unito.it/} {\bibinfo {title} {For program
  description, see http://www.crystal.unito.it/}}\BibitemShut {NoStop}%
\bibitem [{\citenamefont {Moser}(2017)}]{MOSER201729}%
  \BibitemOpen
  \bibfield  {author} {\bibinfo {author} {\bibfnamefont {S.}~\bibnamefont
  {Moser}},\ }\bibfield  {title} {\bibinfo {title} {An experimentalist's guide
  to the matrix element in angle resolved photoemission},\ }\href
  {https://doi.org/https://doi.org/10.1016/j.elspec.2016.11.007} {\bibfield
  {journal} {\bibinfo  {journal} {Journal of Electron Spectroscopy and Related
  Phenomena}\ }\textbf {\bibinfo {volume} {214}},\ \bibinfo {pages} {29}
  (\bibinfo {year} {2017})}\BibitemShut {NoStop}%
\bibitem [{fla()}]{flapwmbpt}%
  \BibitemOpen
  \href {https://www.bnl.gov/cmpmsd/flapwmbpt/} {\bibinfo {title} {For program
  description, see https://www.bnl.gov/cmpmsd/flapwmbpt/}}\BibitemShut
  {NoStop}%
\bibitem [{\citenamefont {Perdew}\ \emph {et~al.}(1996)\citenamefont {Perdew},
  \citenamefont {Burke},\ and\ \citenamefont
  {Ernzerhof}}]{perdew1996generalized}%
  \BibitemOpen
  \bibfield  {author} {\bibinfo {author} {\bibfnamefont {J.~P.}\ \bibnamefont
  {Perdew}}, \bibinfo {author} {\bibfnamefont {K.}~\bibnamefont {Burke}},\ and\
  \bibinfo {author} {\bibfnamefont {M.}~\bibnamefont {Ernzerhof}},\ }\bibfield
  {title} {\bibinfo {title} {Generalized gradient approximation made simple},\
  }\href@noop {} {\bibfield  {journal} {\bibinfo  {journal} {Physical review
  letters}\ }\textbf {\bibinfo {volume} {77}},\ \bibinfo {pages} {3865}
  (\bibinfo {year} {1996})}\BibitemShut {NoStop}%
\end{thebibliography}
%

\end{document}